\documentclass[10pt,acmsmall,nonacm]{acmart}

\usepackage{amsmath,amsthm,amsfonts}
\usepackage{mathtools}
\usepackage{thm-restate}
\usepackage[utf8]{inputenc}
\usepackage[ruled, lined, linesnumbered, commentsnumbered, longend]{algorithm2e}
\usepackage[noend]{algpseudocode}
\usepackage{graphicx}
\usepackage[english]{babel}
\usepackage{blindtext}
\usepackage{url}

\usepackage{caption}
\usepackage{subcaption}
\usepackage[utf8]{inputenc}
\usepackage{microtype}
\usepackage{enumitem}
\usepackage{scalerel}
\usepackage{tcolorbox}
\usepackage{ulem}
\usepackage{wrapfig}
\usepackage{xcolor}

\usepackage{cancel}
\usepackage{array}
\usepackage{makecell}

\definecolor{darkgreen}{rgb}{0.01, 0.75, 0.24}
\definecolor{darkcyan}{rgb}{0.0, 0.55, 0.55}

\newcommand{\deletion}[1]{}

\DeclareSymbolFont{extraup}{U}{zavm}{m}{n}
\DeclareMathSymbol{\varheart}{\mathalpha}{extraup}{86}
\DeclareMathSymbol{\vardiamond}{\mathalpha}{extraup}{87}

\mathchardef\hyphen="2D

\newtheoremstyle{sltheorem}
{}                
{}                
{\slshape}        
{}                
{\bfseries}       
{.}               
{ }               
{}                
\theoremstyle{sltheorem}

\newtheorem{definition}{Definition}
\newtheorem{corollary}{Corollary}
\newtheorem{Lemma}{Lemma}

\newcommand{\myitem}[1]{\vspace*{0.02in}\noindent\textbf{#1}}

\newcommand{\first}{\emph{(i)}\xspace}
\newcommand{\second}{\emph{(ii)}\xspace}
\newcommand{\third}{\emph{(iii)}\xspace}

\newcommand{\ie}{i.e., \@}
\newcommand{\eg}{e.g., \@}

\newcommand{\name}{\textsc{Mars}\xspace}

\newcommand{\NT}{\ensuremath{n_t}\xspace}
\newcommand{\NU}{\ensuremath{n_u}\xspace}
\newcommand{\NP}{\ensuremath{n_p}\xspace}
\newcommand{\NS}{\ensuremath{n_s}\xspace}

\newcommand{\timedP}{\ensuremath{\delta}\xspace}
\newcommand{\static}{\ensuremath{\mathrm{static}}\xspace}
\newcommand{\temporal}{\ensuremath{\mathrm{temporal}}\xspace}
\newcommand{\delay}{\ensuremath{L}\xspace}

\newcommand{\timeslot}{\ensuremath{\Delta}\xspace}
\newcommand{\reconf}{\ensuremath{\Delta_u}\xspace}

\newcommand{\length}{\ensuremath{n}\xspace}
\newcommand{\etal}{et al.}

\DeclareMathOperator{\len}{len}
\DeclareMathOperator{\arl}{ARL}
\DeclareMathOperator{\ard}{ARD}

\setcopyright{none}
\acmDOI{}
\acmISBN{}

\makeatletter
\let\@authorsaddresses\@empty
\makeatother

\begin{document}

\title{\name: Near-Optimal Throughput with Shallow Buffers \\in Reconfigurable Datacenter Networks}
\titlenote{Authors version. Final version of the paper to appear in ACM SIGMETRICS 2023. This work is part of a project that has received funding from the European Research Council (ERC) under the European Union’s Horizon 2020 research and innovation programme, consolidator project Self-Adjusting Networks (AdjustNet), grant agreement No. 864228, Horizon 2020, 2020-2025.}

\author{Vamsi Addanki}
\affiliation{
  \institution{Faculty of Electrical Engineering and Computer Science, TU Berlin, Germany}
}
\author{Chen Avin}
\affiliation{%
  \institution{School of Electrical and Computer Engineering,  Ben-Gurion University of the Negev, Israel}
}
\author{Stefan Schmid}
\affiliation{%
  \institution{Faculty of Electrical Engineering and Computer Science, TU Berlin, Germany}
}

\renewcommand{\shortauthors}{Addanki.et al.}
\renewcommand{\shorttitle}{\name: Near-Optimal Throughput RDCN with Shallow Buffers}

\sloppy


\begin{abstract}
The performance of large-scale computing systems often critically depends on high-performance communication networks. 
Dynamically reconfigurable topologies, e.g., based on optical circuit switches, are emerging as an innovative new technology to deal with the explosive growth of datacenter traffic.
Specifically, \emph{periodic} reconfigurable datacenter networks (RDCNs) such as RotorNet (SIGCOMM 2017), Opera (NSDI 2020) and Sirius (SIGCOMM 2020) have been shown to provide high throughput, by emulating a \emph{complete graph} through fast periodic circuit switch scheduling.

However, to achieve such a high throughput, existing reconfigurable network designs pay a high price: in terms of potentially high delays, but also, as we show as a first contribution in this paper, in terms of the high buffer requirements.  In particular, we show that under buffer constraints, emulating the high-throughput complete graph is infeasible at scale, and we uncover a spectrum of unvisited and attractive alternative RDCNs, which emulate regular graphs, but with lower node degree than the complete graph.

We present \name, a periodic reconfigurable topology which emulates a $d$-regular graph with near-optimal throughput.
In particular, we systematically analyze how the degree~$d$ can be optimized for throughput given the available buffer and delay tolerance of the datacenter. 
We further show empirically that \name achieves higher throughput compared to existing systems when buffer sizes are bounded.
\end{abstract}

\maketitle
\thispagestyle{plain}
\pagestyle{plain}

\section{Introduction}

With the popularity of data-centric and distributed applications, the traffic in datacenters is growing explosively. Dealing with this traffic however becomes increasingly challenging: while cloud traffic roughly doubles each year~\cite{jupiter}, the capacity increase provided by electrical switches for a given power and cost starts to lag behind. This gap is expected to worsen with the current trend to hardware-driven workloads such as distributed machine learning training~\cite{sirius}. 

As the \emph{throughput} of datacenter networks is becoming more and more critical for application performance, over the last years, great efforts have been made to increase the capacity of datacenter topologies. A particularly innovative architecture to meet the stringent bandwidth requirements of modern datacenters, are \emph{reconfigurable} (optical) datacenter networks~(RDCNs)
~\cite{mellette2017rotornet,opera,sirius, helios, firefly, chen2014osa,projector,cthrough,splaynet,proteus},
e.g., based on optical circuit switches, tunable lasers, and simple passive gratings~\cite{osn21}.
By quickly cycling through a sequence of different topologies---typically matchings between top-of-rack (ToR) switches---RDCNs such as RotorNet~\cite{mellette2017rotornet}, Opera~\cite{opera} or Sirius~\cite{sirius} can provide periodic direct connectivity between 
rack pairs, at microsecond or even nanosecond granularity. 
A common property of these systems is that they emulate a \emph{complete graph}, and so   
avoid the ``bandwidth tax'' of multi-hop forwarding~\cite{sigmetrics22cerberus,opera}. Indeed, empirical studies show that periodic reconfigurable datacenter topologies can achieve significantly higher throughput compared to cost-equivalent traditional datacenters based on \emph{static} topologies~\cite{mellette2017rotornet,sirius,opera}. 

\medskip
\emph{This paper is motivated by the observation that the  existing approach of using reconfigurable technologies to emulate complete graphs  
comes at a price: long delays and large buffer requirements.}

\medskip
First, at scale, emulating a complete graph can entail \emph{long delays}: the denser the emulated network, the longer the periodic reconfiguration cycle and hence the longer a given rack pair has to wait to be connected again. Second, as we show analytically in this paper, the resulting long reconfiguration cycles require excessive buffering at the ToR switches and end-hosts. 
The required buffer can intuitively be viewed as \emph{bandwidth-delay} product of dynamic topologies similar to the corresponding notion for static topologies in TCP literature. 
In practice though, datacenter switches are equipped with shallow buffers. Further, several studies in the recent past show an increasing gap between switch capacity and buffer sizes~\cite{9288775,276958}. 

\medskip
\emph{Our main insight in this paper is that accounting for buffer constraints can
significantly change the design considerations of reconfigurable datacenter networks.}

\medskip
In particular, we initiate the study of---and make the case for---reconfigurable networks which emulate graphs \emph{of lower node degree}, uncovering an entire spectrum of possible topology designs.
We present a systematic and formal analysis of the design spectrum and the tradeoffs that these topologies introduce in terms of throughput, delay, and buffer requirements. Our perspective and the main tradeoffs are intuitively visualized in Figure~\ref{fig:tradeoffs} (left), where on the x-axis (at the bottom) we show the topologies according to the node  degree of the graph they emulate. The x-axis (at the top) also represents increasing delay from left to right. The y-axis shows the throughput of the system from low to~high. 
Let us elaborate on the important points in the figure (summarizes in the table on the right):

\begin{figure}[t]
\centering
\begin{tabular}{cc}
\includegraphics[width=0.54\linewidth]{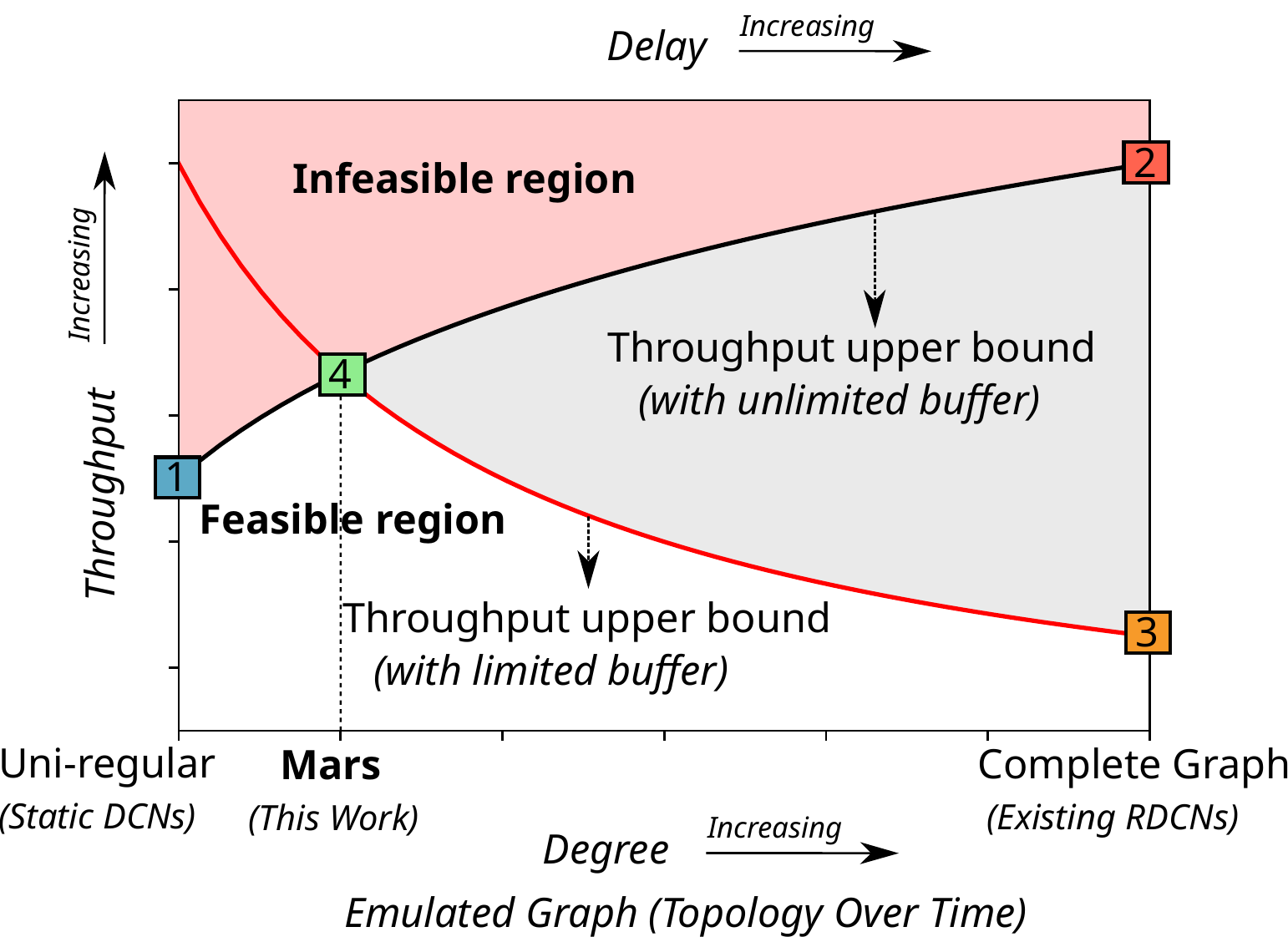}
& 
\footnotesize
\raisebox{3cm}{
\begin{tabular}{|c|c|c|c|}
\hline
\textbf{Topo.} & \textbf{Throughput} & \textbf{Delay} & \textbf{Buffer}\\
\hline
\noindent \tcbox[on line,boxsep=0pt,left=0pt,right=0pt,top=0pt,bottom=0pt,colback=white]{\includegraphics[height=0.25cm]{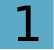}} & \makecell{Low \\ (Optimal)} & Low & Low \\ 
\hline
\noindent \tcbox[on line,boxsep=0pt,left=0pt,right=0pt,top=0pt,bottom=0pt,colback=white]{\includegraphics[height=0.25cm]{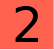}} & \makecell{High \\ (Optimal)} & High & High \\
\hline
\noindent \tcbox[on line,boxsep=0pt,left=0pt,right=0pt,top=0pt,bottom=0pt,colback=white]{\includegraphics[height=0.25cm]{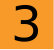}} & \makecell{Low \\ (Not optimal)} &  High & Low \\
\hline \hline
\noindent \tcbox[on line,boxsep=0pt,left=0pt,right=0pt,top=0pt,bottom=0pt,colback=white]{\includegraphics[height=0.25cm]{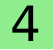}} & \makecell{\textbf{High} \\ \textbf{(Optimal)} } & \textbf{Moderate} & \textbf{Any} \\
\hline
\end{tabular}}
\end{tabular}
\caption{Periodic reconfigurable datacenter topologies pose fundamental tradeoffs across throughput, delay and buffer requirements. Existing designs are at the extremes of a spectrum of optimal designs with lower delay and buffer requirements.}
\label{fig:tradeoffs}
\end{figure}

\medskip
\noindent \tcbox[on line,boxsep=0pt,left=0pt,right=0pt,top=0pt,bottom=0pt,colback=white]{\includegraphics[height=0.3cm]{figures/1.pdf}} \myitem{Static DCNs:} On the very left of the design space are traditional static datacenter networks referred as uni-regular topologies~\cite{tubSigcomm2021} \eg expander based DCNs~\cite{valadarsky2016xpander,singla2012jellyfish,slimfly}\footnote{These works also claimed that static expander-based topologies are similar or better designs (in terms of performance and cost) compared to Clos-based topologies. We henceforth focus on static expanders. A detailed discussion can be found in \S\ref{sec:discussion}.}.
Such a design in principle incurs the lowest delay and requires the least amount of buffer to achieve its ideal throughput, i.e., the optimal throughput under low delay tolerance. However, since the topology remains static, for scalability reasons, each ToR switch can only connect to a limited number of other ToR switches.
This results in long multi-hop paths and hence a high ``bandwidth tax''~\cite{sigmetrics22cerberus}, and lower throughput.

\medskip
\noindent\tcbox[on line,boxsep=0pt,left=0pt,right=0pt,top=0pt,bottom=0pt,colback=white]{\includegraphics[height=0.3cm]{figures/2.pdf}} \myitem{Existing periodic RDCNs:} To reduce the ``bandwidth tax'' and to achieve high throughput, existing designs resort to emulating a complete graph where each rack-pair is connected directly once in every matching cycle~\cite{mellette2017rotornet,sirius}. Such designs achieve high throughput, in fact maximum across all topologies in 
our design space, i.e., the optimal throughput. However, as we will show in this paper, the high throughput offered by emulating a complete graph comes at the cost of high delay and is subject to the availability of large buffers. 

\medskip
\noindent\tcbox[on line,boxsep=0pt,left=0pt,right=0pt,top=0pt,bottom=0pt,colback=white]{\includegraphics[height=0.3cm]{figures/3.pdf}} \myitem{Existing periodic RDCNs under resource constraints: } Given the large buffer requirements of existing designs (emulating a complete graph), we study their throughput under limited buffer. Interestingly, we find that, existing periodic reconfigurable datacenter topologies may perform equally or worse compared to a static uni-regular topology in terms of throughput, when buffer  sizes are bounded.

\medskip
\noindent\tcbox[on line,boxsep=0pt,left=0pt,right=0pt,top=0pt,bottom=0pt,colback=white]{\includegraphics[height=0.3cm]{figures/4.pdf}} \myitem{\name: } Exploiting the fundamental tradeoffs across the design space, we propose \name, a periodic reconfigurable topology that provides near optimal, high throughput with the limited amount of available buffer. Specifically, we parametrize our design based on the delay tolerance and available buffer. We systematically determine the optimal degree $d$ which depending on the resource constraints lies between a static topology and a complete graph.

\medskip
It is interesting to observe from Figure~\ref{fig:tradeoffs} that the throughput-delay relation implies an \emph{infeasible} region for topology design (shown in red shade). The available buffer space further restricts the design space (shown in gray shade), imposing a fundamental tradeoff across the topologies within the feasible region in terms of throughput, delay and buffer.

\medskip
Our analytical approach in this paper is novel and  relies on a reduction of a periodic evolving graph to a specific static graph (Theorem \ref{th:evolving-union}). This enables us to study the throughput maximization problem using well-known graph analysis techniques for static graphs and to analytically evaluate the throughput of both existing dynamic topologies (RotorNet, Opera, and Sirius) as well as possible alternatives (Mars).
We believe that this reduction technique may be of independent interest and could be potentially used to study other properties of dynamic topologies.

\medskip
In summary, our key contributions in this paper are:
\begin{itemize}
\item We provide a throughput-centric view and formal model of the performance of periodic reconfigurable datacenter topologies. In particular, we analytically derive the relation between throughput, delay, and the required amount of buffer in the network which reveals a non-trivial tradeoff in the design of periodic reconfigurable topologies.

\item We present \name, a novel reconfigurable datacenter design that maximizes throughput under limited buffer and delay requirements.

\item  We report on an extensive evaluation showing that \name improves throughput by up to $4$x compared to existing approaches when buffer sizes are bounded. Our evaluation also shows that \name improves the $99$-percentile flow completion time for short flows by up to $75\%$ compared to the state-of-the-art approach~\cite{opera} and by up to $87\%$ compared to Sirius and RotorNet.
\end{itemize}

For ease of presentation and due to space constraints, we explain the main concepts of the paper intuitively, and defer the formal definitions and detailed proofs to the appendix.

\section{Preliminaries \& Model}
\label{sec:model}
We first give a brief background on periodic reconfigurable topologies and introduce our model.

\subsection{Periodic Reconfigurable Topologies}
\label{sec:background-periodic}

The key enablers and building blocks for fast periodic reconfigurations are optical circuit switching technologies such as rotor-switches~\cite{mellette2017rotornet}, AWGR gratings~\cite{cheung2013ultra} and tunable lasers~\cite{sirius,coldren2004tunable}. 
Such technologies are oblivious to the demand, thus avoiding the overhead of measuring and estimating the demand. 
For instance, the overhead is in the order of microseconds for RotorNet and nanoseconds for Sirius which periodically reconfigure the network topology. 

\begin{figure}[!h]
\centering
\includegraphics[width=0.7\linewidth]{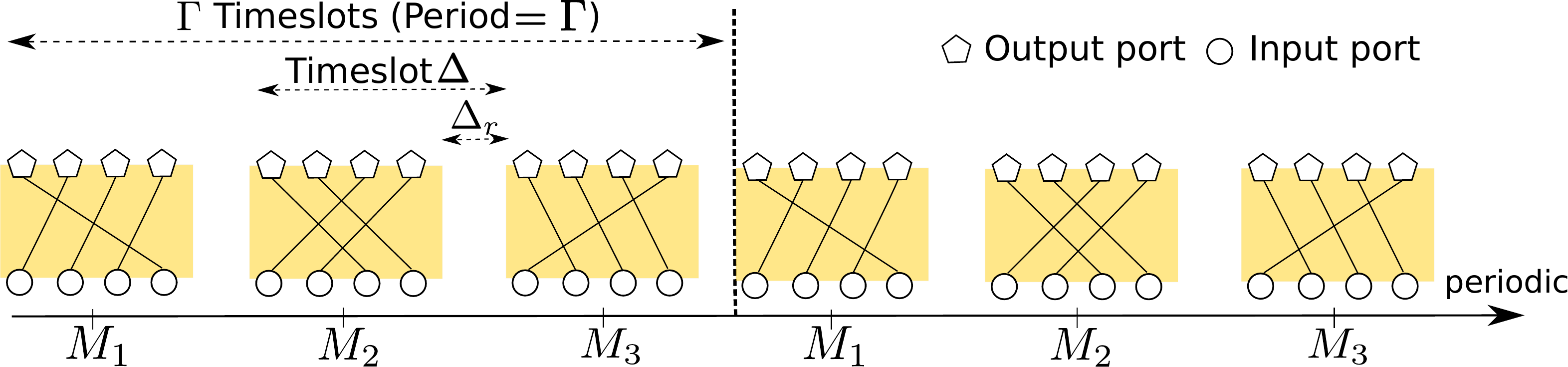}
\caption{The periodic sequence of matchings of an optical circuit switch.}
\label{fig:matchings}
\end{figure}

\medskip
\myitem{Circuit switch model:} As a unified model of periodic circuit switching technologies, a \emph{rotor-switch} serves as a building block for us in the rest of the paper. 
Rotor-switches perform \emph{circuit switching} and implement a sequence of matchings in a periodic schedule with period $\Gamma$ where a matching is a map from the switch's input ports to its output ports. The packets arriving at an input port of a rotor-switch are forwarded to the matched output port. Specifically, rotor-switches do not process packets and the forwarding is dictated by the matching at any given instance of time. Figure~\ref{fig:matchings} shows an example of a rotor-switch with four input and output ports with a sequence of three matchings.
We use the term \emph{timeslot} denoted by \timeslot to refer to the total time spent by rotor-switch in each matching including the reconfiguration time. We denote the reconfiguration time by $\Delta_r$. The utilization time (when traffic is sent) is then $\Delta-\Delta_r$. In essence, a rotor-switch pays a ``latency tax'' of $\reconf=\frac{\Delta_r}{\Delta}$.

\begin{figure*}
    \centering
    \begin{tabular}{cc}

    \begin{subfigure}{0.41\linewidth}
    \centering
    \includegraphics[width=.99\linewidth]{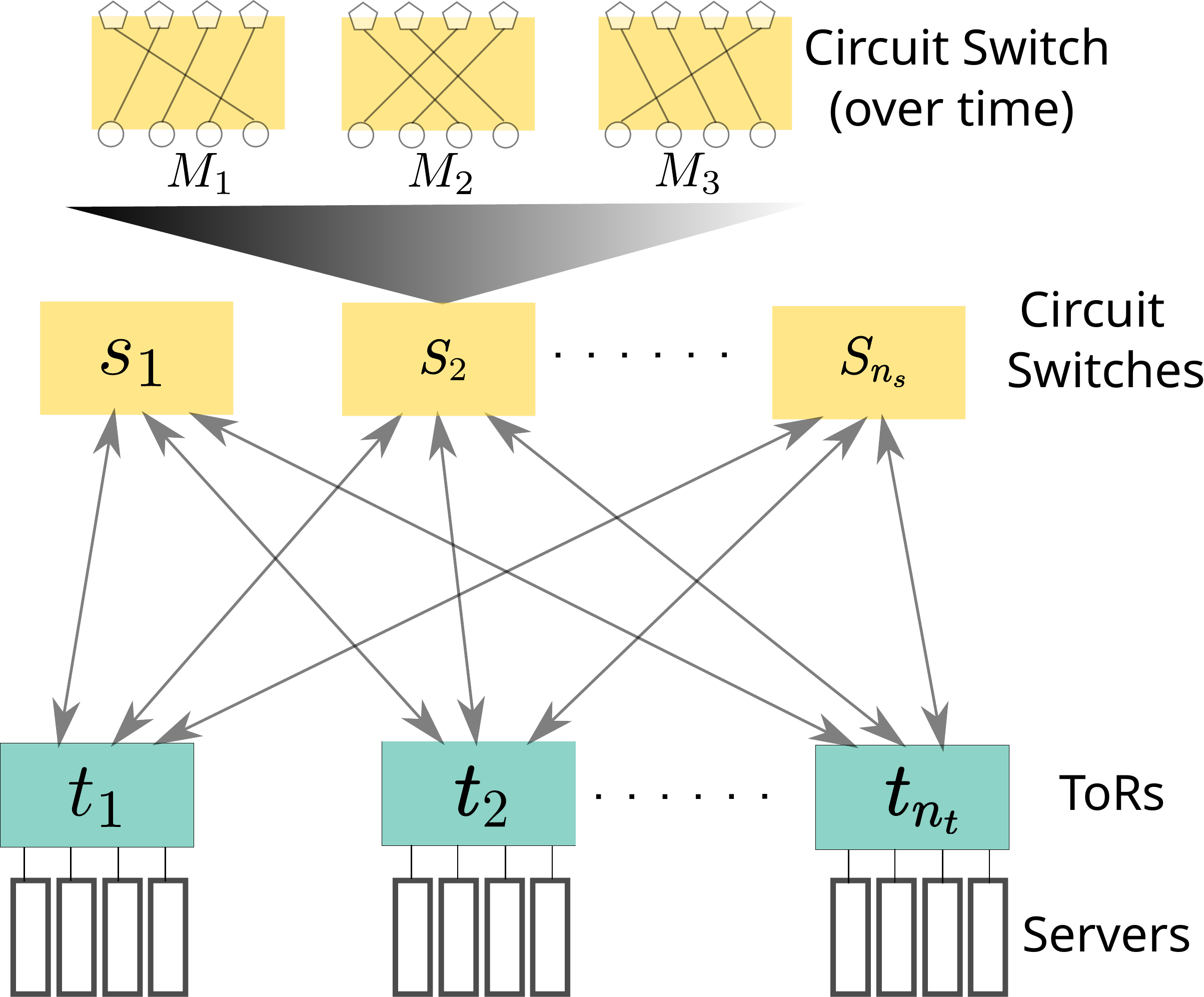}
    \caption{A 2-tier periodic RCDN.}
    \label{fig:typical-topo}
    \end{subfigure}
    &
    \begin{subfigure}{0.57\linewidth} 
    \includegraphics[width=1\linewidth]{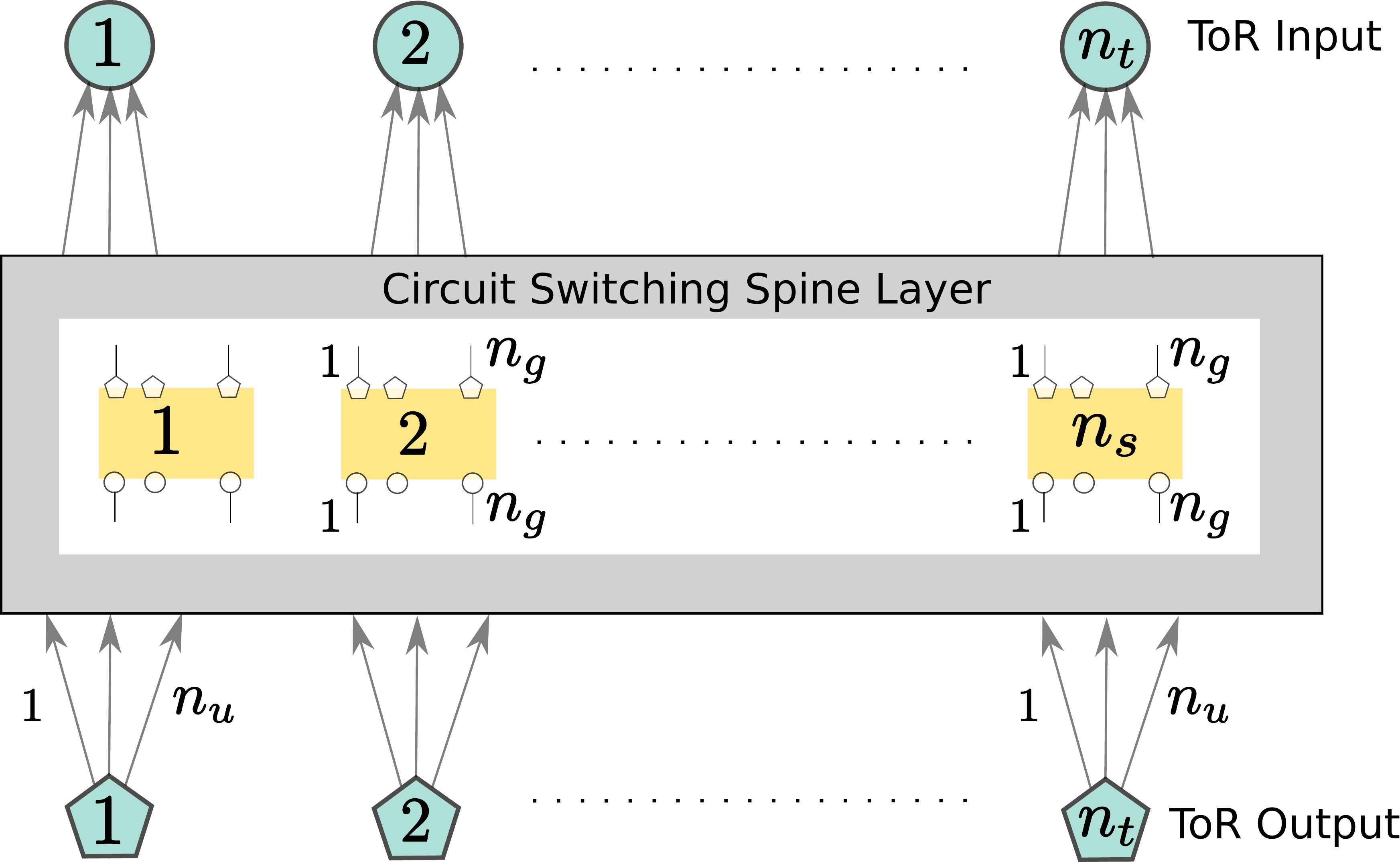}
    \caption{An abstract view of periodic RCDN topologies.}
    \label{fig:generalized-topo}
    \end{subfigure}
    \end{tabular}
    \caption{Illustration of periodic RCDN model we consider in this paper, consisting of \NU ToR switches each with \NU uplinks. The datacenter is interconnected by a circuit switching spine layer consisting of \NS circuit switches, each with \NP input and output ports.}
    \label{fig:topology}
\end{figure*}

Note that, newer proposals such as Sirius~\cite{sirius} use optical gratings and tunable lasers to achieve fast periodic reconfigurations at nanosecond scale. In contrast to rotor-switches, Sirius uses optical gratings which are the building blocks and the ToR switches (or end-hosts) connect to the gratings via tunable lasers which tune the wavelength of emitted light periodically. However, both RotorNet and Sirius logically capture a rotor-switch that rotates periodically across a fixed number of matchings. To this end, Sirius can be abstracted as a topology with rotor-switches and the difference arises in the system level parameters such as reconfiguration time $\Delta_r$. We do not model system level parameters in this paper and only focus on the topological aspects.
In the following, we will sometimes use the more general term \emph{optical switch} when referring to the \emph{rotor-switch}.

\medskip
\myitem{Topology:} 
Following the literature on existing proposals~\cite{mellette2017rotornet,sirius,opera}, we consider 2-tier reconfigurable datacenter topologies with a set of ToR switches (leaf layer) and a set of circuit switches (spine layer) as shown in Figure~\ref{fig:typical-topo}, where each circuit switch functions as a rotor-switch. The uplinks of the ToR switches connect to the spine layer which interconnects the datacenter\footnote{Each ToR uplink is a set of two SERDESes~\cite{sirius,serdes} which are uni-directional corresponding to ToR output (pentagons) and input (circles) as depicted in Figure~\ref{fig:generalized-topo}.}. Concretely, as illustrated in Figure~\ref{fig:generalized-topo}, we consider a periodic reconfigurable datacenter with \NT ToR switches each with \NU uplinks; \NS circuit switches each with \NP input and output ports. We assume that a link $e$ in the topology has capacity $c(e)$.
We will later show that this generalized view reveals an entire spectrum of topologies where the existing systems (\eg RotorNet and Sirius) are special instances of the spectrum which emulate a complete graph \ie each ToR connects to every other ToR in a period.

\subsection{Graph Theoretic Model of Periodic ToR-to-ToR Connectivity}\label{sec:evolving-model}

In contrast to packet switched networks, circuit switched networks do not buffer packets. To this end, since periodic reconfigurable topologies are circuit switched, we are mainly interested in the connectivity between the end points of the circuit switched network, namely the ToR-to-ToR connectivity. Hence, circuit availability between a pair of ToR switches can be considered as a direct link between the pair. In the following, we model the periodically reconfigurable ToR-to-ToR connectivity over time as a \emph{periodic evolving graph}.

\begin{figure*}
    \centering
    \begin{subfigure}{0.66\linewidth}
    \centering
    \includegraphics[width=1\linewidth]{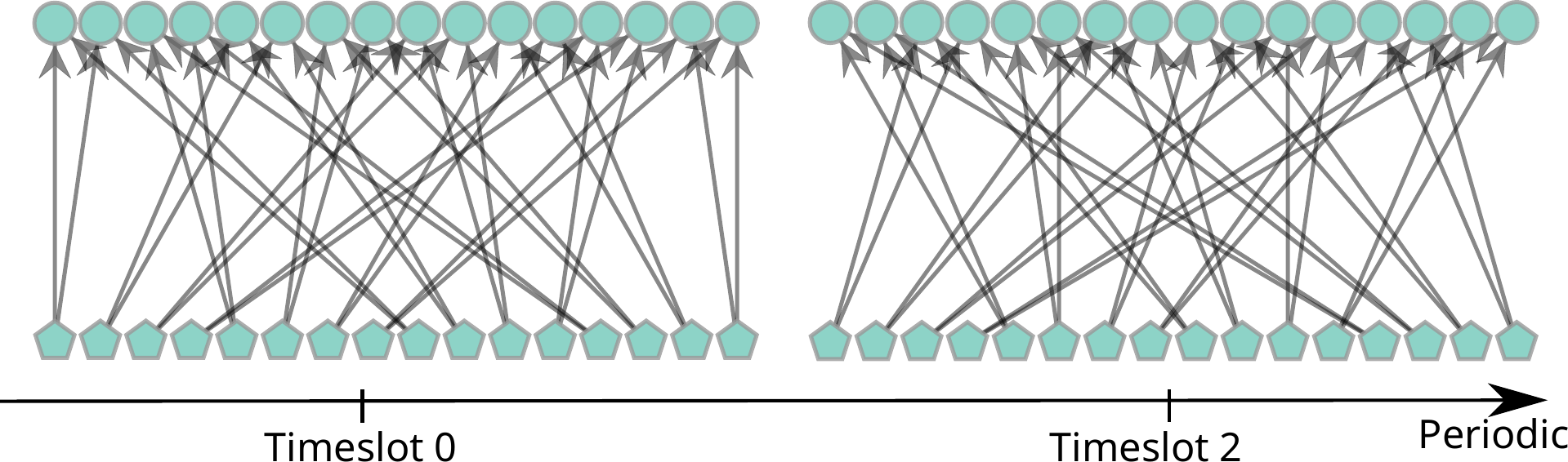}
    \caption{Example of a periodic evolving graph with period $\Gamma=2$ depicting the ToR-to-ToR connectivity over time; $\NT=16$, $\NU=2$, $\NS=2$ and $\NP=16$.  }
    \label{fig:evolving}
    \end{subfigure}\hfill
    \begin{subfigure}{0.31\linewidth}
    \centering
    \includegraphics[width=1\linewidth]{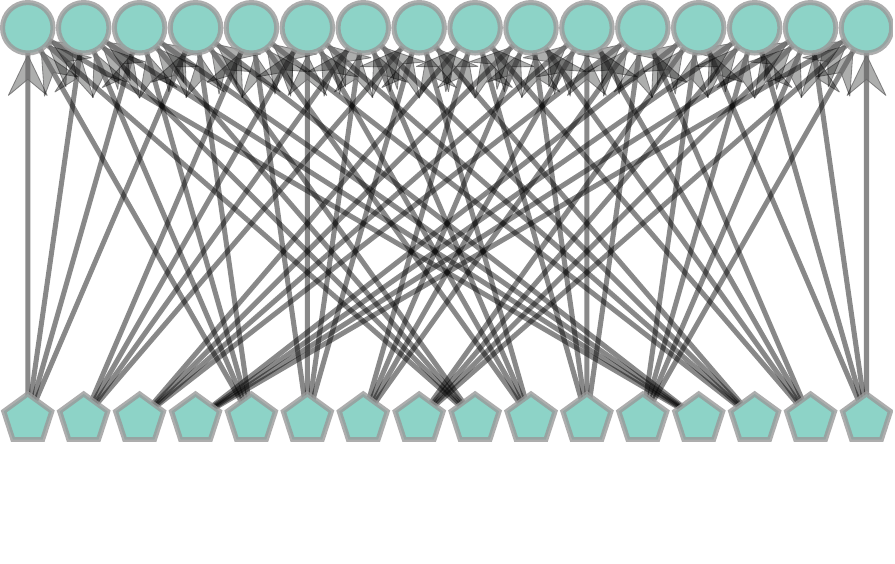}
    \caption{The corresponding \emph{Emulated} graph (static) over time.}
    \label{fig:uniongraph}
    \end{subfigure}
    \caption{Given the number of ToR switches \NT with uplinks \NU and a periodic sequence of matchings corresponding to each circuit switch, the topology is a periodic evolving graph which \emph{emulates} a static graph over~time.}
    \label{fig:evolving-union}
\end{figure*}

\medskip
\myitem{Periodic evolving graph:} Consider the periodic reconfigurable topology described in \S\ref{sec:background-periodic} (shown in Figure~\ref{fig:generalized-topo}). ToR switches connect to the optical switches via uplinks. The optical spine layer in turn establishes a circuit between ToR pairs \emph{periodically} as shown in Figure~\ref{fig:evolving} since the optical switches reconfigure periodically according to a fixed schedule. 
To this end, we represent the ToR-to-ToR connectivity as \emph{periodic evolving graph} denoted by $\mathcal{G} = (V,\mathcal{E})$. Specifically, $\mathcal{G}$ is a periodic sequence of directed graphs defined for timeslots $t\in [0,\infty)$ where each timeslot is of $\Delta$ duration. 
We denote the graph at time $t$ as $\mathcal{G}_t=(V,\mathcal{E}_t)$, where~$V$ is the set of all ToR switches and the edge set $\mathcal{E}_t$ represents the circuit availability between the ToR switches at time~$t$. The sequence of edge sets $\mathcal{E}_t$ and consequently the evolving graph $\mathcal{G}$ are periodic with period of $\Gamma$~timeslots. The period in turn relates to the periodic switching schedule of the optical switch. We denote the capacity of circuit between ToR pairs at any time $t$ as $c_t(e)$ for all $e\in \mathcal{E}_t$. We explicitly set $c_t(e)=0$ if $e\not\in\mathcal{E}_t$. 
Further, we model the ``latency tax'' $\reconf$ due to reconfigurations by limiting the access to an edge $e\in \mathcal{E}_t$ to only a $(1-\reconf)$ fraction of a timeslot (from the start of every timeslot).
This model serves as an entry point for our formal analysis of throughput, delay,  and buffer requirements of reconfigurable topologies in the next section.

\section{Motivation: Fundamental Tradeoffs of Periodic RDCNs}\label{sec:motivation}
We now provide a more detailed motivation for our work. Specifically, we analytically derive the throughput of periodic RDCNs (\S\ref{sec:prdcn-throughput}) and its relation to delay (\S\ref{sec:prdcn-delay}). We further study the buffer requirements (\S\ref{sec:prdcn-buffer}) and highlight that the topologies introduce a fundamental tradeoff between throughput, delay and buffer. We then present remarks and practical implications of our results~(\S\ref{sec:faq}). Finally, we discuss the optimization opportunities in the design of periodic reconfigurable topologies (\S\ref{sec:prdcn-tradeoff}).
\subsection{Throughput of Periodic RDCNs}
\label{sec:prdcn-throughput}

We first study the throughput of periodic reconfigurable topologies. Specifically, we prove that the throughput of a periodic evolving graph is equivalent to that of a unique static graph. Following this result, we re-investigate the throughput of static graphs and obtain a more general result which is in line with the definitions and intuitions discussed in the literature. Our result asserts that the throughput of periodic reconfigurable topologies as well as static topologies is inversely proportional to \textit{average route length} as opposed to shortest path lengths used in the most recent work~\cite{tubSigcomm2021}.

In the following, we briefly introduce our definitions in the context of periodic evolving graphs.
Prior works~\cite{jyothi2016measuring,highthroughputSingla,tubSigcomm2021} define throughput of a static topology given a demand matrix $\mathcal{M}$ as the maximum scaling factor $\theta$ such that there exists a feasible flow that satisfies the scaled demand $\theta\cdot \mathcal{M}$. A more formal definition appears in Appendix~\ref{sec:motivation-throughput}.
While prior works studied the throughput of static topologies, we are not aware of a formal definition and study of the throughput of general dynamic topologies. Initial studies on dynamic topologies informally define throughput of existing systems (\eg RotorNet) which emulate a complete graph, specifically in terms of demand completion times~\cite{sigmetrics22cerberus}.  
We emphasize that our definitions and results in this paper hold for \emph{any} periodic reconfigurable topology.
Next, we define the throughput of dynamic topologies using the notion of \emph{temporal flows} which are formally defined in Appendix~\ref{sec:throughput-reconf}.

\medskip
\noindent
\myitem{Demand matrix $\mathcal{M}$:} Given a set of vertices $V$, a demand matrix specifies the demand rate between every pair of vertices in bits per second defined as $\mathcal{M}=\{ m_{u,v} \mid u\in V, v\in V \}$ where $m_{u,v}$ is the demand between the pair $u,v$. A saturated demand matrix is such that the total demand originating at a source~$s$ equals its outgoing capacity and the total demand terminating at a destination~$d$ equals its incoming capacity. We consider \emph{saturated} demand matrices since we focus on the maximum achievable throughput. Specifically, saturated demand matrices allow for studying the maximum demand that can be routed in a topology within the capacity constraints.

\medskip
\noindent
\myitem{Legal flow:} A legal flow in a static graph obeys capacity constraints and satisfies flow conservation at all times (Appendix \ref{sec:motivation-throughput}, Definition~\ref{def:flow}). In the evolving graph, a legal \emph{temporal} flow (Appendix \ref{sec:throughput-reconf}, Definition~\ref{def:time-flow}) is subject to capacity constraints at all times and conserves flow in every period ($\Gamma$ timeslots).

\medskip
\myitem{Throughput $\theta(\mathcal{M},\mathcal{F})$ of a flow:}
Following the seminal works on throughput~\cite{shahrokhi1990maximum,leighton1989approximate,10.1145/331524.331526}\footnote{The maximum concurrent flow problem~\cite{shahrokhi1990maximum} considers that the demand between each pair $s,d$ is equal. We relax this assumption for the throughput maximization problem in our networking context similar to prior work~\cite{jyothi2016measuring}.}, given a demand matrix $\mathcal{M}$ specified in bits per second, we define the throughput achieved by a \emph{temporal flow} $\mathcal{F}$ in a periodic evolving graph as the scaling factor $\theta(\mathcal{M},\mathcal{F})$ such that the flow $\mathcal{F}$ satisfies the scaled demand matrix $\theta(\mathcal{M},\mathcal{F})\cdot\mathcal{M}$ where $\mathcal{F}$ is subject to capacity constraints at all times and flow conservation in every period ($\Gamma$ timeslots). We assume that certain amount of buffer is available at each node to hold traffic until it can be forwarded to the next hop or destination. 

\medskip
\myitem{Throughput $\theta(\mathcal{M})$ of the graph for a~demand~matrix:}
Since we are interested in the ideally achievable throughput, following prior work~\cite{jyothi2016measuring,tubSigcomm2021,highthroughputSingla}, we define the throughput of a periodic evolving graph as follows: given a demand matrix~$\mathcal{M}$, throughput is the maximum scaling factor $\theta(\mathcal{M})$ such that the scaled demand matrix $\theta(\mathcal{M})\times \mathcal{M}$ is feasible in the network \ie $\theta(\mathcal{M})$ is the maximum throughput across all feasible flows: $\theta(\mathcal{M}) = \max_{\mathcal{F}} \theta(\mathcal{M},\mathcal{F})$. 

\medskip
\myitem{Throughput $\theta^*$ of the graph:}
The \emph{throughput} $\theta^*$ is the maximum scaling factor $\theta(\mathcal{M})$ for a worst-case demand matrix \ie $\theta^*$ is the minimum throughput across all the \emph{saturated} demand matrices: $\theta^* = \min_{\mathcal{M}} \theta(\mathcal{M})$. In essence, the network can always support $\theta^*\cdot \mathcal{M}$ demand for \emph{any} saturated demand matrix, where a saturated 
demand matrix is such that the total demand originating at any source, and received by any destination equals the total node capacity.

\medskip
We study the throughput of periodic RDCNs in relation to the graph emulated by the periodic topology which we call the \emph{emulated graph}, formally defined below.

\medskip
\myitem{Emulated graph:} Consider a periodic evolving graph (see \S\ref{sec:evolving-model}) $\mathcal{G}=(V,\mathcal{E})$ with edge capacities $c_t(e)$ for every edge~$e$ at time~$t$. Consider a static (directed, multi, edge-labeled) graph which is a function of the evolving graph $\mathcal{G}$ defined as $G(\mathcal{G})=(V,E)$, where $E= \{ (e,\ell)\mid e\in \mathcal{E}_\ell,\ \ell\in[0,\Gamma) \}$ is the union of edges of the evolving graph over one period of time-interval and the labels correspond to the time in the evolving graph. Let the capacity of an edge $(e,\ell)\in E$ in the static graph be $\hat{c}(e,\ell)$ where $\hat{c}(e,\ell) = \frac{(1-\reconf)}{\Gamma}\cdot c_\ell(e)$. We refer to this static graph as \emph{emulated graph}. Figure~\ref{fig:uniongraph} illustrates an example of the emulated graph corresponding to the periodic evolving graph in Figure~\ref{fig:evolving}. Our formal definition appears in Appendix~\ref{sec:union-graph}. 

\medskip
Our first main result, which is the basis for our methodology, is that 
the throughput  of a periodic evolving graph is equivalent to the throughput of its corresponding emulated graph.

\begin{restatable}[Relation to Emulated Graph]{theorem}{relationtoStatic}\label{th:evolving-union}
Given a demand matrix $\mathcal{M}$, any legal temporal flow $\mathcal{F}$ in the periodic evolving graph $\mathcal{G}$ with throughput $\theta(\mathcal{M},\mathcal{F})$ can be converted to a legal flow $F$ in the emulated graph $G(\mathcal{G})$ with the same throughput $\theta(\mathcal{M},F)=\theta(\mathcal{M},\mathcal{F})$ and vice versa. Consequently the periodic evolving graph and the emulated graph have the same throughput $\theta(\mathcal{M}) = \max_{F} \theta(\mathcal{M},F)$: the maximum scaling factor given a demand matrix $\mathcal{M}$; they further have the same throughput $\theta^*$ for a worst-case demand matrix, where $\theta^* = \min_{\mathcal{M}} \theta(\mathcal{M})$.
\end{restatable}

\textit{Proof sketch.} We present the formal proof of Theorem~\ref{th:evolving-union} in Appendix~\ref{sec:throughput-analysis} and provide a proof sketch here. 
We first define what we call \emph{extended paths} for the static emulated graph that allows a conversion between paths in the periodic evolving graph and the corresponding static emulated graph. While the standard definition of path in a static graph is a sequence of edges connecting a source and a destination, our extended path definition differs in two aspects. Given a temporal path in the periodic evolving graph, we define the extended path in the static emulated graph as the sequence of edges traversed in the temporal path where each edge is associated with a label corresponding to the time when the edge was accessed in the temporal path. Further, each extended path consists of a unique identifier. Since the evolving graph is periodic, we mainly focus on the foundation set of temporal paths that start in the first period. As a result, there exists a one-to-one mapping between the foundation set of temporal paths in the periodic evolving graph and the extended set of paths in the static emulated graph. Exploiting this relation, we prove the following: \first we prove in Lemma~\ref{lem:static-to-evolving} that if a legal flow in the emulated graph has throughput $\theta$ over the set of extended paths, then there exists a legal temporal flow in the periodic evolving graph with the same throughput; \second we prove in Lemma~\ref{lem:evolving-to-static} that if a legal temporal flow achieves throughput $\theta$ in the evolving graph, then there exists a legal flow in the emulated graph with the same throughput over the set of extended paths; \third finally, we prove in Lemma~\ref{lem:extendedJustif} that a flow in the emulated graph over the set of \emph{all} paths has the same throughput as that of a flow over the set of extended paths. Using the above three results, we prove our main claim in Theorem~\ref{th:evolving-union}.

\medskip
\textit{Discussion.} Theorem~\ref{th:evolving-union} enables us to study the throughput of periodic reconfigurable topologies with the techniques used in static graphs. 
We emphasize that our result is \emph{general} for periodic reconfigurable topologies and is not specific to the existing systems; for instance the emulated graph could indeed be an expander graph, or even a non-regular graph, but our result still holds. 
We believe that our proof renders useful not only to study throughput maximization problem but also other variants of flow problems. For example, by using our technique to convert flows from the static to the periodic graph and vice versa one could easily derive a relation between the metric of interest in each graph. We leave this exploration for future work.

We now shift our focus to obtain the throughput upper bound of static topologies for two reasons. First, from Theorem~\ref{th:evolving-union}, the throughput of a periodic reconfigurable topology is equivalent to its corresponding emulated graph (a \emph{static graph}). Second, the best known throughput upper bound~\cite{tubSigcomm2021} for static topologies is specific to certain datacenter topologies and is not general enough (see Appendix~\ref{sec:motivation-throughput}).
To this end, we revisit the problem and analyze it again using the formal definitions typically used in the literature. We find that, the throughput of a static topology is directly proportional to the total capacity and inversely proportional to the \emph{average route length}, which generalizes the recently derived throughput upper bound~\cite{tubSigcomm2021} for shortest paths. 

\begin{restatable}[Throughput]{theorem}{throughputUpperBound}\label{th:throughput-upper-bound}
Given a demand matrix $\mathcal{M}$ and a flow $F$, the throughput $\theta(\mathcal{M},F)$ of a static graph represented as $G=(V,E)$ is given by Equation~\ref{eq:throughput-upper-bound}.
The graph $G$ has throughput $\theta(\mathcal{M}) = \max_{F} \theta(\mathcal{M},F)$: the maximum scaling factor given a demand matrix $\mathcal{M}$; and has throughput $\theta^* = \min_{\mathcal{M}} \theta(\mathcal{M})$ for a worst-case demand matrix.
\begin{equation}\label{eq:throughput-upper-bound}
\theta(\mathcal{M},F) \le \frac{\hat{C}}{M \cdot \arl(\mathcal{M},F)}
\end{equation}
where $\hat{C}=\sum_{e\in E} \hat{c}(e)$ is the total capacity of the network, 
$M= \sum_{s,d\in V} m_{s,d}$ is the total demand for the network and
$\arl(\mathcal{M},F) = \sum_{s,d\in V}\sum_{p \in P_{s,d}}  \frac{m_{s,d}}{M} \cdot r_p \cdot \len(p)$ is the \emph{average route length} for $\mathcal{M}$ and $F$,
where $r_p$ is the fraction of demand transmitted on the path $p$. 
\end{restatable}

\textit{Proof sketch.} Our proof of Theorem~\ref{th:throughput-upper-bound} appears in Appendix~\ref{sec:improved-bound}. We rely on \emph{average route length} formally defined in Definition~\ref{def:arl}, \ie the weighted sum of all path lengths where the weight corresponding to each path is the fraction of total demand transmitted on the path. The rest of the proof follows by capacity constraints and flow conservation rules.

\medskip
\textit{Discussion.} Theorem~\ref{th:throughput-upper-bound} indeed expresses that the throughput maximization problem essentially boils down to minimizing \emph{average route length} as opposed to shortest path lengths~\cite{tubSigcomm2021} or average shortest paths~\cite{highthroughputSingla}.
Both Theorem~\ref{th:evolving-union} and Theorem~\ref{th:throughput-upper-bound} provide a concrete understanding on the achievable throughput of periodic reconfigurable topologies. First, Theorem~\ref{th:evolving-union} provides a relation between the throughput of periodic reconfigurable topologies and the throughput of static topologies. Second, Theorem~\ref{th:throughput-upper-bound} provides the throughput of static topologies more concretely.  Finally, for the worst-case demand matrix, we rely on an important result stated in~\cite{tubSigcomm2021} (and even earlier informally in~\cite{jyothi2016measuring,highthroughputSingla}) and claims 
that the worst-case demand matrix in a static topology is a specific permutation matrix (longest-matching).
Consequently the worst-case demand matrix for \emph{any} periodic reconfigurable topology, due to Theorem~\ref{th:evolving-union},
is a specific permutation matrix in the corresponding emulated graph\footnote{A similar but less general result was recently claimed for specific RDCN designs~\cite{sigmetrics22cerberus}.
In particular, it holds only for designs which emulate a complete graph, while our result holds for \emph{any} emulated graph due to our new result in Theorem~\ref{th:evolving-union}.}.

However, so far we still assumed that \first a certain amount of buffer is available at each intermediate node and \second any amount of delay is tolerable. Hence, we study two important questions in the remainder of this section.
We first seek to understand whether high throughput provided by periodic reconfigurations comes at the cost of inflated delay.

\medskip
\noindent\textit{\textbf{(Q1) Delay:} What is the relation between throughput and delay in a periodic reconfigurable topology?}

\medskip
Given the increasing gap between capacity growth and switch buffers, we are further interested in investigating whether periodic reconfigurable topologies address the near-end of Moore's law w.r.t buffer requirements.

\medskip
\noindent\textit{\textbf{(Q2) Buffer:} How much buffer is required at each node to achieve the throughput upper bound stated in Theorem~\ref{th:throughput-upper-bound}?}

\subsection{Delay of Periodic RDCNs}
\label{sec:prdcn-delay}

The periodic reconfigurations naturally introduce a certain delay. In our context, delay is the time it takes for a packet sent from a source to reach its destination in the periodic evolving graph, \emph{without experiencing congestion from other packets.}
The maximum delay for a path $\delta$ is bounded by $\len(\delta)\cdot \Gamma$ since every consecutive edge in the path is available within $\Gamma$ timeslots (period).
We are mainly interested in the maximum delay incurred by feasible paths which typically reflects in the tail latencies observed in a datacenter. 
In the following, we state the relation between throughput $\theta^*$ and delay.

\begin{restatable}[Delay]{theorem}{delayTheorem}\label{th:delay} Given a demand matrix $\mathcal{M}$, a \NU-regular periodic evolving graph $\mathcal{G}$ emulating a $d$-regular graph with a period of $\Gamma$ timeslots each of duration $\Delta$ and a flow $\mathcal{F}$ that achieves throughput $\theta(\mathcal{M},\mathcal{F})$,
the average route delay $\ard(\mathcal{M},\mathcal{F})$ and the maximum delay $L_{max}$ are given by\footnote{$\Omega$ is the asymptotic lower bound notation.},
\begin{align}\label{eq:delaytheorem}
L_{max} \ge \ard(\mathcal{M},\mathcal{F}) &= \arl(\mathcal{M},\mathcal{F}) \cdot \Gamma\cdot \Delta  \nonumber \\
&\ge \Omega\left(\frac{d\cdot \Delta}{\NU\cdot \theta(\mathcal{M},\mathcal{F})}\right)
\end{align}
where $\ard(\mathcal{M},\mathcal{F}) = \sum_{s,d\in V}\sum_{\delta}  \frac{m_{s,d}}{M} \cdot r_\delta \cdot \delay(\delta)$ is the \emph{average route delay} for $\mathcal{M}$ and $\mathcal{F}$,
where $r_\delta$ is the fraction of demand transmitted on the legal temporal path~$\delta$, and $L(\delta)$ is the delay of the path $\delta$, $M=\sum_{s\in V}\sum_{d\in V} m_{s,d}$ is the total demand.  In particular, for a worst-case demand matrix, the maximum latency is bounded by
\begin{equation}\label{eq:delayMax}
L_{max} \ge \Omega\left(\frac{d\cdot \Delta}{\NU\cdot \theta^*} \right)
\end{equation}
\end{restatable}

\textit{Proof sketch.} We use the argument of flow conservation to prove our claim. Our full proof appears in Appendix~\ref{sec:buffer-analysis}. Specifically, consider a temporal path $\delta$ with delay $L(\delta)$ between a source $s$ and a destination $d$; $r_\delta$ be the fraction of demand $m_{s,d}$ transmitted on the path $\delta$. Notice that the destination $d$ does not receive the flow initially until $L(\delta)$ duration due to the path delay. From then on, in order to satisfy the demand, source $s$ transmits $\theta (\mathcal{M},\mathcal{F})\cdot m_{s,d}\cdot r_{\delta}$ amount of flow in every period and destination $d$ receives $\theta (\mathcal{M},\mathcal{F})\cdot m_{s,d}\cdot r_{\delta}$ amount of flow is every period. Due to conservation of flow, the initially transmitted data (flow volume) until $\delay(\delta)$ time \ie $\theta(\mathcal{M},\mathcal{F})\cdot m_{s,d}\cdot r_\delta\cdot \delay(\delta)$ circulates (moves between $s\hyphen d$) in the network in every period. As a result, the overall capacity consumed can be written as $\theta(\mathcal{M},\mathcal{F})\cdot \frac{ARD(\mathcal{M},\mathcal{F})}{\Gamma\cdot \Delta}$ where $\ard$ is the average route delay (Definition~\ref{def:ard}). However, the overall capacity utilized can also be written in terms of path lengths \ie $\theta(\mathcal{M},\mathcal{F})\cdot \arl(\mathcal{M},\mathcal{F})$ where $\arl$ is the average route length (Definition~\ref{def:arl}). By equating the two, we obtain our result in Eq.~\eqref{eq:delaytheorem}. Eq.~\eqref{eq:delayMax} follows by considering the throughput $\theta^*$ for a worst-case demand matrix.

\medskip
\textit{Discussion.}
Using Theorem~\ref{th:delay}, we illustrate the throughput and delay relation for different values of $d$ in Figure~\ref{fig:tradeoffs}. Here $d$ is the degree of the graph emulated by a periodic reconfigurable topology. Note that the worst-case demand matrix (a permutation matrix) specifies non-zero demands between ToR pairs at maximum distance in the graph~\cite{tubSigcomm2021} and hence the average route lengths $\arl$ is close to the diameter which is bounded by $\log_d(\NT)$ for $d$-regular graphs. Hence the throughput $\theta^*$ is inversely proportional to $\log_d(\NT)$ whereas the delay based on Eq.~\eqref{eq:delayMax} is proportional to $d\cdot \log_d(\NT)$. Specifically, we notice that the existing designs which emulate a complete graph achieve the highest throughput but at the cost of high delay.

\subsection{Buffer Requirements of Periodic RDCNs}
\label{sec:prdcn-buffer}

With the understanding of the relation between throughput and delay of periodic reconfigurable topologies, we now study their buffer requirements. We assume that each node is equipped with a memory region shared across the entire device and hence buffer is an aggregate value per node in our analysis. This model is similar to shared memory architectures with complete sharing~\cite{abm}. Our analysis reveals an interesting inequality for the required buffer, which is conceptually  of the well-known ``bandwidth-delay product'' form. 

\begin{restatable}[Buffer]{theorem}{bufferBound}\label{th:buffer}
    Given a demand matrix $\mathcal{M}$, a periodic evolving graph $\mathcal{G}$ requires at least $\hat{B}$ total amount of buffer in the network in order to achieve throughput $\theta(\mathcal{M},\mathcal{F})$.
\begin{align}\label{eq:buffer}
&\hat{B} \ge (\theta(\mathcal{M},\mathcal{F}) \cdot M) \cdot \ard(\mathcal{M},\mathcal{F})
\end{align}
where $\hat{B}=\sum_{u\in V} B(u)$ is the total buffer of the network and $B(u)$ is the available buffer at a node $u$; 
$M= \sum_{s,d\in V} m_{s,d}$ is the total demand for the network; $\ard(\mathcal{M},\mathcal{F}) = \sum_{s,d\in V}\sum_{\delta}  \frac{m_{s,d}}{M} \cdot r_\delta \cdot \delay(\delta)$ is the \emph{average route delay} for $\mathcal{M}$ and $\mathcal{F}$,
where $r_\delta$ is the fraction of demand transmitted on the legal temporal path~$\delta$. 
\end{restatable}

\textit{Proof sketch.} While the proof of Theorem~\ref{th:delay} accounts for flow conservation with in one period time interval, the proof of Theorem~\ref{th:buffer} follows by carefully accounting the flow that is in transit within one period time interval before conservation applies. Our full proof appears in Appendix~\ref{sec:buffer-analysis}.

\medskip
\textit{Discussion.} Theorem~\ref{th:buffer} reveals a non-trivial tradeoff across the spectrum of periodic reconfigurable topologies which we discuss in detail in \S\ref{sec:prdcn-tradeoff}. Our result for the required buffer in Theorem~\ref{th:buffer} intuitively resembles the ``bandwidth-delay product'' commonly used in the TCP literature. A ``pipe'' must have at least a bandwidth-delay product amount of bytes in transit (or inflight) in order to achieve full utilization. Similarly, our result shows that the total buffer in a periodic reconfigurable network must be at least the product of total demand (in bits per second) and the average route delay. The required buffer stems from the waiting times at intermediate nodes along a path due to the periodic reconfigurations. We believe this analogy may also render useful for instance to predict the overall graph throughput based on the configuration of transport protocols.  For example, the window size\footnote{The window size of a transport protocol is typically the maximum amount of bytes allowed to be in transit at any point in time.} of a transport protocol typically relates to the amount of buffering, and the maximum window size may relate to the overall periodic graph throughput. We leave it for future work to study such relations in detail.

\subsection{Remarks and Discussion on Theorems \ref{th:evolving-union}-\ref{th:buffer}}\label{sec:faq}
Our results in the previous sections allow us to answer and shed light on several basic questions also discussed in the literature~\cite{mellette2017rotornet,reactor,sirius}.

\noindent \textbf{\textit{Can buffering at the end-hosts instead of ToRs alleviate the problems of excessive buffering?}}
Our result in Theorem~\ref{th:buffer} expresses the \emph{total} amount of buffering required in the network. We note that this buffering can either be done at the ToR switches or can be moved to the end-hosts.
Several works in the past~\cite{mellette2017rotornet,reactor} informally discussed the need for in-network buffers but outside the circuit switching layer\footnote{Recall that circuit switches are bufferless.}. Two-hop and in general multi-hop routing further worsens the need for additional in-network buffers in periodic reconfigurable networks~\cite{mellette2017rotornet}. To this end, prior works proposed to move the buffering from ToR switches to the end-hosts with specialized synchronization and packet pause/unpause techniques between the end-hosts and the ToR switches~\cite{reactor}. These techniques are motivated based on the fact that end-host DRAM memory is cheap and abundant while on-chip ToR buffers are costly and limited in size.
Nevertheless, excessive buffering at the end-hosts still poses the well-known problems of queuing delay, although it alleviates the problems of scarce buffer resource at the ToRs. The question of buffering at the end-hosts vs ToR switches poses an inherent tradeoff between \emph{a)} simplicity by buffering at the ToR but at the cost of large on-chip costly buffers and \emph{b)} cost-effectiveness by buffering at the end-hosts but at the cost of complex techniques that consume additional CPU resources and potentially added processing delays.

\noindent \textbf{\textit{Can nanosecond reconfiguration mask the problems of delay and buffer needs?}}

\noindent Notice that delay (from Theorem~\ref{th:delay}) is directly proportional to the timeslot duration $\timeslot$. Naturally, if the reconfiguration delay $\Delta_r$ is small, the timeslot duration $\timeslot$ can also be reduced. As a result, the maximum delay and the average route delay can be significantly reduced. However, notice that the buffer requirements (from Theorem~\ref{th:buffer}) are directly proportional to the timeslot duration $\Delta$ as well as the total capacity \eg the buffer requirements would remain the same if all the links in the network are increased in capacity by $2$x even if the reconfiguration time is reduced by $2$x.

\noindent \textbf{\textit{What is the difference between the degree of the emulated graph and the physical topology?}}

\noindent We emphasize that the emulated graph is obtained by the union of edges of the periodic evolving graph (ToR-to-ToR connectivity) over one period of a  time-interval. Hence, the node degree of the emulated graph is the number of ToR switches that are directly connected to a given ToR switch (over one period). In contrast, the node degree of the physical topology is the number of circuit switches that are directly connected to a given ToR switch at any time. In particular, the node degree in the physical topology is given by $\NU$: the number of uplinks. However, for example, both RotorNet and Sirius emulate a complete graph and hence have the same node degree in the emulated graph given by $\NT$: the total number of ToR switches. Hereafter in this paper, the emulated graph and its degree play an important role (\S\ref{sec:main}) especially as seen in Theorem~\ref{th:evolving-union} and Theorem~\ref{th:delay} where $d$ is the degree.

\noindent \textbf{\textit{Are the results applicable to Valiant routing?}}

\noindent We stress that our results in Theorem~\ref{th:evolving-union},~\ref{th:throughput-upper-bound},~\ref{th:delay},~\ref{th:buffer} are general for an ideal routing scheme and are \emph{not} based on Valiant load balancing. We only assume the specific case of Valiant load balancing later in~\S\ref{sec:main}.

\subsection{Tradeoffs \& Optimization Opportunity}
\label{sec:prdcn-tradeoff}

Interestingly, the relation between throughput and delay is a concave function without buffer constraints (Theorem~\ref{th:throughput-upper-bound}) but is a convex function for a fixed amount of buffer (Theorem~\ref{th:buffer}). Figure~\ref{fig:tradeoffs} illustrates the tradeoffs that arise across the design space of periodic reconfigurable topologies. Specifically, for a certain fixed amount of buffer, existing designs emulating a complete graph are no longer throughput optimal as the available buffer simply cannot hold enough traffic to achieve the ideally achievable maximum. Further, throughput and delay pose a fundamental tradeoff in terms of performance.

Although periodic reconfigurable datacenter designs were developed in view of bandwidth scaling problem posed by the gradual end of Moore's law, the required buffer with linear relation to throughput does not indicate high scalability in future. Several studies in the recent past show an increasing gap between the capacity growth and switch buffers~\cite{9288775,276958}. Specifically, the top-of-the-rack buffers are extremely shallow and the buffer per port per Gbps has been gradually decreasing over the recent years. In the following, we summarize our key motivation to explore the design space of topologies.

\begin{itemize}
    \item \textbf{Throughput:} If buffer is not a concern, then emulating a complete graph (existing designs) results in high throughput but at the cost of high delay. 
    \item \textbf{Delay:} If delay is a concern, even without buffer concerns, emulating a $d$-regular graph where $d<\NT$ results in lower delay but at the cost of throughput.
    \item \textbf{Buffer:} Finally, if buffer is a concern, there exists a $d$-regular graph that maximizes throughput within the buffer limits.
\end{itemize}

Our goal in this work is to design a periodic reconfigurable topology that maximizes throughput given the buffer and delay requirements.

\section{\name: Near-Optimal Throughput RDCN with Shallow Buffers}\label{sec:main}
Reflecting on our observations in \S\ref{sec:motivation}, we study the family of periodic reconfigurable topologies that emulate a $d$-regular graph. Specifically, our aim is to systematically determine the high throughput topology within the underlying buffer limits.

\subsection{Overview}\label{sec:overview}
We first give an intuition on our design. Recall that the network consists of $\NT$ ToR switches each with $\NU$ in/out ports interconnected via optical circuit switches.

\medskip
\myitem{\name emulates a $d$-regular graph with near-optimal throughput:} Given the importance of the graph emulated by the topology, \name emulates a ``good'' $d$-regular graph with the degree $d$ optimized for throughput with the limited available buffer.
Specifically, the degree $d$ of the emulated graph influences two factors \first average path lengths which relates to throughput and \second delay which relates to the buffer requirements. 
Among the set of all $d$-regular graphs, we are interested in the ``good'' graphs with diameter close to $\log_d(\NT)$, where \NT is the number of ToR switches. Note that the lower bound for diameter of any $d$-regular graph is $\log_d(\NT)$.
Following prior work~\cite{sirius}, we assume Valiant load balancing~\cite{valiant1981universal} for the routing scheme for simplicity. This inflates the average route length by a factor of two \ie $2\cdot \log_d(\NT)$. Notice that emulating a $K_{\NT}$ complete graph with large degree would result in shorter average route length. However, as we will see later, emulating a degree $d\le \NT$ results in better delay and better throughput under limited buffer.

\medskip
\myitem{A parametrized approach:} We parametrize our design based on two values \first delay requirement $L$ and \second buffer limit $B$. The degree $d$ of the emulated graph of \name depends on the delay and buffer requirements.

\medskip
\myitem{Delay and buffer requirements:} 
\name finds a balance between throughput, delay and buffer. Specifically in Figure~\ref{fig:tradeoffs}, \name is the topology at the intersection of throughput upper bound with and without buffer restrictions. To this end, \name maximizes the throughput while finding a balance in delay and buffer requirements.

\subsection{Properties of \name}
We first express the throughput of \name without delay and buffer constraints which gives an intuition on the ideally achievable throughput. 

\begin{restatable}[Unconstrained Throughput Upper Bound]{theorem}{unconstrainedThroughput}\label{th:mars-throughput}
The throughput of \name connecting $\NT$ ToR switches each with $\NU$ in/out ports, emulating a $d$-regular graph without any delay and buffer constraints, under Valiant load balancing, is given~by:
\begin{equation*}
\theta^* = \frac{1}{ARL} \approx
\frac{1}{2\cdot \log_d(\NT)} 
\end{equation*}
\end{restatable}

First the diameter of \name's emulated graph is close to $\log_d(\NT)$. Second, we then argue that the worst-case permutation demand matrix specifies non-zero demands between ToR pairs which are separated by a distance close to diameter. Further, Valiant load balancing inflates the \emph{route lengths} by a factor of~$2$ \eg similar to existing designs~\cite{sirius} \ie $\arl = 2\cdot \log_d(\NT)$ under worst-case demand matrix and hence $\theta^* = \frac{1}{2\cdot \log_d(\NT)}$. For instance, in the case of $d=\NT$ (emulating a complete graph), we obtain $\theta^*=\frac{1}{2}$. 

With the understanding of the throughput of \name, we can now relate its throughput to delay. Using our result in Theorem~\ref{th:delay}, the delay $L_{max}$ of \name is related to throughput as $\Omega(\frac{d\cdot \Delta}{\NU\cdot \theta^*}) = \Omega(\frac{2\cdot (\log_d \NT)\cdot d\cdot \Delta}{\NU})$. Given a constraint on delay \ie the topology must ideally incur a delay less than $L$, we state the optimal degree for \name in the following.

\begin{restatable}[Optimal degree $d$ with delay constraints]{theorem}{optdegreeTheorem}\label{th:opt-degree-delay}
The optimal $d$-regular graph emulated by \name that maximizes throughput (given the delay requirement $L$ and under Valiant load balancing) has a degree $d$ given by,
\[
d = \lfloor e^{-\mathcal{W}(k)} \rfloor
\]
where $k = \frac{-2\cdot \ln(\NT) \cdot \Delta}{\NU\cdot L}$; $\mathcal{W}$ is the Lambert W function~\cite{corless1996lambertw}; $\ln(.)$~is the natural logarithm and $e$ is the Euler's number.
\end{restatable}
We provide a proof sketch here. We begin by equating the delay of \name and the desired delay $L$. We then use the property that if $y^y = k$ where $k$ is some constant, then $y = e^{\mathcal{W}(ln(k))}$. Our full proof appears in Appendix~\ref{sec:mars-analyis}.

Our observations in \S\ref{sec:motivation} reveal that a key tradeoff arises across the buffer requirements and the achievable throughput. We are now interested in determining the optimal degree $d$ of \name for a limited buffer $B$ at each node. 
As an intuition, a topology requires more buffer as the product of the scaled demand and the average route delay increases.

\begin{restatable}[Optimal degree $d$ with buffer constraints]{theorem}{optDegreeBuffer}\label{th:opt-degree-buffer}
The optimal degree $d$ for the emulated graph of \name independent of the specific flow that maximizes throughput, given a limited buffer $B\le \NT \cdot \Delta$ at each node, is given by,
\[
d = \lfloor\frac{B}{c\cdot \Delta}\rfloor
\]
where $c$ is the capacity of every edge in the topology and $\Delta$ is the timeslot value.
\end{restatable}

Our proof is a straight-forward extension using our results in Theorem~\ref{th:delay},~\ref{th:buffer}. The full proof appears in~\S\ref{sec:mars-analyis}. From Theorem~\ref{th:opt-degree-buffer}, emulating a complete graph like the existing designs would require a buffer of $\frac{\NT}{\NU} \cdot (\NU \cdot c)\cdot \Delta = \NT\cdot c \cdot \Delta$ where $\NU\cdot c \cdot \Delta$ is the amount of data that can be sent out or received in a single timeslot and $\frac{\NT}{\NU}$ is the period for a complete graph. In contrast, a $d$-regular graph would only require $d\cdot c\cdot \Delta$ amount of buffer and achieves better throughput under limited buffer compared to a periodic topology emulating complete graph.

\subsection{Interconnect}\label{sec:interconnect}

\myitem{Switch size:} Given \NT number of ToR switches each with \NU uplinks, \name requires a switch size of at least $\NP=\NT$ similar to prior work~\cite{mellette2017rotornet,opera}. We need $\NU$ such circuit switches to interconnect the datacenter. We leave it for future work to study the feasibility of using circuit switches with lower port count.

\medskip
\myitem{Wiring:} Each ToR uses its \NU uplinks to connect to each of the \NU circuit switches.

\medskip
\myitem{Matchings:}
We first generate a $d$-regular directed graph.
Specifically, in generating the emulated graph, we choose the $d$-regular graph for which the diameter approaches the lower bound of $\lceil \log_d(\NT) \rceil$~\cite{bollobas1982diameter,imase1985connectivity,imase1983design} \eg deBruijn digraphs~\cite{debruijn}.
A straight-forward argument shows that a $d$-regular directed graph can be decomposed into $d$ perfect matchings. The idea is to recursively find a $1$-factor and delete it from the graph where each $1$-factor by definition is a perfect matching~\cite{NADDEF1981283,petersen1891theorie}. We shuffle the resulting $d$ matchings and randomly assign $\frac{d}{\NU}$ matchings to each of the $\NU$ circuit switches in the topology. Each circuit switch then cycles through the $\frac{d}{\NU}$ matchings periodically. We note that decomposing the $d$-regular emulated graph into matchings can be computationally expensive. However, this is performed only once at the time of deployment and the circuit switches cycle through the installed matchings periodically thereafter.

\emph{Note that the interconnect for \name can be implemented using the same hardware as in prior work \cite{mellette2017rotornet,opera,sirius}; in fact, the only change required in \name compared to these systems concerns the matchings configurations.} 

\subsection{Example: deBruijn-based Emulated Graph}\label{sec:example}
\begin{wrapfigure}{l}{0.5\textwidth}
  \begin{center}
    \includegraphics[width=0.49\textwidth]{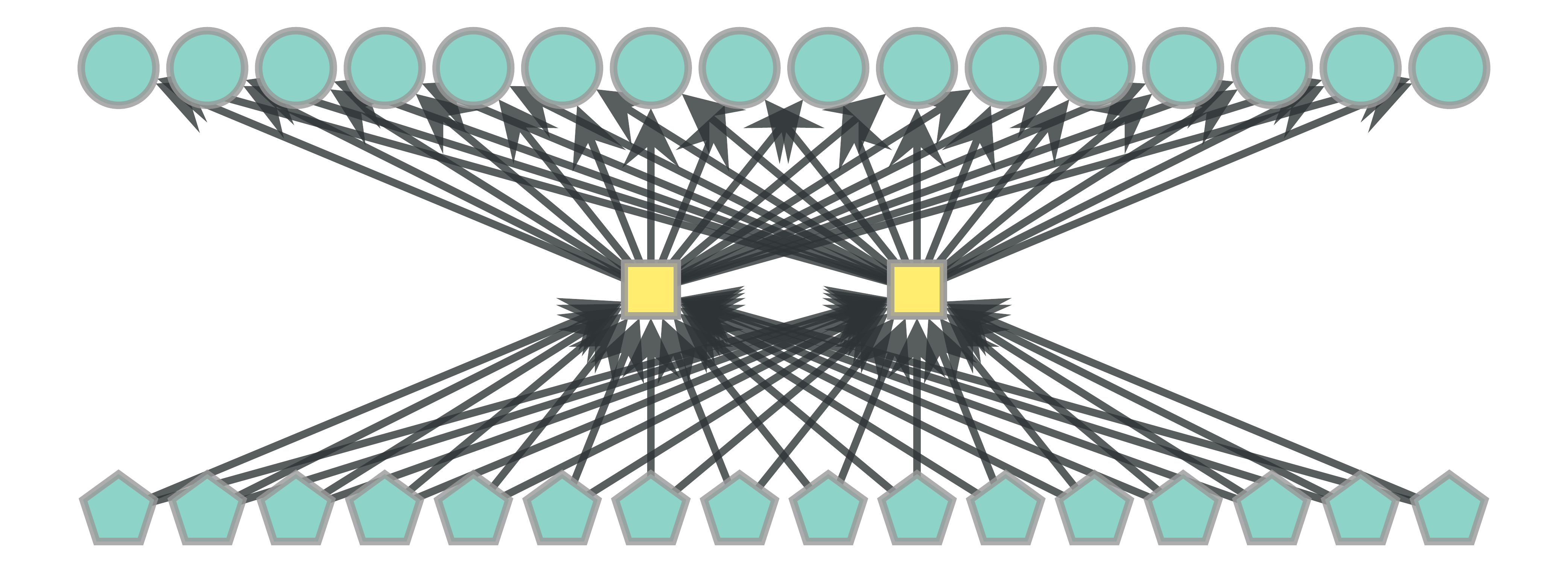}
  \end{center}
  \caption{An example of a periodic RDCN interconnect with $16$ ToR switches each with $2$ in/out ports; $2$ circuit switches each with $16$ in/out ports. For better visualization, the ToR switch outputs are depicted as pentagons and inputs are depicted as circles. Circuit switches are depicted as squares.}
  \vspace{-2mm}
  \label{fig:example-interconnect}
\end{wrapfigure}
We walk through an example to better illustrate the topology construction and the inherent tradeoffs.
We consider an example with $\NT=16$ ToR switches each with $\NU=2$ in/out ports. The interconnect requires $\NS=\NU=2$ optical circuit switches each of size $\NP=\NT=16$. Each ToR uses its uplinks to connect to each of the two circuit switches as shown in Figure~\ref{fig:example-interconnect}. Circuits are set up for $90\mu s$ duration and it takes $10\mu s$ to reconfigure (timeslot $\Delta = 100\mu s$ for each matching) similar to RotorNet switches~\cite{mellette2017rotornet}. All the links have a capacity of $c=400$Gbps.

\begin{figure*}
\centering
\begin{subfigure}{0.32\linewidth}
    \centering
    \includegraphics[width=1\linewidth]{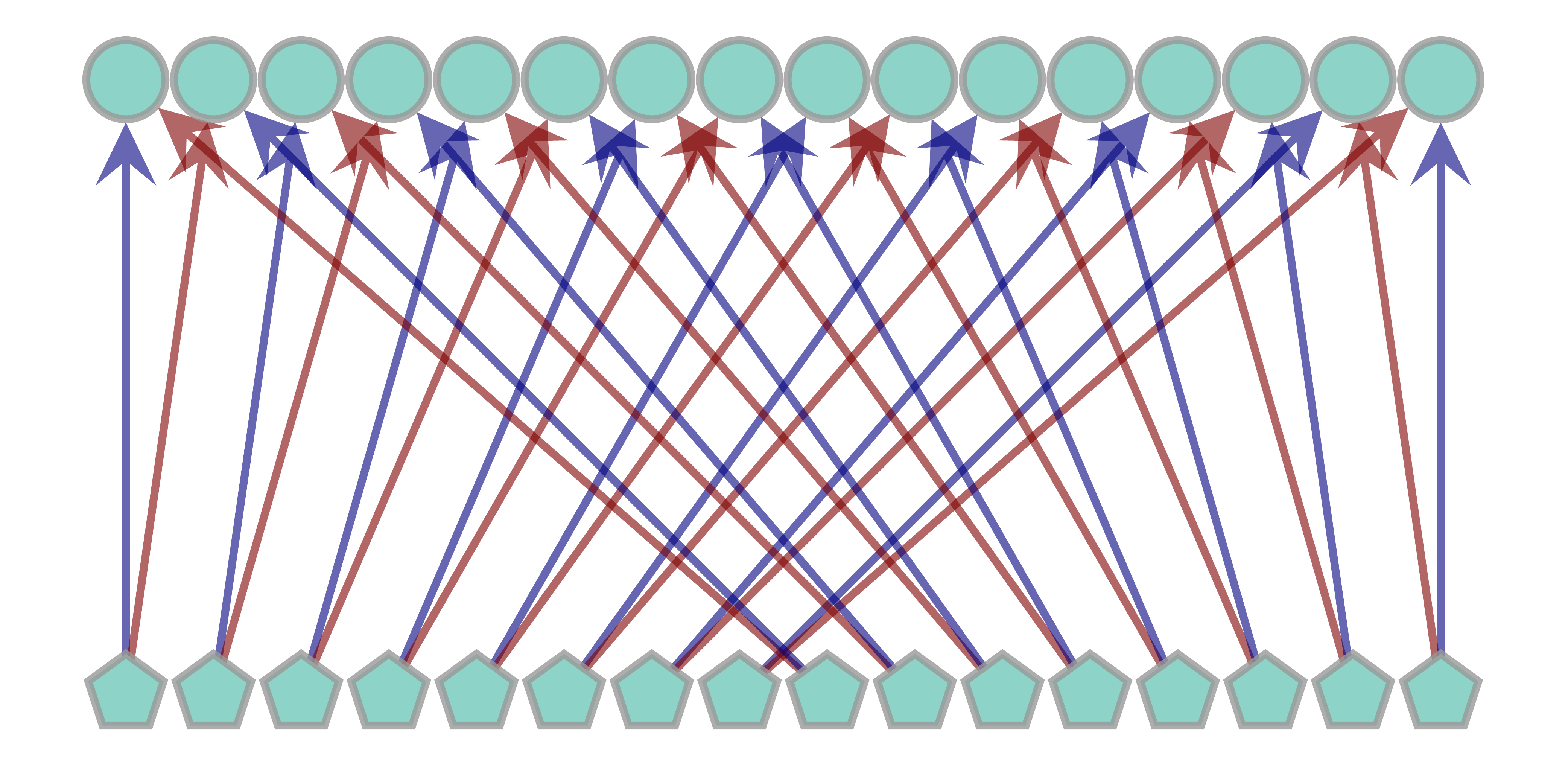}
    \caption{Static $2$-regular \\ Emulated Graph (2 Matchings)}
    \label{fig:static-example}
\end{subfigure}
\begin{subfigure}{0.32\linewidth}
    \centering
    \includegraphics[width=1\linewidth]{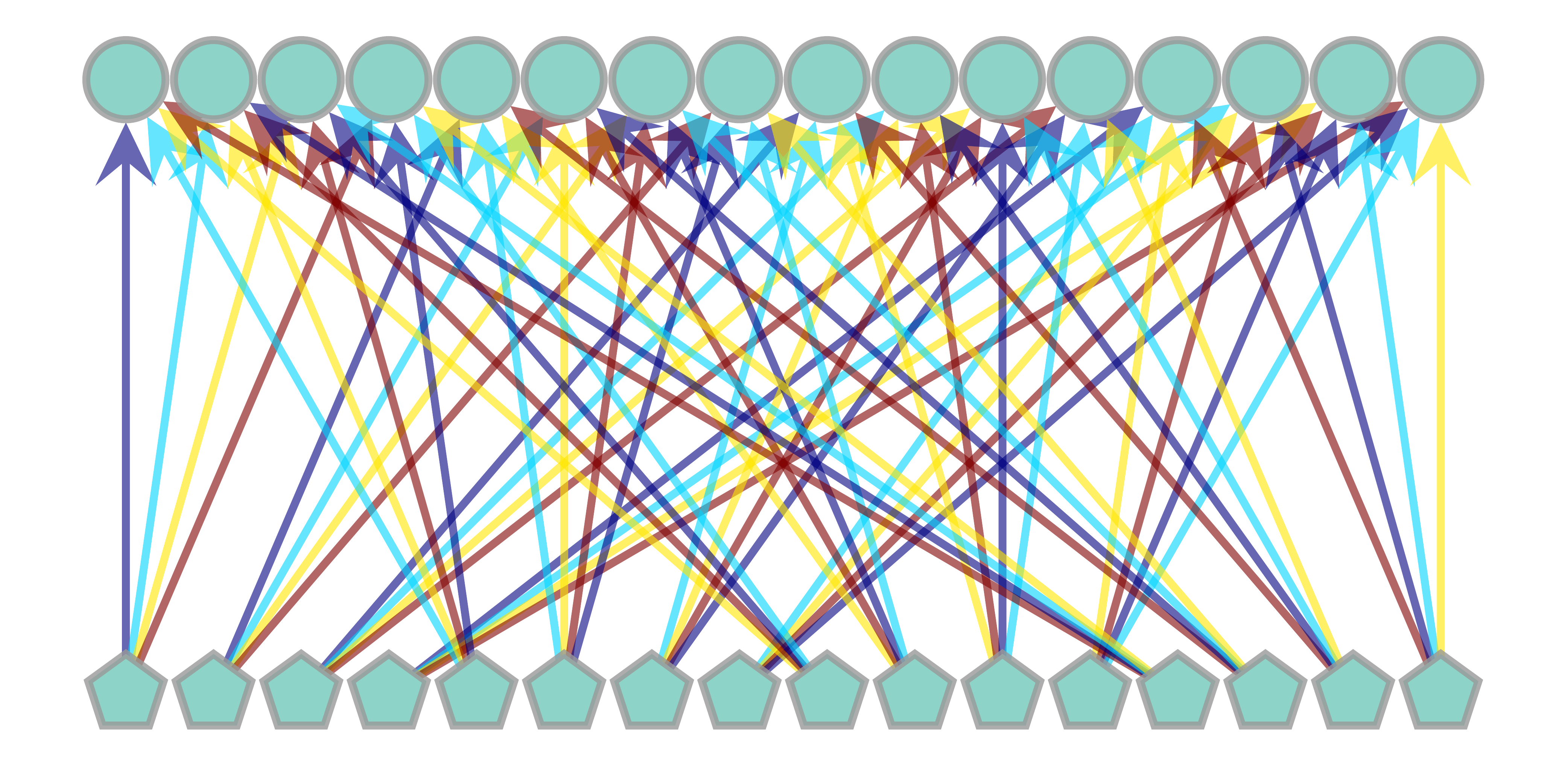}
    \caption{$4$-regular deBruijn \\ Emulated Graph (4 Matchings)}
    \label{fig:mars-example}
\end{subfigure}
\begin{subfigure}{0.32\linewidth}
    \centering
    \includegraphics[width=1\linewidth]{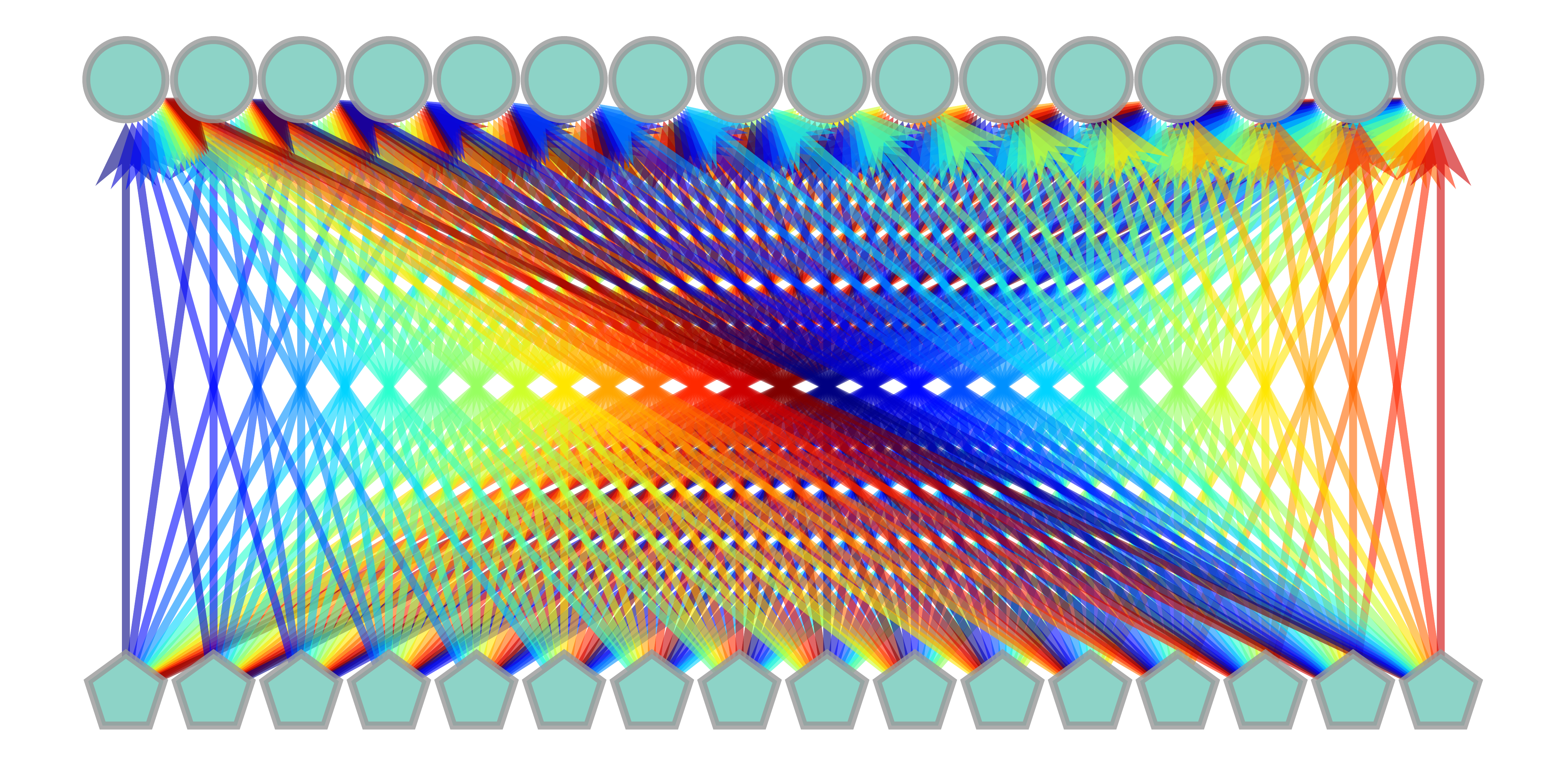}
    \caption{Complete Emulated Graph \\ (16 Matchings)}
    \label{fig:complete-example}
\end{subfigure}
\caption{A periodic reconfigurable topology (shown in Figure~\ref{fig:example-interconnect}) can emulate a spectrum of graphs depending on the matchings schedule in the circuit switches. \textbf{(a)} A static emulated graph has the least delay but at the cost of throughput. \textbf{(b)} \name finds the best degree of the emulated graph in order to achieve near-optimal throughput based on the available buffer and delay tolerance. \textbf{(c)} Existing designs emulate a complete graph trading delay for throughput and suffer in terms of throughput with shallow~buffers. Each color in the figure corresponds to a matching. Since there are two circuit switches (shown in Figure~\ref{fig:example-interconnect}), at any point in time, the union of two matchings corresponds to the ToR-to-ToR connectivity.}
\label{fig:example}
\end{figure*}

Before constructing the sequence of matchings, we take one of the factors into consideration \first buffer size of ToR switches and \second latency tolerance. Using Theorem~\ref{th:opt-degree-delay} and Theorem~\ref{th:opt-degree-buffer}, we determine the optimal degree $d$ to maximize throughput. It remains to generate a $d$-regular emulated graph with optimal diameter and decompose it to a sequence of matchings which are then deployed in the circuit switches.

We consider the spectrum of deBruijn digraphs in generating the emulated graph. Specifically, for $\NT$ ToR switches (vertices), $G=(V,E)$ is a deBruijn graph (emulated graph), where $V=\{0,1,...,\NT-1\}$ is the set of vertices and $E=\{ (u,v)\mid v \equiv (u\cdot d +a) \ \textbf{mod} \ \NT, a\in\{ 0,1,...,d-1\} \}$ is the edge set~\cite{imase1985connectivity}. It is known that deBruijn graphs have a low diameter of $\lceil \log_d(\NT) \rceil$ which is close to Moore's bound for diameter. Our emphasis on the graph diameter is due to Theorem~\ref{th:throughput-upper-bound} as the average route length gets closer to diameter under a worst-case demand matrix which specifies non-zero demand between pairs at maximum distance in the graph.

Using the optimal degree $d$ based on Theorem~\ref{th:opt-degree-delay} and Theorem~\ref{th:opt-degree-buffer}, we generate a $d$-regular directed deBruijn graph and decompose the edge set into $d$ matchings. We shuffle the $d$-matchings and assign $\frac{d}{2}$ number of matchings to each of the two circuit switches. In Figure~\ref{fig:example}, we show three emulated graph instances which differ in the degree $d$ for $\NT=16$ ToR switches each with $\NU=2$ in/out ports. We note that the standard complete graph $K_{\NT}$ has a degree of $\NT-1$ without self-loops. For simplicity, we consider that the optical circuit switches match every ToR to every ToR including one self-loop \ie degree is $\NT$ instead of $\NT-1$.  Table~\ref{table:example-summary} summarizes our example. We next walkthrough the throughput, delay and buffer requirements of different design choices including \name.

\begin{table}[!h]
\begin{center}
\begin{tabular}{
    |p{0.15\linewidth}|p{0.2\linewidth}|p{0.2\linewidth}|p{0.1\linewidth}|
}
\hline
\textbf{Topology} & \textbf{Throughput} & \textbf{Delay} & \textbf{Buffer}\\
\hline
\noindent \tcbox[on line,boxsep=0pt,left=0pt,right=0pt,top=0pt,bottom=0pt,colback=white]{\includegraphics[height=0.3cm]{figures/1.pdf}} & $0.125$ & $\approx0$ & $\approx0$ \\
\noindent \tcbox[on line,boxsep=0pt,left=0pt,right=0pt,top=0pt,bottom=0pt,colback=white]{\includegraphics[height=0.3cm]{figures/2.pdf}} & $0.5$ & $1600\mu s$ & $80$MB \\
\noindent \tcbox[on line,boxsep=0pt,left=0pt,right=0pt,top=0pt,bottom=0pt,colback=white]{\includegraphics[height=0.3cm]{figures/3.pdf}} & $0.125$ &  $1600\mu s$ & $20$MB \\
\noindent \tcbox[on line,boxsep=0pt,left=0pt,right=0pt,top=0pt,bottom=0pt,colback=white]{\includegraphics[height=0.3cm]{figures/4.pdf}} & $0.25$ & $850\mu s$ & $20$MB \\

\hline
\end{tabular}
\end{center}
\caption{Tradeoffs across different design choices based on the degree of the emulated graph for the example topology shown in Figure~\ref{fig:example},~\ref{fig:example-interconnect}.}
\label{table:example-summary}
\end{table}

\medskip
\noindent \tcbox[on line,boxsep=0pt,left=0pt,right=0pt,top=0pt,bottom=0pt,colback=white]{\includegraphics[height=0.3cm]{figures/1.pdf}} \myitem{Static uni-regular DCNs:} With two circuit switches in the topology, each deployed with only one matching results in a static topology\footnote{In this case, since there are no reconfigurations, we set the timeslot value as $\Delta\approx 0$, the latency tax due to reconfigurations $\reconf=0$ and the period $\Gamma=1$.} as shown in Figure~\ref{fig:static-example}. While this is an extreme choice given the reconfigurable circuit switches, it is interesting to observe the throughput vs buffer requirements. Specifically, the diameter is $4$ resulting in a throughput of $\frac{1}{8}$ (from Theorem~\ref{th:mars-throughput}) for a worst-case permutation demand matrix. However, the delay is nearly zero theoretically since the topology remains static assuming negligible transmission and propagation delays. Since the delay is near-zero, the topology also requires near-zero buffers at each node to achieve the ideal throughput of $\frac{1}{8}$.

\medskip
\noindent\tcbox[on line,boxsep=0pt,left=0pt,right=0pt,top=0pt,bottom=0pt,colback=white]{\includegraphics[height=0.3cm]{figures/2.pdf}} \myitem{Existing designs:} Using $16$ matchings in total, with $8$ matchings deployed in each of the two circuit switches, the topology emulates a complete graph as shown in Figure~\ref{fig:complete-example} which is similar to existing designs~\cite{mellette2017rotornet,sirius}. Emulating a complete graph results in optimal throughput of $\frac{1}{2}$ but at the expense of high delay as much as $16\cdot \Delta = 1600\mu s$ (from Theorem~\ref{th:delay}). In order to achieve the optimal throughput of $\frac{1}{2}$, each ToR switch would require significantly large buffers as much as $80$MB in our small scale example topology.

\medskip
\noindent\tcbox[on line,boxsep=0pt,left=0pt,right=0pt,top=0pt,bottom=0pt,colback=white]{\includegraphics[height=0.3cm]{figures/3.pdf}} \myitem{Existing designs under resource constraints:} While emulating a complete graph as shown in Figure~\ref{fig:complete-example} may result in optimal throughput, the achievable throughput critically depends on the available buffer at each ToR switch. For instance, if each ToR switch is equipped with only $20$MB buffer, then the throughput drops to $\frac{1}{8}$ (from Theorem~\ref{th:buffer}). Notice that even with $10$MB buffer, a simple static topology achieves similar throughput with significantly lower delay (see above).

\medskip
\noindent\tcbox[on line,boxsep=0pt,left=0pt,right=0pt,top=0pt,bottom=0pt,colback=white]{\includegraphics[height=0.3cm]{figures/4.pdf}} \myitem{\name:} Our approach leverages the insights from \S\ref{sec:motivation}. Based on the available buffer (say $B=20$MB) and the delay requirements (say $L=850\mu s$), we systematically determine the degree $d$ of the emulated graph to be emulated by the reconfigurable network. In this case, the optimal degree turns out to be $d=4$ both in terms of delay $L$ (from Theorem~\ref{th:opt-degree-delay}) and available buffer $B$ (from Theorem~\ref{th:opt-degree-buffer}). The topology has a diameter of $\log_4 (16)=2$ and achieves much higher throughput: $\frac{1}{4}$ in our example.

\section{Evaluation}

We evaluate \name and compare against the existing approaches in the design space. Our evaluation aims at answering four main questions.

\begin{figure*}
\centering
\begin{subfigure}{0.32\linewidth}
\centering
\includegraphics[width=1\linewidth]{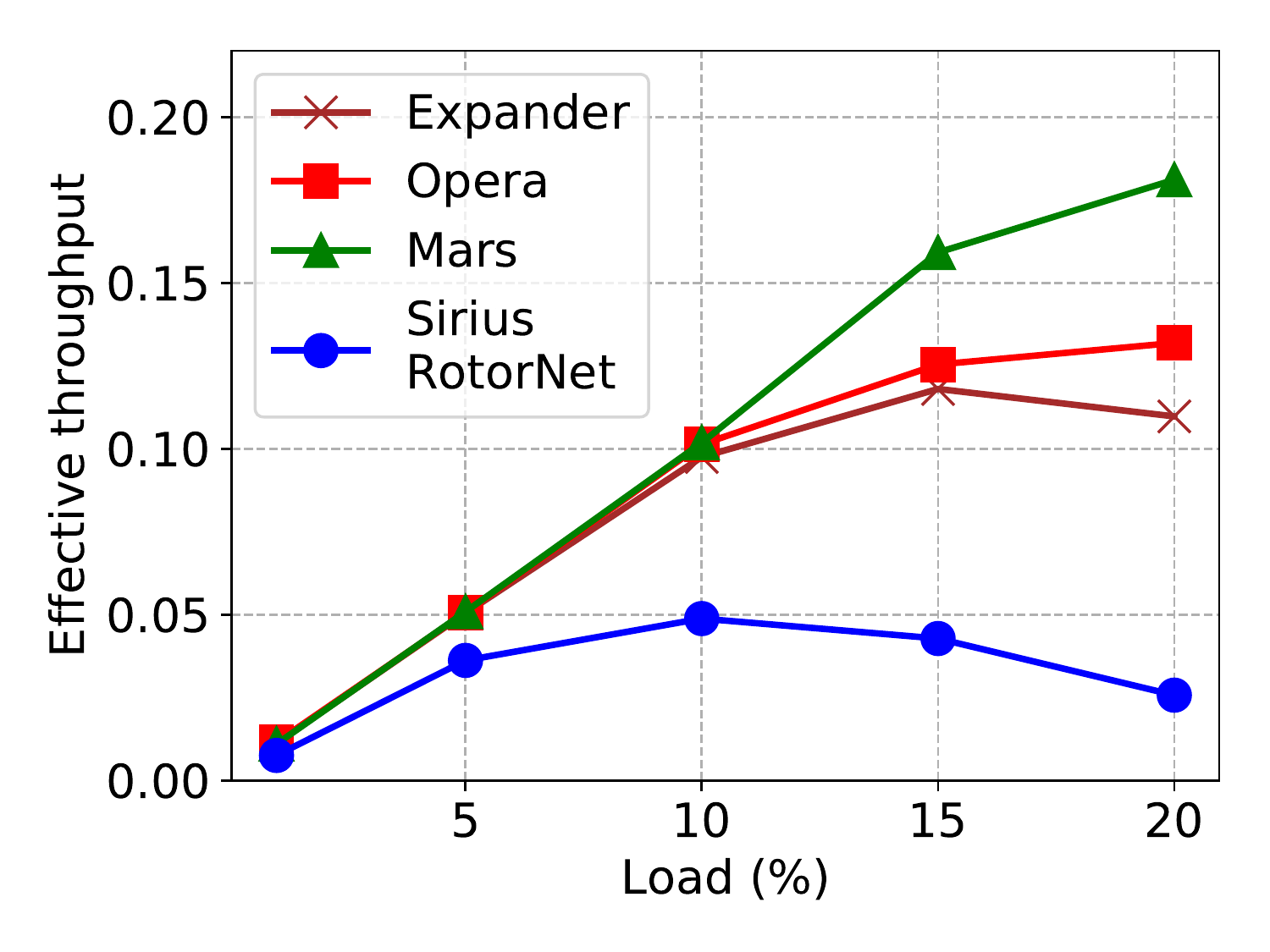}
\subcaption{Effective throughput under all-to-all demand matrix and websearch workload with $8$ packets per port buffer size.}
\label{fig:th-load-a2a}
\end{subfigure}\hfill
\begin{subfigure}{0.32\linewidth}
\centering
\includegraphics[width=1\linewidth]{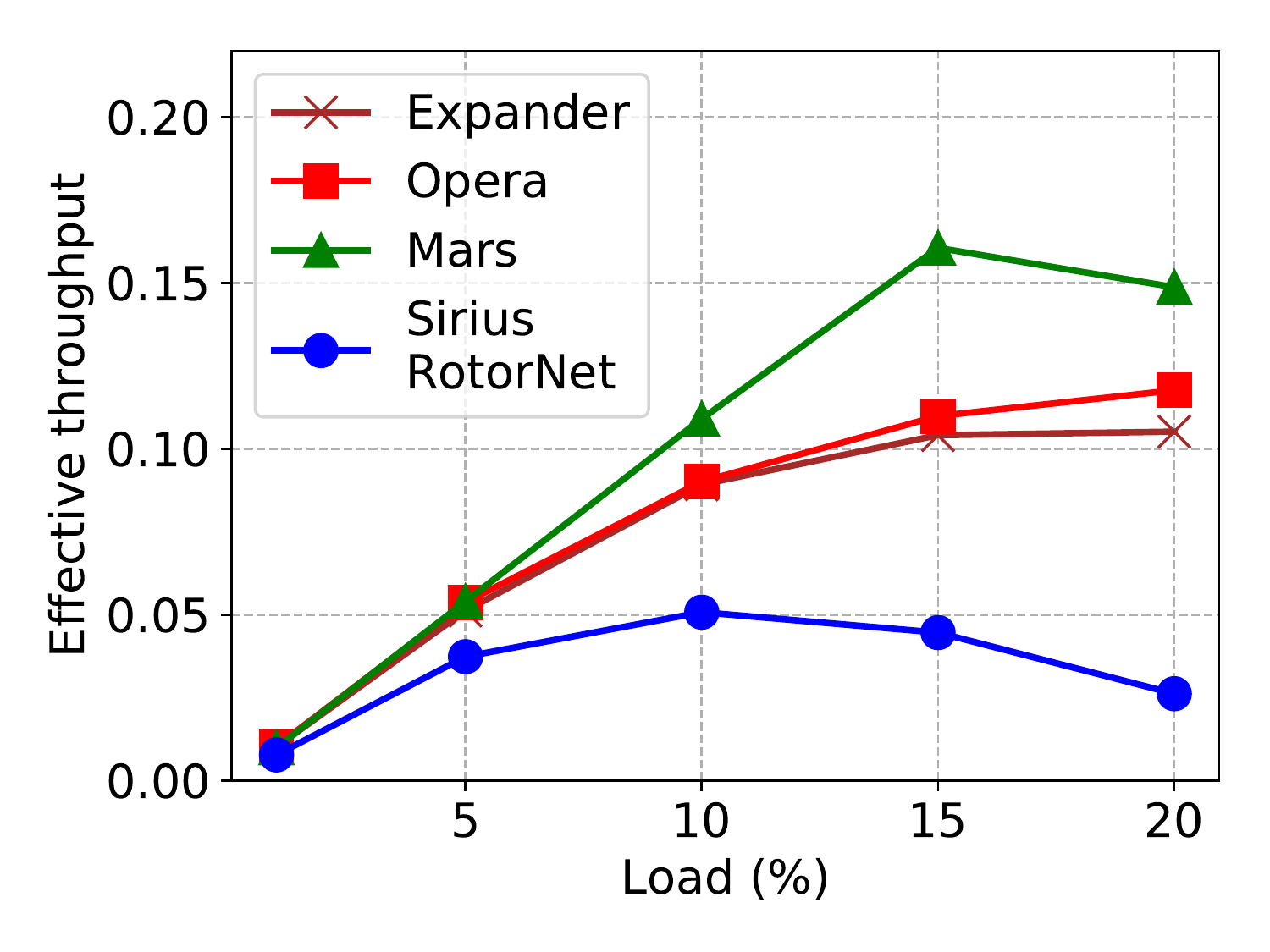}
\subcaption{Effective throughput under random permutation demand matrix and websearch workload with $8$ packets per port buffer size.}
\label{fig:th-load-perm}
\end{subfigure}\hfill
\begin{subfigure}{0.32\linewidth}
\centering
\includegraphics[width=1\linewidth]{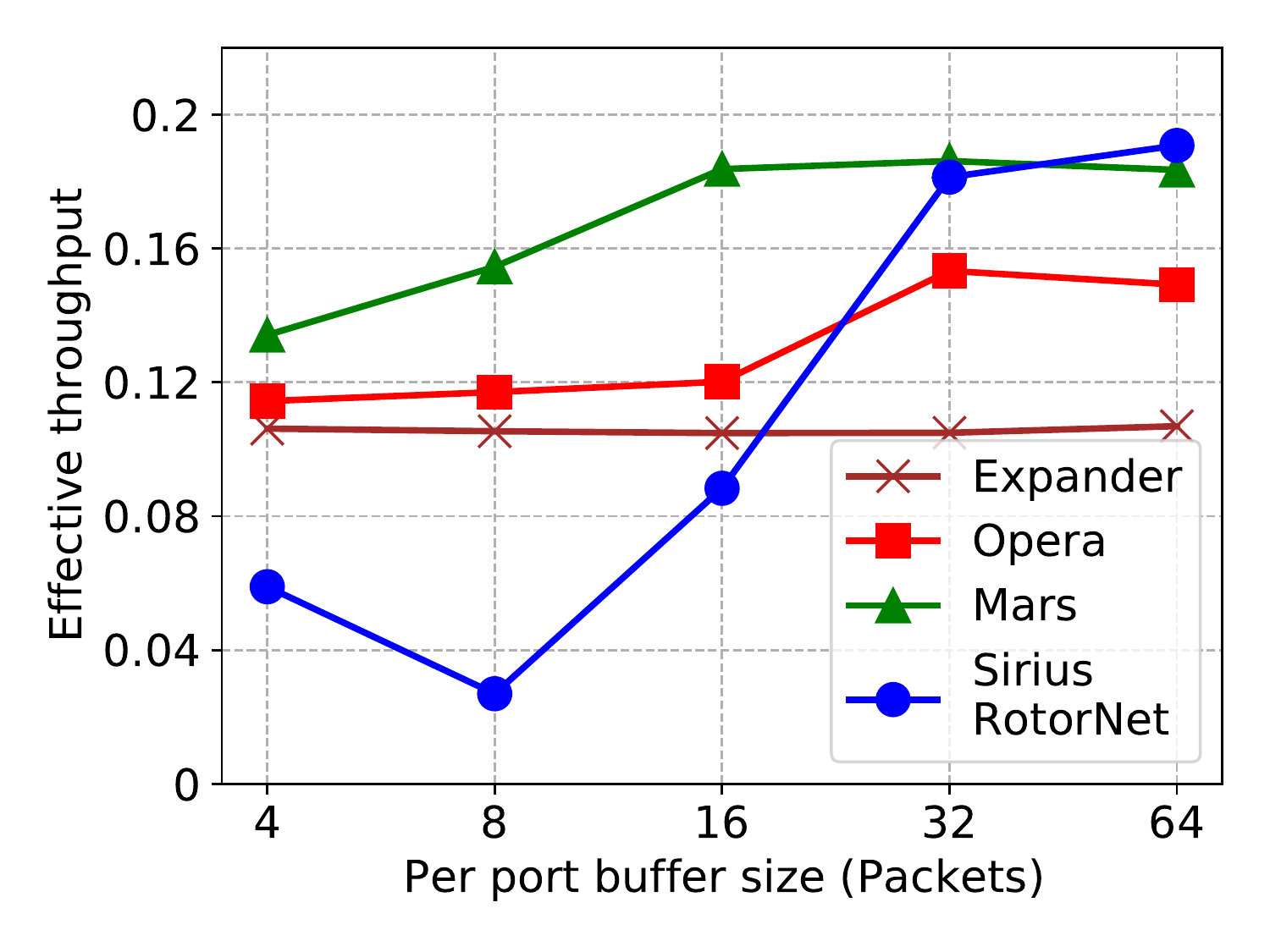}
\subcaption{Effective throughput across buffer sizes under random permutation demand matrix and websearch workload.}
\label{fig:th-buffer-perm}
\end{subfigure}
\caption{\name achieves the best throughput under shallow buffers. Existing approaches which emulate a complete graph (Sirius/RotorNet) suffer significantly under limited buffer (a,b) but can gain throughput if buffers are large (c).}
\label{fig:th-load-buffer}
\end{figure*}

\medskip
\noindent \textbf{\textit{(Q1)} Can \name improve the throughput in datacenters?}
Our evaluation shows that, \name improves the throughput of existing datacenter designs by up to $64\%$ compared to an expander DCN, by up to $37\%$ compared to Opera~\cite{opera} and by more that $4$x compared to Sirius~\cite{sirius} and RotorNet~\cite{mellette2017rotornet} under Valiant load balancing.

\medskip
\noindent \textbf{\textit{(Q2)} How does \name perform under shallow buffers?}
Even under extremely shallow buffers, \name significantly outperforms existing approaches by improving the throughput by up to $6$x at moderate load conditions. At low loads, \name performs similar to existing approaches.

\medskip
\noindent \textbf{\textit{(Q3)} Can \name improve the FCTs of short flows?}

\noindent We find that \name does not trade latency for throughput. Indeed \name's low buffer requirements to achieve high throughput also contribute to better latency even under permutation demand matrices.
Our evaluation shows that \name can improve the FCTs of short flows by up to $96\%$ compared to expander DCN, by up to $75\%$ compared to Opera and by up to $87\%$ compared to RotorNet and Sirius under Valiant load balancing.

\medskip
\noindent \textbf{\textit{(Q4)} Do long flows benefit from \name?}
Our results show that \name reduces the FCTs of long flows for various workloads by up to $4\%$ on average compared to existing approaches under various loads and demand matrices.

\begin{figure*}
\centering
\begin{subfigure}{0.48\linewidth}
\centering
\includegraphics[width=1\linewidth]{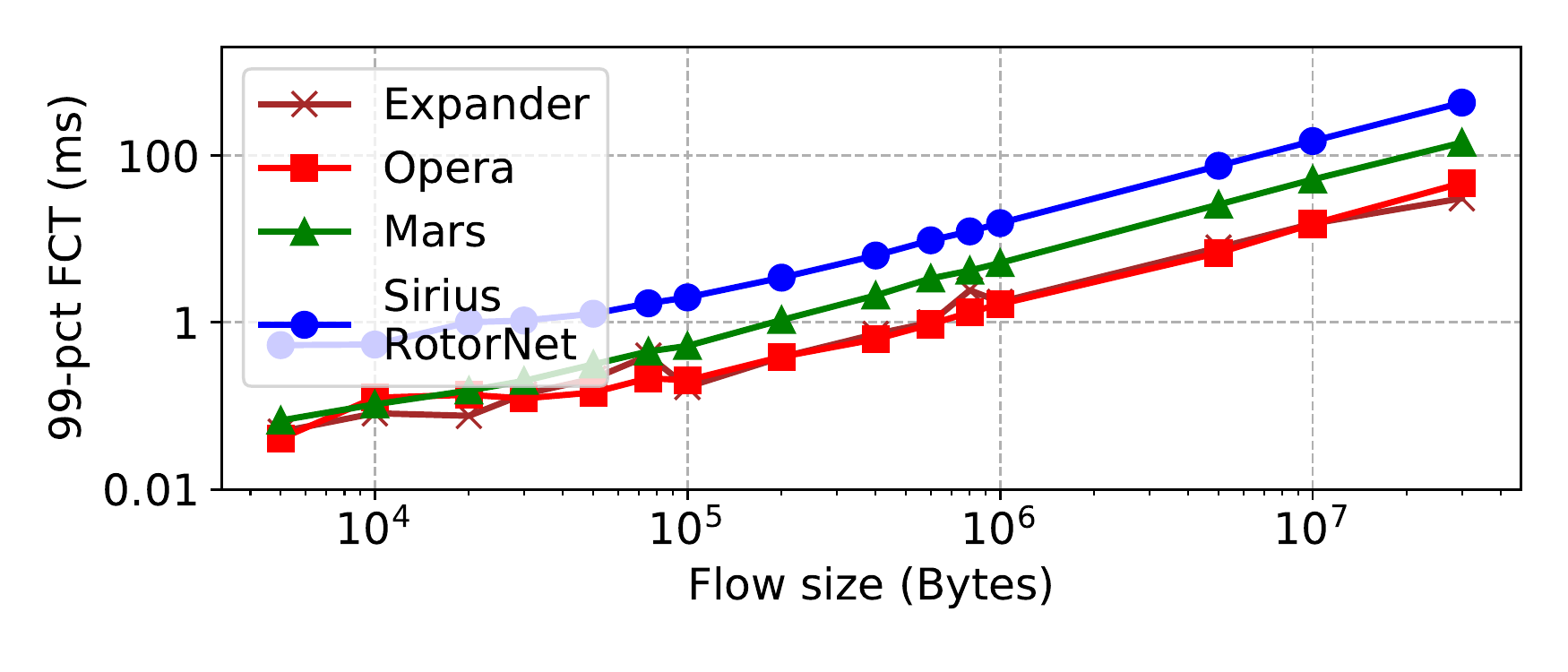}
\subcaption{$1\%$ load}
\label{fig:fcts-0.01-perm}
\end{subfigure}
\begin{subfigure}{0.48\linewidth}
\centering
\includegraphics[width=1\linewidth]{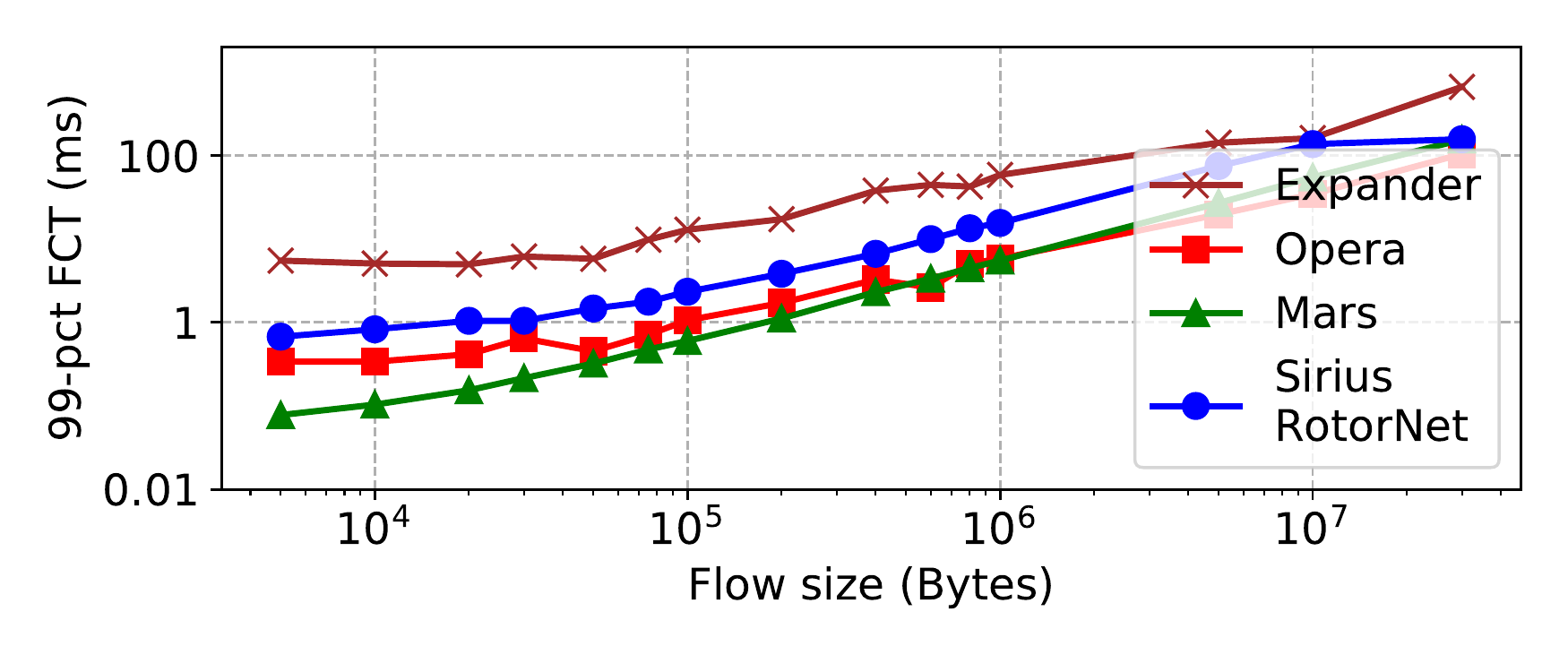}
\subcaption{$5\%$ load}
\label{fig:fcts-0.05-perm}
\end{subfigure}
\begin{subfigure}{0.48\linewidth}
\centering
\includegraphics[width=1\linewidth]{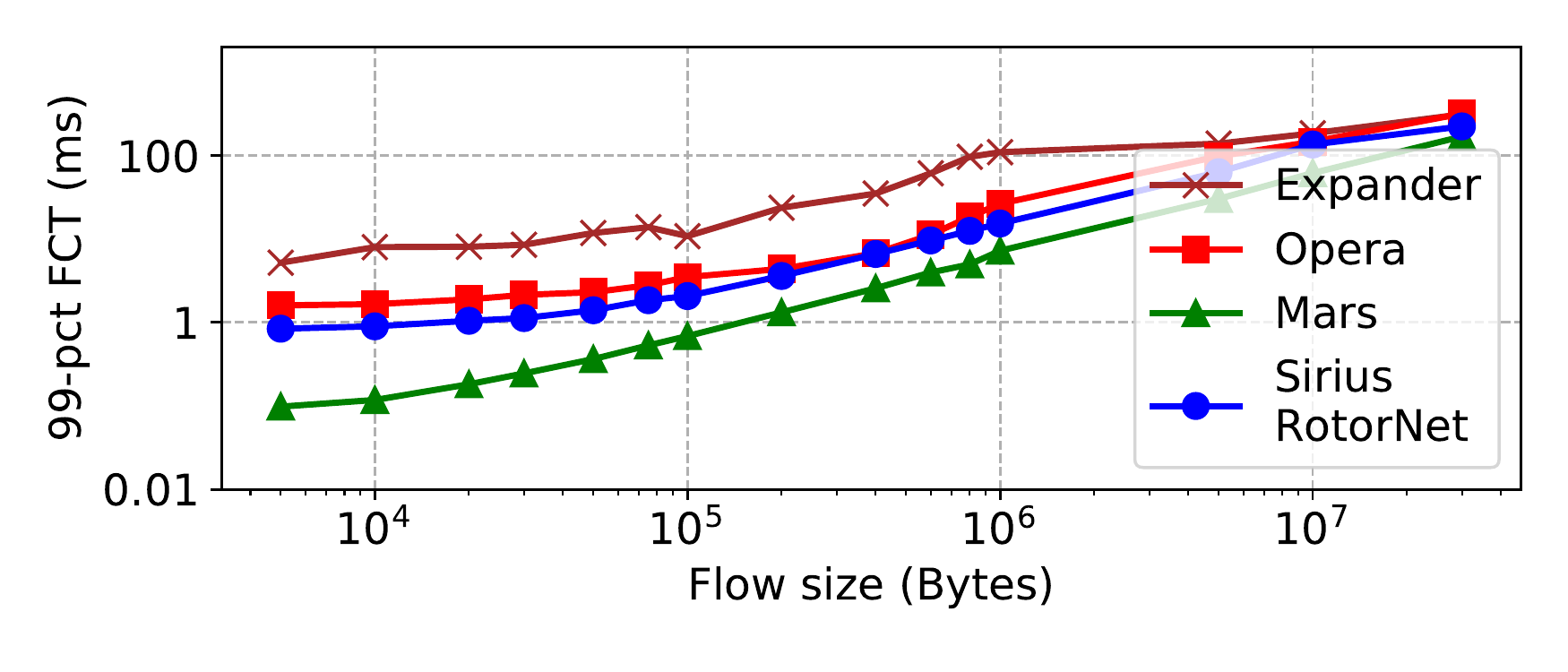}
\subcaption{$10\%$ load}
\label{fig:fcts-0.1-perm}
\end{subfigure}
\begin{subfigure}{0.48\linewidth}
\centering
\includegraphics[width=1\linewidth]{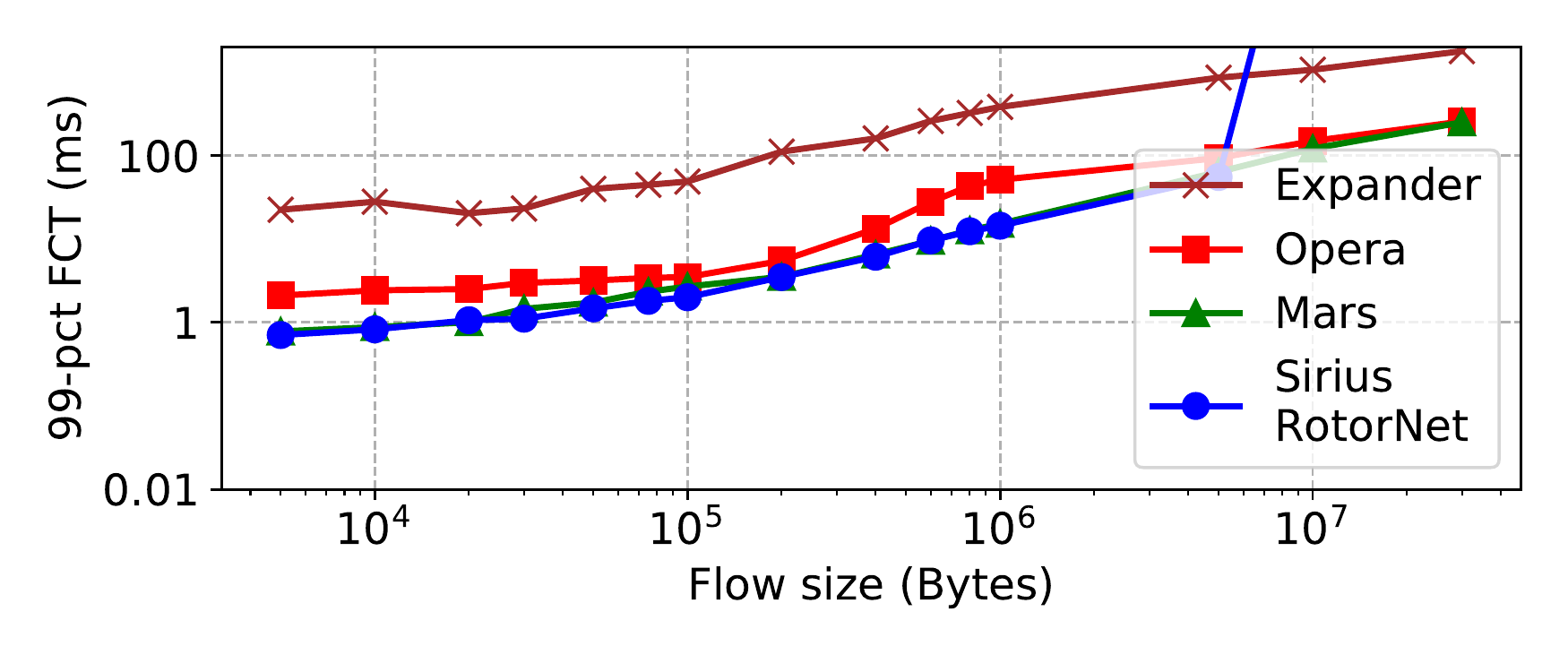}
\subcaption{$20\%$ load}
\label{fig:fcts-0.2-perm}
\end{subfigure}
\caption{For the websearch workload under a random permutation demand matrix, \name improves the $99$-percentile flow completion times for short flows without sacrificing the flow completion times of long flows. Note the log scale on both axes.}
\vspace{-5mm}
\label{fig:fctsperm}
\end{figure*}

\subsection{Setup}\label{sec:setup}

Our evaluation is based on packet-level simulations using \textsc{htsim} used in prior work~\cite{handley2017re,opera}.

\medskip
\myitem{Interconnect:} We consider a datacenter network with $256$ servers organized into $64$ ToR switches. Each ToR switch has $4$ uplinks that connect to $4$ rotor-switches~\cite{mellette2017rotornet} with a reconfiguration delay of $1\mu s$. All the links have a capacity of $10$Gbps and $500$ nanoseconds delay. 

\medskip
\myitem{Workload:} We generate traffic using websearch~\cite{alizadeh2011data} and datamining~\cite{alizadeh2013pfabric} workloads. We evaluate across various loads on the server-ToR links in the range $1-20\%$ for all-to-all and random permutation demand matrices. Note that $20\%$ load is already close to the maximum load that an expander topology can sustain\footnote{Atleast $\frac{1}{\arl}$ capacity is sacrificed due to multi-hop routing; where $\arl$ is the average route length. An alternative in order to further increase throughput is to simply``undersubscribe'' \ie with oversubscription $<$ 1. We leave it for future work to analyze such designs which requires a comprehensive cost-analysis for a fair comparison.}. Flows arrive according to a Poisson process and we control the mean inter-arrival time to achieve the desired load. In the interest of space, we report our results for the datamining workload in the Appendix and only discuss our results for websearch workload in this section. 

\medskip
\myitem{Comparisons:} We compare \name to Opera~\cite{opera}, Sirius~\cite{sirius}, RotorNet~\cite{mellette2017rotornet} and static expander networks. For expander, we generate $d=4$ (number of uplinks) random regular graphs $G$ with $64$ ToRs as vertices such that $\lambda(G) \le 2\sqrt{u-1}$ where $\lambda$ is the second eigen value and $2\sqrt{u-1}$ is the Ramanujan constant. This gives us a Ramanujan graph which is known to be excellent expander.

\medskip
\myitem{Metrics:} We report server downlink utilization indicating the effective throughput. We also report $99$-percentile flow completion times (FCTs) and buffer occupancies.

\medskip
\myitem{Configuration:} We construct \name which emulates a deBruijn directed graph of degree $8$ (see~\S\ref{sec:interconnect}) optimized for a shallow buffer size of $8$ packets per port in the interconnect described above. Switches are configured with routing information statically at initialization time and packets are sprayed across all equal cost paths. \name, Sirius~\cite{sirius} and RotorNet~\cite{mellette2017rotornet} use Valiant load balanced paths; expander uses all (edge-disjoint) shortest paths; and Opera uses all (edge-disjoint) shortest paths in each topology slice~\cite{opera} for short flows and single hop paths for long flows. The long flow cutoff is set to $15$MB based on~\cite{opera}. We use NDP~\cite{handley2017re} as the transport protocol and set the trimming threshold per port to $8$ packets by default based on~\cite{handley2017re} unless explicitly stated. We vary the packet trimming threshold to vary the maximum buffer size values in our evaluation. Finally, we set minRTO to $1$ms.

\subsection{Results}
\myitem{\name significantly improves the throughput:}
In Figure~\ref{fig:th-load-a2a} and Figure~\ref{fig:th-load-perm}, we show the effective throughput achieved by \name and existing approaches across various loads of websearch workload. Specifically, in Figure~\ref{fig:th-load-a2a} for the All-to-All demand matrix, we see that \name improves the effective throughput at $20\%$ load by $1.37$x compared to Opera, by $1.64$x compared to expander and by $7.02$x compared to Sirius and RotorNet under Valiant load balancing. In Figure~\ref{fig:th-load-perm}, for random permutation demand matrix, we see that \name improves the effective throughput by up to $1.33$x on average compared to Opera and expander; and by $5.66$x compared to Sirius and RotorNet. At low loads, \name achieves similar throughput compared to existing approaches.

\begin{figure*}
\centering
\begin{subfigure}{0.48\linewidth}
\centering
\includegraphics[width=1\linewidth]{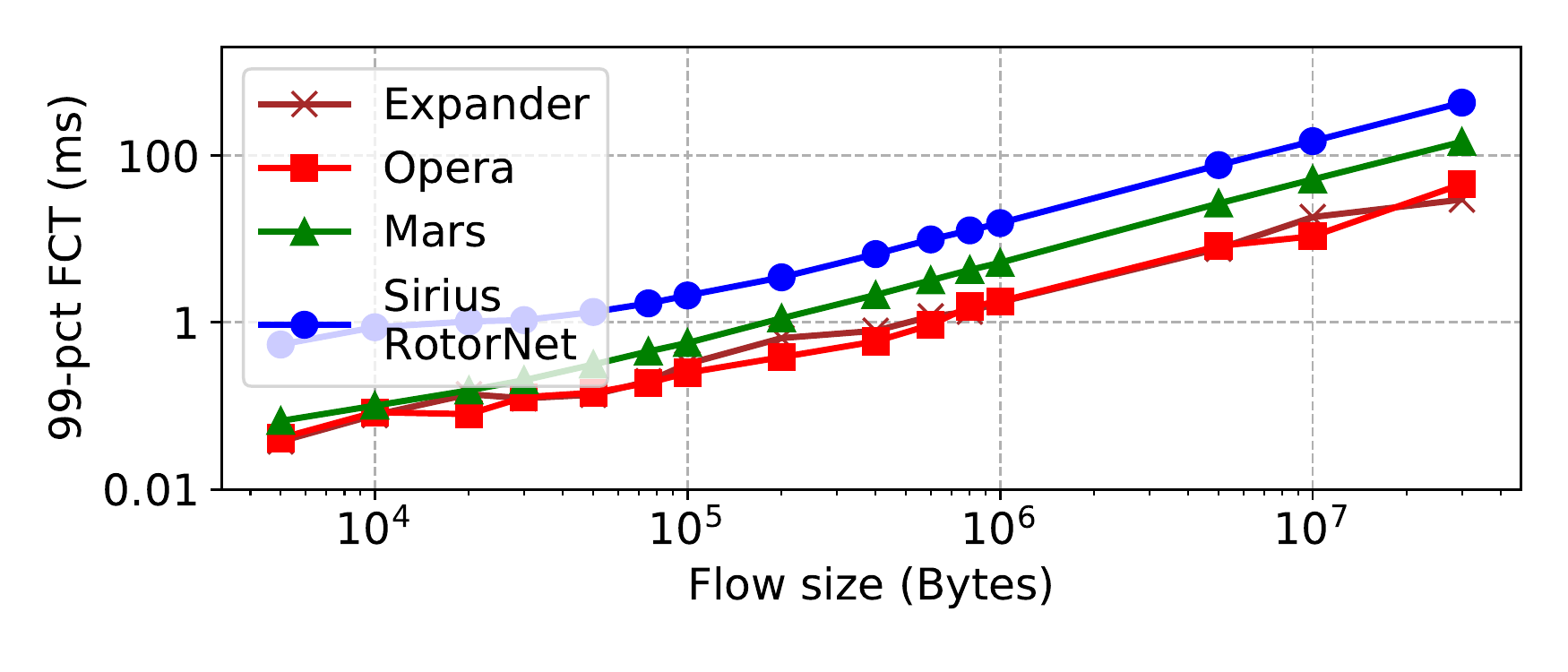}
\subcaption{$1\%$ load}
\label{fig:fcts-0.01-a2a}
\end{subfigure}
\begin{subfigure}{0.48\linewidth}
\centering
\includegraphics[width=1\linewidth]{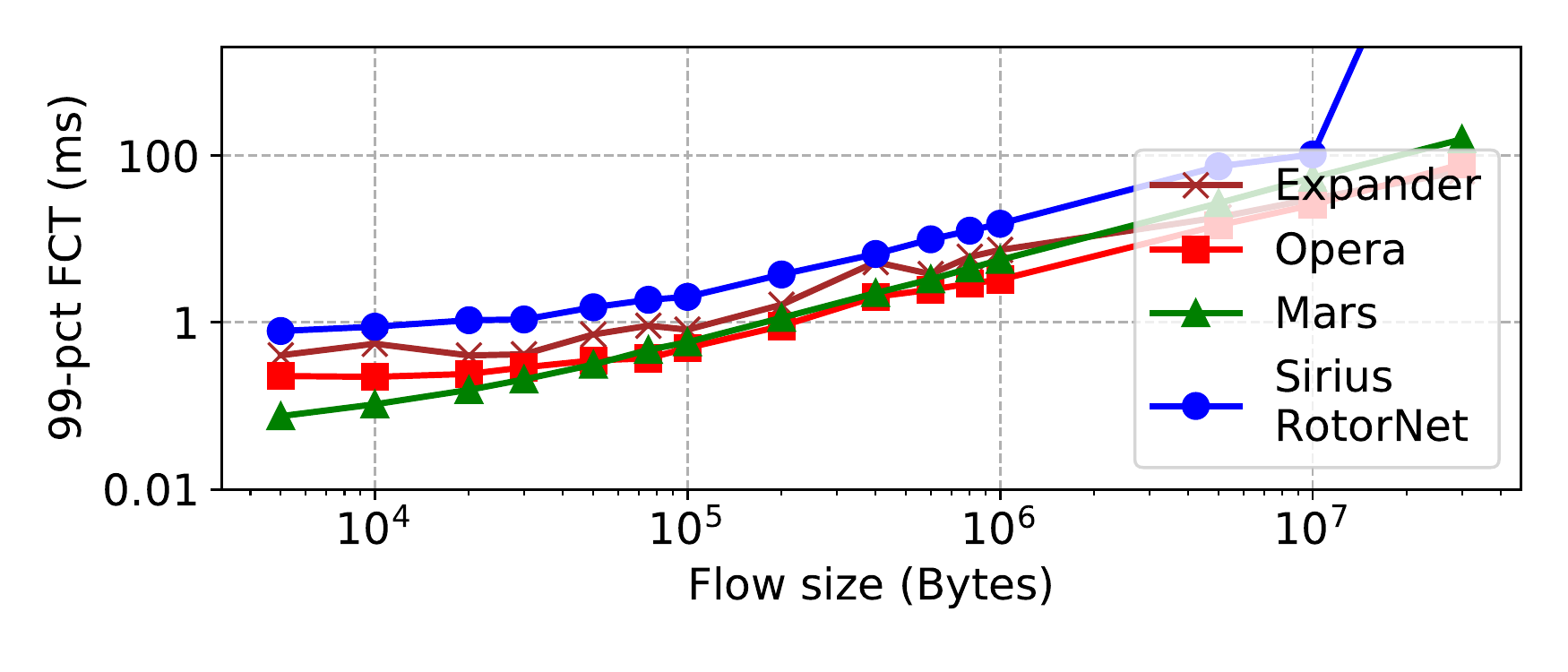}
\subcaption{$5\%$ load}
\label{fig:fcts-0.05-a2a}
\end{subfigure}
\begin{subfigure}{0.48\linewidth}
\centering
\includegraphics[width=1\linewidth]{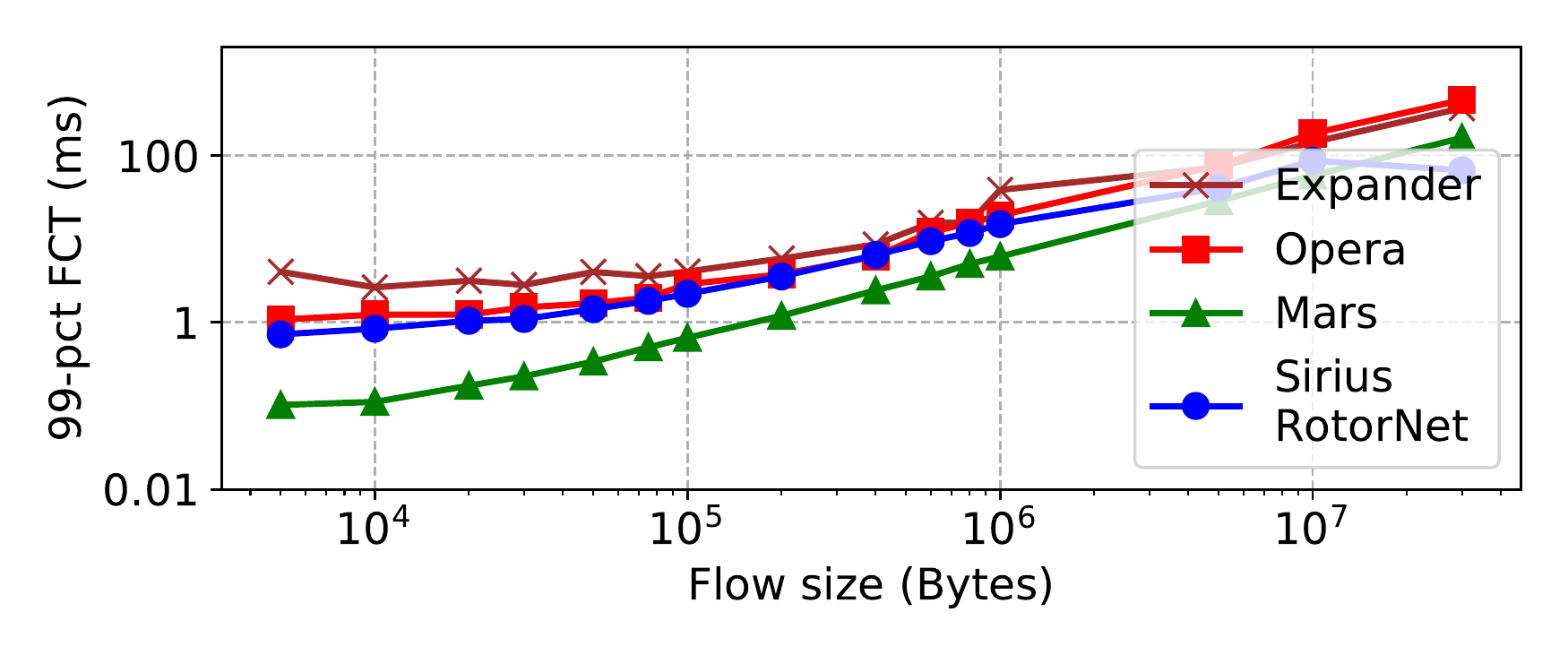}
\subcaption{$10\%$ load}
\label{fig:fcts-0.1-a2a}
\end{subfigure}
\begin{subfigure}{0.48\linewidth}
\centering
\includegraphics[width=1\linewidth]{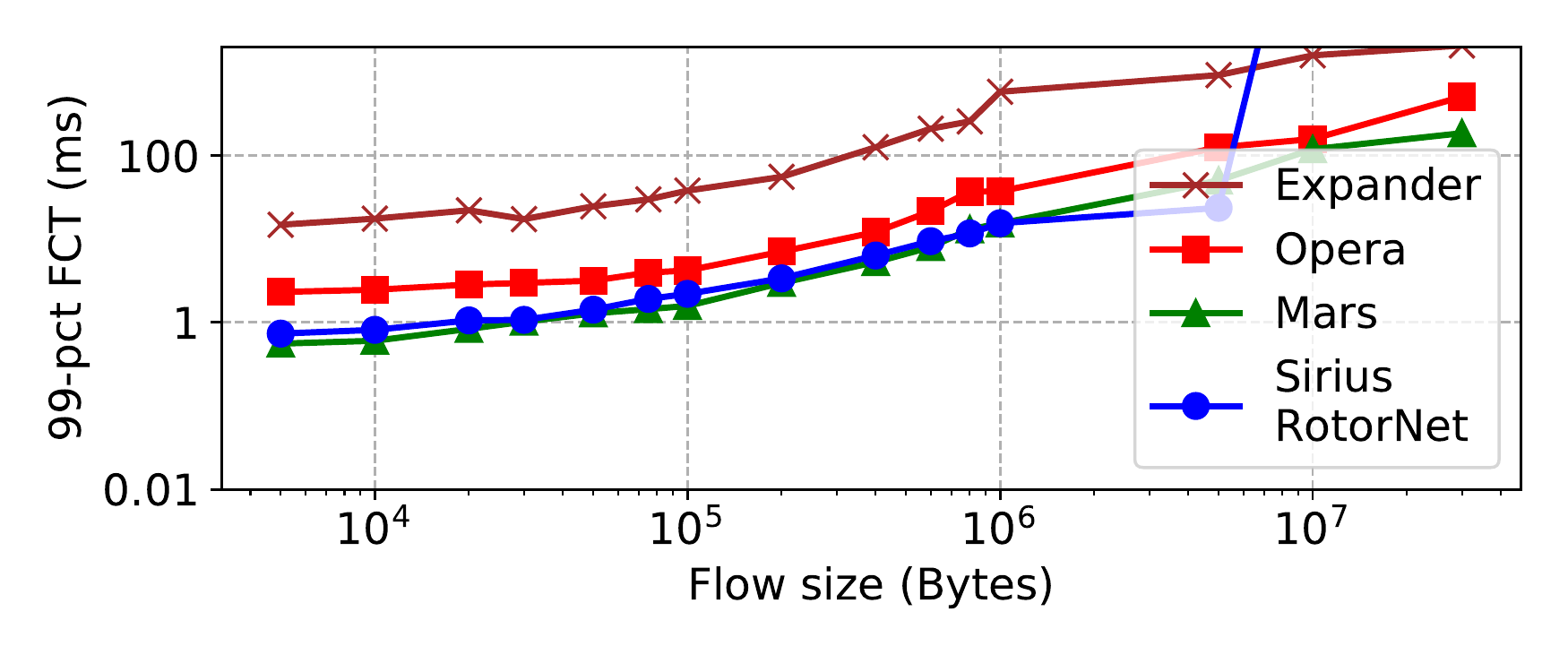}
\subcaption{$20\%$ load}
\label{fig:fcts-0.2-a2a}
\end{subfigure}
\caption{For the websearch workload under all-to-all demand matrix, as the load increases, \name outperforms in flow completion times for both short and long flows. Note the log scale on both the axes.}
\label{fig:fctsa2a}
\end{figure*}

\medskip
\myitem{\name outperforms under shallow buffers:}
Given the high throughput of \name, we evaluate its performance under various buffer sizes at the ToR switches and compare against existing approaches. In Figure~\ref{fig:th-buffer-perm}, we see that even at extremely shallow buffers such as $4$ packets per port, \name achieves $1.57$x better throughput on average compared to Opera, expander, Sirius and RotorNet. However, as we show in Figure~\ref{fig:tradeoffs}, existing systems require significantly more buffer to achieve high throughput. As expected, in Figure~\ref{fig:th-buffer-perm}, we see that Opera, Sirius and RotorNet increase in throughput with large buffers. Indeed, Sirius and RotorNet can sustain even higher loads up to $50\%$ since they emulate a complete graph with shorter path lengths but this requires extremely large buffers. We omit these results for brevity, given that Sirius and RotorNet require buffer sizes as large as $64$ packets per port to achieve $20\%$ throughput in Figure~\ref{fig:th-buffer-perm}.

\begin{wrapfigure}{r}{0.5\textwidth}
  \begin{center}
    \includegraphics[width=0.49\textwidth]{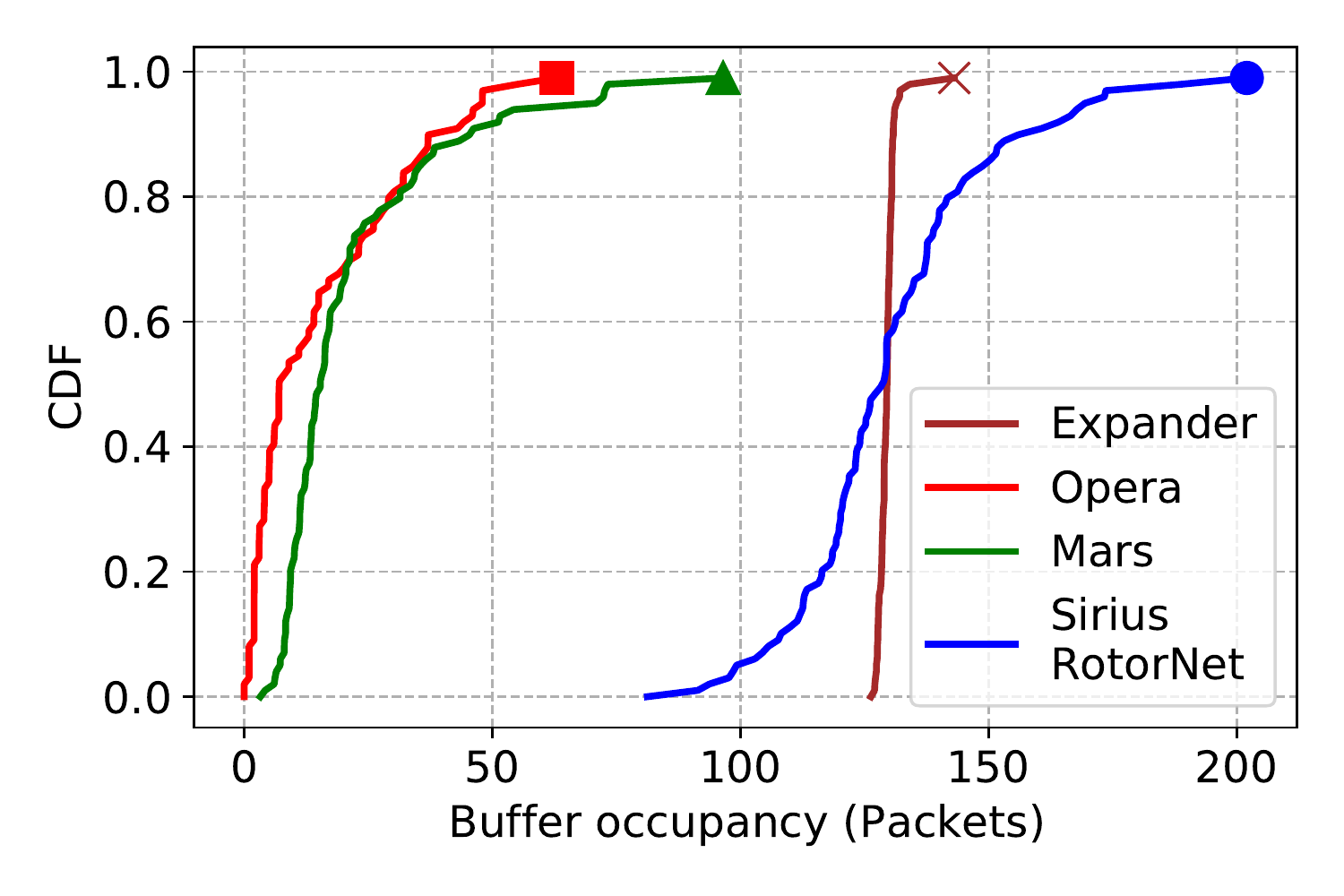}
  \end{center}
  \caption{\name achieves high throughput but requires significantly less buffer compared to Expander, Sirius and RotorNet; and similar buffers as Opera.}
  \label{fig:buf-cdf}
\end{wrapfigure}

\medskip
\myitem{\name does not require large buffers:}
In Figure~\ref{fig:buf-cdf}, we show the CDF of buffer occupancies at the ToR switches for $20\%$ websearch load and $64$ packet per port buffers. We see that \name and Opera require significantly lower buffer compared to expander, Sirius and RotorNet (under Valiant load balancing). Opera selectively buffers packets at the end-hosts based on their flow size and only routes short flows across multi-hops which contributes its very low buffer requirements as seen in Figure~\ref{fig:buf-cdf}. However, Opera achieves $1.22$x \emph{lower} throughput compared to \name as seen in Figure~\ref{fig:th-buffer-perm}. We also observe that Sirius and RotorNet emulating a complete graph, consume significantly more buffer compared to \name: $2.09$x higher at the tail.

\medskip
\myitem{\name significantly improves FCTs of short flows:}
The high throughput as well as low buffer requirements of \name significantly improve the FCTs of short flows. Figure~\ref{fig:fctsperm} shows the $99$-percentile FCTs across various loads for random permutation demand matrix under websearch workload. At $1\%$ load (Figure~\ref{fig:fcts-0.01-perm}), we see that \name achieves similar FCTs for short flows compared to Opera and expander networks and improves upon Sirius and RotorNet by $87.37\%$. As the load increases, \name outperforms existing approaches. Specifically, at $5\%$ load (Figure~\ref{fig:fcts-0.05-perm}), \name reduces the FCTs for short flows by $77\%$ compared to Opera, by $98.11\%$ compared to expander and by $88.43\%$ compared to Sirius and RotorNet.
Further, at $10\%$ load (Figure~\ref{fig:fcts-0.1-perm}), \name improves the FCTs for short flows by $93.88\%$ compared to Opera, by $98.11\%$ compared to expander and by $83.47\%$ compared to Sirius and RotorNet.

We observe similar improvements under an all-to-all demand matrix as shown in Figure~\ref{fig:fctsa2a}. At $5\%$ ($10\%$) load, \name reduces the FCTs for short flows by $66.96\%$ ($90.15\%$) compared to Opera, by $81.35\%$ ($97.46\%$) compared to expander and by $87.6\%$ ($83.90\%$) compared to Sirius and RotorNet.

It is interesting to observe that the $99\%$-percentile flow completion times for short flows at $1\%$ load follows the delay trend shown in Figure~\ref{fig:tradeoffs}: a static expander has the least path delay and similarly Opera since it routes all the short flows over an expander; Sirius and RotorNet which emulate a complete graph experience the highest delays. However, as the load increases, multi-hop routing and queueing delays also impact the FCTs of short flows.

\medskip
\myitem{\name does not trade long flow FCTs for short flows:}
\name not only improves the FCTs for short flows but also achieves on-par $99$-percentile FCTs for long flows compared to existing approaches.
Under random permutation demand matrix (Figure~\ref{fig:fctsperm}), at $20\%$ load, \name reduces the $99$-percentile FCTs for long flows by $2.37\%$ compared to Opera, by $85.98\%$ compared to expander and by $99.9\%$ compared to Sirius and RotorNet. Further, under an all-to-all demand matrix (Figure~\ref{fig:fctsa2a}), at $20\%$ load, \name reduces the FCTs for long flows by $63.42\%$ compared to Opera, by $81.96\%$ compared to expander and by $99.2\%$ compared to Sirius and RotorNet.
Interestingly, at $20\%$ load, under random permutation (Figure~\ref{fig:fcts-0.2-perm}) and all-to-all (Figure~\ref{fig:fcts-0.2-a2a}) demand matrices, Sirius and RotorNet cannot sustain long flow FCTs given the shallow buffers used in our setup. This result further strengthens our motivation on the performance of existing approaches under resource constraints.

\medskip
Overall, our evaluation confirms that \name outperforms existing approaches by improving throughput and reducing the flow completion times for short flows as well as long flows. 

\section{Discussion}
\label{sec:discussion}

While our model and choice of topologies are in line with the assumptions and conclusions in the literature, they have certain practical implications which we discuss below.

\medskip
\noindent
\myitem{Demand Matrix:} Our model assumes that a demand matrix is fixed and does not change over time. In contrast, real-world demand matrices evolve. This would indeed complicate our analysis framework by introducing time variables to the demand. However, our analysis still provides insights within the duration in which the demand matrix remains constant. As demand matrices do not change rapidly over time~\cite{10.1145/3544216.3544265}, one could use our analysis to find the throughput between two time instances when the demand matrix changes significantly.

\medskip
\noindent
\myitem{Congestion control and load balancing:} Both congestion control and load balancing significantly impact the achievable throughput. Our main results in this paper rely on the theoretical definition of flow and throughput maximization problem. As a result, we make a simplifying assumption on the underlying system: congestion control, load balancing and routing have a central view of the entire network and perform ideally. Our theoretical results can be useful in understanding the ideal scenario and provides insights into the performance gap of a deployed system.

\medskip
\noindent
\myitem{Worst-case analysis:} Literature defines throughput of topologies based on worst-case demand matrices and corresponding maximum flow. On one hand, worst-case demand matrices help in understanding the performance bound of a topology under \emph{any} demand matrix (within the scope of our definition). In other words, a topology optimized for the worst-case demand matrix achieves strictly greater throughput under any other demand matrix. On the other hand, the theoretical definition of ``flow'' relates to fluid transmissions with ideal congestion control, load balancing and routing (see above). In essence, our analysis captures the performance bounds of an ideal transport under worst-case demand. Our key insights from theory (summarized in \S\ref{sec:prdcn-tradeoff}) drove the design of \name which is optimized for the worst-case. However, our evaluation of \name incorporates stochastic flow arrival process (not the worst-case) based on real-world flow size distributions~\cite{alizadeh2011data,alizadeh2013pfabric} at a given load. The significantly better performance of \name (optimized for the worst-case) in our evaluation setup (not the worst-case) shows that our analysis indeed finds itself useful even under realistic settings.

\medskip
\noindent
\myitem{Cost-equivalent Clos topologies:} 
We emphasize that our results in Theorem~\ref{th:evolving-union},~\ref{th:throughput-upper-bound},~\ref{th:delay},~\ref{th:buffer} and our evaluation concerns 
uni-regular topologies \ie every switch is connected to servers and generates traffic. In contrast, traditional datacenters built on Clos topologies add additional layers of switches (additional capacity) that do not generate traffic in order to maximize throughput. This paper \emph{does not} argue that uni-regular topologies are better than Clos topologies. However, most recent works~\cite{mellette2017rotornet,sirius,opera,projector,beyondfattress} demonstrate the significant benefits of reconfigurable (uni-regular) topologies over Clos. Furthermore, comparing any other topologies outside the spectrum shown in Figure~\ref{fig:tradeoffs} requires a cost analysis which is volatile in nature. We hence focus strictly on the spectrum of topologies with periodic reconfigurations and their tradeoffs. It would be interesting to determine the buffer size at which the performance of a throughput-optimal periodic reconfigurable topology drops below a cost-equivalent Clos topology. 

\medskip
We leave it for future work to theoretically study the throughput of periodic reconfigurable topologies under dynamic demand matrices and more practical design considerations.

\section{Related Work}

Indeed, only little is known today about how existing reconfigurable
topologies fare against each other, and whether alternative designs could improve the throughput further. In the following, we discuss the related work in network throughput both in the context of traditional static datacenter topologies~\cite{valadarsky2016xpander,singla2012jellyfish,clos,greenberg2009vl2,227667,slimfly,jupiter} and emerging reconfigurable datacenter topologies~\cite{sigmetrics23duo,sirius,zhou2012mirror, kandula2009flyways, mellette2017rotornet, opera, helios, firefly, megaswitch, quartz, chen2014osa,projector,cthrough,splaynet,venkatakrishnan2018costly,schwartz2019online,proteus,100times,fleet}.

\medskip
\myitem{Throughput as a metric:} Bisection bandwidth has been used extensively as a metric in the networking community. A classic result in graph theory suggests that the max-flow can be an $\mathcal{O}(\log(n))$ factor lower than the sparsest cut~\cite{10.1145/331524.331526,leighton1989approximate}. The \emph{throughput} maximization problem has been studied in the context of max-flow multicommodity flow and maximum concurrent flow problems~\cite{doi:10.1137/S0097539794285983,shahrokhi1990maximum,doi:10.1137/S0097539793243016}. A fully polynomial-time approximation scheme exists for arbitrary demands and uniform capacity~\cite{shahrokhi1990maximum}. Recently, Jyothi \etal~\cite{jyothi2016measuring} revisited the relation between cut-based metrics and max-flow in our networking context and proposed throughput under a worst-case demand matrix as metric. While it still remains an active area of research to efficiently compute throughput of a static topology~\cite{tubSigcomm2021,jyothi2016measuring,highthroughputSingla}, initial studies on dynamic and demand-aware topologies attempt to characterize throughput in terms of demand-completion times~\cite{sigmetrics22cerberus}.
We are not aware of a formal definition and analysis of throughput in the context of periodic reconfigurable topologies.

\medskip
\myitem{Static topologies:} Clos-based topologies~\cite{clos,greenberg2009vl2} have been shown to provide optimal throughput~\cite{tubSigcomm2021}. Bcube~\cite{bcube} proposes a server-centric architecture and provides high capacity for all-to-all traffic patterns but may not be optimal in terms of the throughput metric. JellyFish~\cite{singla2012jellyfish} argues that random graphs are highly flexible for datacenters in-terms of heterogeneous expansion and fault-tolerance while achieving high throughput. SlimFly~\cite{slimfly} optimizes for diameter of the topology (consequently throughput) but imposes strict conditions on the size of switches. Xpander~\cite{valadarsky2016xpander} focuses on incremental deployability while achieving high throughput. F10~\cite{f10} on fault-tolerance, and FatClique~\cite{227667} on the cost, incremental expansion and management in a datacenter.

\medskip
\myitem{Reconfigurable topologies:} In contrast to static topologies based on costly, power-intensive electrical packet switches, reconfigurable topologies rely on cost-effective technologies such as optical circuit switches and tunable lasers. Reconfigurable topologies can be broadly classified into two types \first demand-aware and \second demand-oblivious. Demand-aware topologies such as Duo~\cite{sigmetrics23duo}, ReNets~\cite{apocs21renets} and others~\cite{projector,cthrough,splaynet,spaa21rdcn} adjust the topology based on the traffic patterns. However, such networks incur high ``latency tax'' due to the added complexity of measuring and calculating the demand via control-plane.     Demand-oblivious topologies such as RotorNet~\cite{mellette2017rotornet,opera,sirius} rely on a pre-defined schedule for circuit setup and have been shown to provide high-throughput with low ``latency tax'' due to reconfigurations. More recently in a parallel work, the relation between throughput and delay of periodic reconfigurable networks has been studied in detail~\cite{optimalOblivious}. Our main focus in this work has been on demand-oblivious designs under resource constraints. In particular, we reveal the fundamental tradeoffs across throughput, delay \emph{and} buffer requirements for such designs. We further propose \name, a throughput-optimal periodic (demand-oblivious) reconfigurable topology.

\medskip
\myitem{Reducing the buffer requirements:} A vast literature in buffer management~\cite{choudhury1998dynamic,apostolaki2019fab,abm} and active queue management~\cite{codel,6602305,red,westphalpacket}, scheduling~\cite{alizadeh2013pfabric,hong2012finishing,sharma2020programmable} and end-host congestion control~\cite{alizadeh2011data,timely,montazeri2018homa,handley2017re,li2019hpcc,powertcp} focuses on reducing the queueing at a bottleneck link specifically in static datacenter topologies. However, as we rigorously discussed in this paper (\S\ref{sec:motivation},~\S\ref{sec:main}), periodic reconfigurable topologies fundamentally require certain amount of buffering and incur significant throughput loss under limited amount of buffer. Our approach exploits this tradeoff to systematically find the throughput-optimal design for any given delay and buffer constraints.

\section{Conclusion}
\label{sec:conclusion}

This paper was motivated by the observation that 
while dynamically reconfigurable datacenter networks can greatly improve throughput, 
existing RCDN designs which emulate complete graphs can entail high delays and buffer requirements. 
Based on our analysis of the underlying performance tradeoffs, we presented a more scalable network design, \name, which achieves near-optimal throughput by emulating a $d$-regular graph and ensuring shallow buffers. 

We understand our work as a first step and believe that it opens several interesting
avenues for future research. In particular, naturally, the optimal RDCN topology will also depend on the price, and it will be interesting to conduct an economic study of the viability of different architectures. Furthermore, we have so far focused on flat networks; while such networks are common in the literature and have several advantages, it will be interesting to extend our study of buffer-aware RDCN designs to multi-tier networks as well.

\section*{Acknowledgements}
\noindent This work is part of a project that has received funding from the European Research Council (ERC) under the European Union’s Horizon 2020 research and innovation programme, consolidator project Self-Adjusting Networks (AdjustNet), grant agreement No. 864228, Horizon 2020, 2020-2025.
\begin{figure}[!h]
    \centering
   \includegraphics[width=0.3\linewidth]{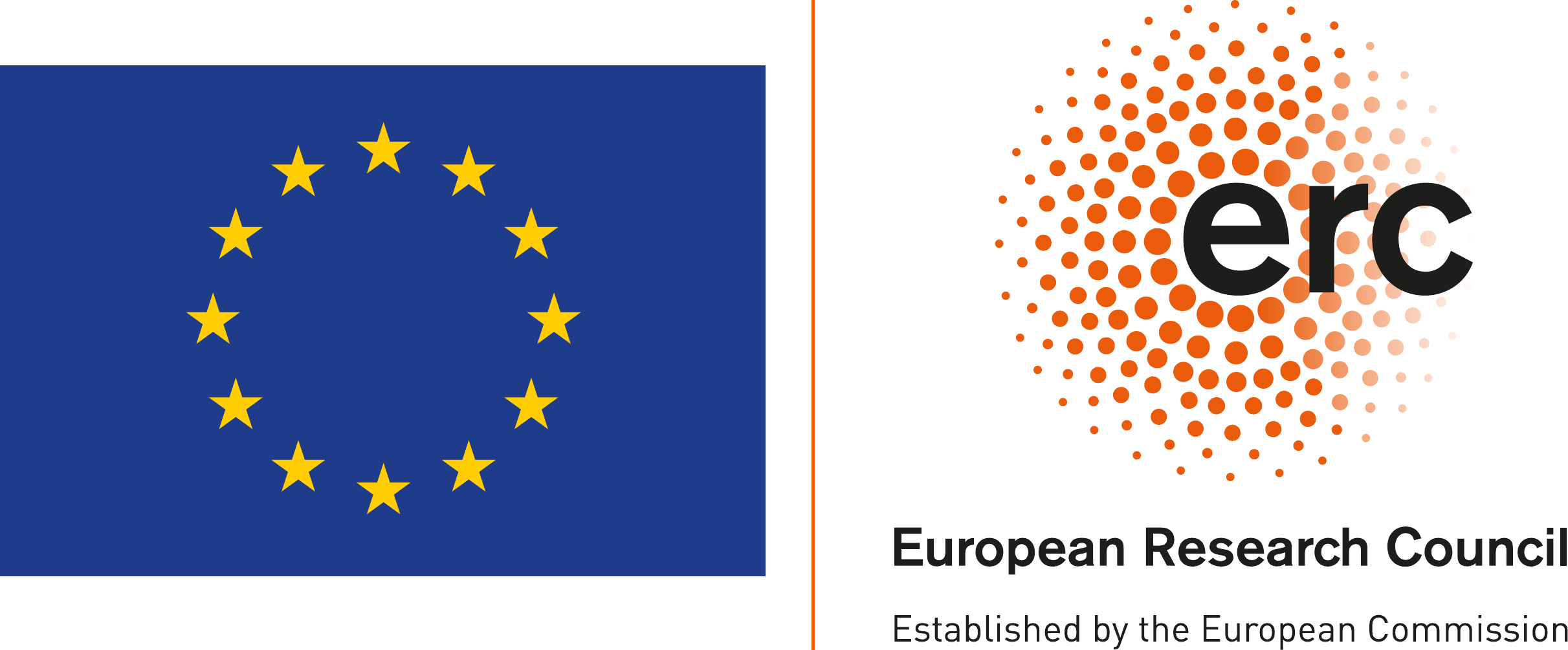}
\end{figure}

\label{bodyLastPage}
\bibliographystyle{unsrt}
\bibliography{references}

\begin{thebibliography}{10}

\bibitem{jupiter}
Arjun Singh, Joon Ong, Amit Agarwal, Glen Anderson, Ashby Armistead, Roy
  Bannon, Seb Boving, Gaurav Desai, Bob Felderman, Paulie Germano, Anand
  Kanagala, Jeff Provost, Jason Simmons, Eiichi Tanda, Jim Wanderer, Urs
  H\"{o}lzle, Stephen Stuart, and Amin Vahdat.
\newblock Jupiter rising: A decade of clos topologies and centralized control
  in google's datacenter network.
\newblock In {\em Proceedings of the ACM SIGCOMM 2015 Conference}, page
  183–197, 2015.

\bibitem{sirius}
Hitesh Ballani, Paolo Costa, Raphael Behrendt, Daniel Cletheroe, Istvan Haller,
  Krzysztof Jozwik, Fotini Karinou, Sophie Lange, Kai Shi, Benn Thomsen, and
  Hugh Williams.
\newblock Sirius: A flat datacenter network with nanosecond optical switching.
\newblock In {\em Proceedings of the ACM SIGCOMM 2020 Conference}, page
  782–797, 2020.

\bibitem{mellette2017rotornet}
William~M. Mellette, Rob McGuinness, Arjun Roy, Alex Forencich, George Papen,
  Alex~C. Snoeren, and George Porter.
\newblock Rotornet: A scalable, low-complexity, optical datacenter network.
\newblock In {\em Proceedings of the ACM SIGCOMM 2017 Conference}, page
  267–280, 2017.

\bibitem{opera}
William~M. Mellette, Rajdeep Das, Yibo Guo, Rob McGuinness, Alex~C. Snoeren,
  and George Porter.
\newblock Expanding across time to deliver bandwidth efficiency and low
  latency.
\newblock In {\em 17th USENIX Symposium on Networked Systems Design and
  Implementation (NSDI 20)}, pages 1--18, Santa Clara, CA, February 2020.
  USENIX Association.

\bibitem{helios}
Nathan Farrington, George Porter, Sivasankar Radhakrishnan, Hamid~Hajabdolali
  Bazzaz, Vikram Subramanya, Yeshaiahu Fainman, George Papen, and Amin Vahdat.
\newblock Helios: A hybrid electrical/optical switch architecture for modular
  data centers.
\newblock In {\em Proceedings of the ACM SIGCOMM 2010 Conference}, page
  339–350, 2010.

\bibitem{firefly}
Navid Hamedazimi, Zafar Qazi, Himanshu Gupta, Vyas Sekar, Samir~R. Das, Jon~P.
  Longtin, Himanshu Shah, and Ashish Tanwer.
\newblock Firefly: A reconfigurable wireless data center fabric using
  free-space optics.
\newblock In {\em Proceedings of the ACM SIGCOMM 2014 Conference}, page
  319–330, 2014.

\bibitem{chen2014osa}
Kai Chen, Ankit Singla, Atul Singh, Kishore Ramachandran, Lei Xu, Yueping
  Zhang, Xitao Wen, and Yan Chen.
\newblock Osa: An optical switching architecture for data center networks with
  unprecedented flexibility.
\newblock {\em IEEE/ACM Transactions on Networking}, 22(2):498--511, 2014.

\bibitem{projector}
Monia Ghobadi, Ratul Mahajan, Amar Phanishayee, Nikhil Devanur, Janardhan
  Kulkarni, Gireeja Ranade, Pierre-Alexandre Blanche, Houman Rastegarfar,
  Madeleine Glick, and Daniel Kilper.
\newblock Projector: Agile reconfigurable data center interconnect.
\newblock In {\em Proceedings of the ACM SIGCOMM 2016 Conference}, page
  216–229, 2016.

\bibitem{cthrough}
Guohui Wang, David~G. Andersen, Michael Kaminsky, Konstantina Papagiannaki,
  T.S.~Eugene Ng, Michael Kozuch, and Michael Ryan.
\newblock C-through: Part-time optics in data centers.
\newblock {\em ACM SIGCOMM Computer Communication Review}, 40(4):327–338, aug
  2010.

\bibitem{splaynet}
Stefan Schmid, Chen Avin, Christian Scheideler, Michael Borokhovich, Bernhard
  Haeupler, and Zvi Lotker.
\newblock Splaynet: Towards locally self-adjusting networks.
\newblock {\em IEEE/ACM Transactions on Networking}, 24(3):1421--1433, 2016.

\bibitem{proteus}
Ankit Singla, Atul Singh, Kishore Ramachandran, Lei Xu, and Yueping Zhang.
\newblock Proteus: A topology malleable data center network.
\newblock In {\em Proceedings of the 9th ACM SIGCOMM Workshop on Hot Topics in
  Networks}, Hotnets-IX, 2010.

\bibitem{osn21}
Matthew {Nance Hall}, Klaus-Tycho Foerster, Stefan Schmid, and Ramakrishnan
  Durairajan.
\newblock A survey of reconfigurable optical networks.
\newblock {\em Optical Switching and Networking}, 41:100621, 2021.

\bibitem{sigmetrics22cerberus}
Chen Griner, Johannes Zerwas, Andreas Blenk, Manya Ghobadi, Stefan Schmid, and
  Chen Avin.
\newblock Cerberus: The power of choices in datacenter topology design - a
  throughput perspective.
\newblock {\em Proc. ACM Meas. Anal. Comput. Syst.}, 5(3), dec 2021.

\bibitem{9288775}
Wei Bai, Shuihai Hu, Kai Chen, Kun Tan, and Yongqiang Xiong.
\newblock One more config is enough: Saving (dc)tcp for high-speed extremely
  shallow-buffered datacenters.
\newblock {\em IEEE/ACM Transactions on Networking}, 29(2):489--502, 2021.

\bibitem{276958}
Prateesh Goyal, Preey Shah, Kevin Zhao, Georgios Nikolaidis, Mohammad Alizadeh,
  and Thomas~E Anderson.
\newblock Backpressure flow control.
\newblock In {\em 19th USENIX Symposium on Networked Systems Design and
  Implementation (NSDI 22)}, Renton, WA, April 2022. USENIX Association.

\bibitem{tubSigcomm2021}
Pooria Namyar, Sucha Supittayapornpong, Mingyang Zhang, Minlan Yu, and Ramesh
  Govindan.
\newblock A throughput-centric view of the performance of datacenter
  topologies.
\newblock In {\em Proceedings of the ACM SIGCOMM 2021 Conference}, page
  349–369, 2021.

\bibitem{valadarsky2016xpander}
Asaf Valadarsky, Gal Shahaf, Michael Dinitz, and Michael Schapira.
\newblock Xpander: Towards optimal-performance datacenters.
\newblock In {\em Proceedings of the 12th International on Conference on
  emerging Networking EXperiments and Technologies}, pages 205--219, 2016.

\bibitem{singla2012jellyfish}
Ankit Singla, Chi-Yao Hong, Lucian Popa, and P.~Brighten Godfrey.
\newblock Jellyfish: Networking data centers randomly.
\newblock In {\em 9th USENIX Symposium on Networked Systems Design and
  Implementation (NSDI 12)}, pages 225--238, San Jose, CA, April 2012. USENIX
  Association.

\bibitem{slimfly}
Maciej Besta and Torsten Hoefler.
\newblock Slim fly: A cost effective low-diameter network topology.
\newblock In {\em SC'14: Proceedings of the International Conference for High
  Performance Computing, Networking, Storage and Analysis}, pages 348--359.
  IEEE, 2014.

\bibitem{cheung2013ultra}
Stanley Cheung, Tiehui Su, Katsunari Okamoto, and SJB Yoo.
\newblock Ultra-compact silicon photonic 512$\times$ 512 25 ghz arrayed
  waveguide grating router.
\newblock {\em IEEE Journal of Selected Topics in Quantum Electronics},
  20(4):310--316, 2013.

\bibitem{coldren2004tunable}
Larry~A Coldren, Gregory~Alan Fish, Y~Akulova, JS~Barton, L~Johansson, and
  CW~Coldren.
\newblock Tunable semiconductor lasers: A tutorial.
\newblock {\em Journal of Lightwave Technology}, 22(1):193, 2004.

\bibitem{serdes}
Broadcom. 2020. 25.6 tb/s strataxgs tomahawk 4 ethernet switch series.

\bibitem{jyothi2016measuring}
Sangeetha~Abdu Jyothi, Ankit Singla, P~Brighten Godfrey, and Alexandra Kolla.
\newblock Measuring and understanding throughput of network topologies.
\newblock In {\em SC'16: Proceedings of the International Conference for High
  Performance Computing, Networking, Storage and Analysis}, pages 761--772.
  IEEE, 2016.

\bibitem{highthroughputSingla}
Ankit Singla, P.~Brighten Godfrey, and Alexandra Kolla.
\newblock High throughput data center topology design.
\newblock In {\em 11th USENIX Symposium on Networked Systems Design and
  Implementation (NSDI 14)}, pages 29--41, Seattle, WA, April 2014. USENIX
  Association.

\bibitem{shahrokhi1990maximum}
Farhad Shahrokhi and David~W Matula.
\newblock The maximum concurrent flow problem.
\newblock {\em Journal of the ACM (JACM)}, 37(2):318--334, 1990.

\bibitem{leighton1989approximate}
Tom Leighton and Satish Rao.
\newblock An approximate max-flow min-cut theorem for uniform multicommodity
  flow problems with applications to approximation algorithms.
\newblock Technical report, Massachusetts Inst of Tech Cambridge Microsystems
  Research Center, 1989.

\bibitem{10.1145/331524.331526}
Tom Leighton and Satish Rao.
\newblock Multicommodity max-flow min-cut theorems and their use in designing
  approximation algorithms.
\newblock {\em J. ACM}, 46(6):787–832, nov 1999.

\bibitem{abm}
Vamsi Addanki, Maria Apostolaki, Manya Ghobadi, Stefan Schmid, and Laurent
  Vanbever.
\newblock Abm: Active buffer management in datacenters.
\newblock In {\em Proceedings of the ACM SIGCOMM 2022 Conference}, 2022.

\bibitem{reactor}
He~Liu, Feng Lu, Alex Forencich, Rishi Kapoor, Malveeka Tewari, Geoffrey~M.
  Voelker, George Papen, Alex~C. Snoeren, and George Porter.
\newblock Circuit switching under the radar with {{REACToR}}.
\newblock In {\em 11th USENIX Symposium on Networked Systems Design and
  Implementation (NSDI 14)}, pages 1--15, Seattle, WA, April 2014. USENIX
  Association.

\bibitem{valiant1981universal}
Leslie~G Valiant and Gordon~J Brebner.
\newblock Universal schemes for parallel communication.
\newblock In {\em Proceedings of the thirteenth annual ACM symposium on Theory
  of computing}, pages 263--277, 1981.

\bibitem{corless1996lambertw}
Robert~M Corless, Gaston~H Gonnet, David~EG Hare, David~J Jeffrey, and Donald~E
  Knuth.
\newblock On the lambertw function.
\newblock {\em Advances in Computational mathematics}, 5(1):329--359, 1996.

\bibitem{bollobas1982diameter}
B{\'e}la Bollob{\'a}s and W~Fernandez de~la Vega.
\newblock The diameter of random regular graphs.
\newblock {\em Combinatorica}, 2(2):125--134, 1982.

\bibitem{imase1985connectivity}
Makoto Imase, Terunao Soneoka, and Keiji Okada.
\newblock Connectivity of regular directed graphs with small diameters.
\newblock {\em IEEE Transactions on Computers}, 34(03):267--273, 1985.

\bibitem{imase1983design}
Makoto Imase and Masaki Itoh.
\newblock A design for directed graphs with minimum diameter.
\newblock {\em IEEE Transactions on Computers}, 32(08):782--784, 1983.

\bibitem{debruijn}
D.~Z. Du and F.~K. Hwang.
\newblock Generalized de bruijn digraphs.
\newblock {\em Networks}, 18(1):27--38, 1988.

\bibitem{NADDEF1981283}
D.~Naddef and W.R. Pulleyblank.
\newblock Matchings in regular graphs.
\newblock {\em Discrete Mathematics}, 34(3):283--291, 1981.

\bibitem{petersen1891theorie}
Julius Petersen.
\newblock Die theorie der regul{\"a}ren graphs.
\newblock {\em Acta Mathematica}, 15(1):193--220, 1891.

\bibitem{handley2017re}
Mark Handley, Costin Raiciu, Alexandru Agache, Andrei Voinescu, Andrew~W.
  Moore, Gianni Antichi, and Marcin W\'{o}jcik.
\newblock Re-architecting datacenter networks and stacks for low latency and
  high performance.
\newblock In {\em Proceedings of the ACM SIGCOMM 2017 Conference}, page
  29–42, 2017.

\bibitem{alizadeh2011data}
Mohammad Alizadeh, Albert Greenberg, David~A. Maltz, Jitendra Padhye, Parveen
  Patel, Balaji Prabhakar, Sudipta Sengupta, and Murari Sridharan.
\newblock Data center tcp (dctcp).
\newblock In {\em Proceedings of the ACM SIGCOMM 2010 Conference}, page
  63–74, 2010.

\bibitem{alizadeh2013pfabric}
Mohammad Alizadeh, Shuang Yang, Milad Sharif, Sachin Katti, Nick McKeown,
  Balaji Prabhakar, and Scott Shenker.
\newblock Pfabric: Minimal near-optimal datacenter transport.
\newblock In {\em Proceedings of the ACM SIGCOMM 2013 Conference}, page
  435–446, 2013.

\bibitem{10.1145/3544216.3544265}
Leon Poutievski, Omid Mashayekhi, Joon Ong, Arjun Singh, Mukarram Tariq, Rui
  Wang, Jianan Zhang, Virginia Beauregard, Patrick Conner, Steve Gribble, Rishi
  Kapoor, Stephen Kratzer, Nanfang Li, Hong Liu, Karthik Nagaraj, Jason
  Ornstein, Samir Sawhney, Ryohei Urata, Lorenzo Vicisano, Kevin Yasumura,
  Shidong Zhang, Junlan Zhou, and Amin Vahdat.
\newblock Jupiter evolving: Transforming google's datacenter network via
  optical circuit switches and software-defined networking.
\newblock In {\em Proceedings of the ACM SIGCOMM 2022 Conference}, SIGCOMM '22,
  page 66–85, 2022.

\bibitem{beyondfattress}
Simon Kassing, Asaf Valadarsky, Gal Shahaf, Michael Schapira, and Ankit Singla.
\newblock Beyond fat-trees without antennae, mirrors, and disco-balls.
\newblock In {\em Proceedings of the Conference of the ACM Special Interest
  Group on Data Communication}, pages 281--294. ACM, 2017.

\bibitem{clos}
Mohammad Al-Fares, Alexander Loukissas, and Amin Vahdat.
\newblock A scalable, commodity data center network architecture.
\newblock In {\em Proceedings of the ACM SIGCOMM 2008 Conference}, page
  63–74, 2008.

\bibitem{greenberg2009vl2}
Albert Greenberg, James~R. Hamilton, Navendu Jain, Srikanth Kandula, Changhoon
  Kim, Parantap Lahiri, David~A. Maltz, Parveen Patel, and Sudipta Sengupta.
\newblock Vl2: A scalable and flexible data center network.
\newblock In {\em Proceedings of the ACM SIGCOMM 2009 Conference}, page
  51–62, 2009.

\bibitem{227667}
Mingyang Zhang, Radhika~Niranjan Mysore, Sucha Supittayapornpong, and Ramesh
  Govindan.
\newblock Understanding lifecycle management complexity of datacenter
  topologies.
\newblock In {\em 16th USENIX Symposium on Networked Systems Design and
  Implementation (NSDI 19)}, pages 235--254, Boston, MA, February 2019. USENIX
  Association.

\bibitem{sigmetrics23duo}
Johannes Zerwas, Csaba Gyorgyi, Andreas Blenk, Stefan Schmid, and Chen Avin.
\newblock Duo: A high-throughput reconfigurable datacenter network using local
  routing and control.
\newblock In {\em ACM SIGMETRICS}, 2023.

\bibitem{zhou2012mirror}
Xia Zhou, Zengbin Zhang, Yibo Zhu, Yubo Li, Saipriya Kumar, Amin Vahdat, Ben~Y.
  Zhao, and Haitao Zheng.
\newblock Mirror mirror on the ceiling: Flexible wireless links for data
  centers.
\newblock {\em ACM SIGCOMM Computer Communication Review}, 42(4):443–454, aug
  2012.

\bibitem{kandula2009flyways}
Srikanth Kandula, Jitendra Padhye, and Paramvir Bahl.
\newblock Flyways to de-congest data center networks.
\newblock In {\em HotNets}. {ACM} {SIGCOMM}, 2009.

\bibitem{megaswitch}
Li~Chen, Kai Chen, Zhonghua Zhu, Minlan Yu, George Porter, Chunming Qiao, and
  Shan Zhong.
\newblock Enabling {{Wide-Spread}} communications on optical fabric with
  {{MegaSwitch}}.
\newblock In {\em 14th USENIX Symposium on Networked Systems Design and
  Implementation (NSDI 17)}, pages 577--593, Boston, MA, March 2017. USENIX
  Association.

\bibitem{quartz}
Yunpeng~James Liu, Peter~Xiang Gao, Bernard Wong, and Srinivasan Keshav.
\newblock Quartz: A new design element for low-latency dcns.
\newblock {\em ACM SIGCOMM Computer Communication Review}, 44(4):283–294, aug
  2014.

\bibitem{venkatakrishnan2018costly}
Shaileshh Bojja~Venkatakrishnan, Mohammad Alizadeh, and Pramod Viswanath.
\newblock Costly circuits, submodular schedules and approximate
  carath{\'e}odory theorems.
\newblock {\em Queueing Systems}, 88(3):311--347, 2018.

\bibitem{schwartz2019online}
Roy Schwartz, Mohit Singh, and Sina Yazdanbod.
\newblock Online and offline greedy algorithms for routing with switching
  costs.
\newblock {\em CoRR}, abs/1905.02800, 2019.

\bibitem{100times}
Michelle Hampson.
\newblock Reconfigurable optical networks will move supercomputerdata 100x
  faster,".
\newblock {\em IEEE Spectrum}, 2021.

\bibitem{fleet}
Fred Douglis, Seth Robertson, Eric van~den Berg, Josephine Micallef, Marc
  Pucci, Alex Aiken, Keren Bergman, Maarten Hattink, and Mingoo Seok.
\newblock Fleet—fast lanes for expedited execution at 10 terabits: Program
  overview.
\newblock {\em IEEE Internet Computing}, 25(3):79--87, 2021.

\bibitem{doi:10.1137/S0097539794285983}
Yonatan Aumann and Yuval Rabani.
\newblock An o(log k) approximate min-cut max-flow theorem and approximation
  algorithm.
\newblock {\em SIAM Journal on Computing}, 27(1):291--301, 1998.

\bibitem{doi:10.1137/S0097539793243016}
Naveen Garg, Vijay~V. Vazirani, and Mihalis Yannakakis.
\newblock Approximate max-flow min-(multi)cut theorems and their applications.
\newblock {\em SIAM Journal on Computing}, 25(2):235--251, 1996.

\bibitem{bcube}
Chuanxiong Guo, Guohan Lu, Dan Li, Haitao Wu, Xuan Zhang, Yunfeng Shi, Chen
  Tian, Yongguang Zhang, and Songwu Lu.
\newblock Bcube: A high performance, server-centric network architecture for
  modular data centers.
\newblock In {\em Proceedings of the ACM SIGCOMM 2009 Conference}, page
  63–74, 2009.

\bibitem{f10}
Vincent Liu, Daniel Halperin, Arvind Krishnamurthy, and Thomas Anderson.
\newblock F10: A {{Fault-Tolerant}} engineered network.
\newblock In {\em 10th USENIX Symposium on Networked Systems Design and
  Implementation (NSDI 13)}, pages 399--412, Lombard, IL, April 2013. USENIX
  Association.

\bibitem{apocs21renets}
Chen Avin and Stefan Schmid.
\newblock Renets: Statically-optimal demand-aware networks.
\newblock In {\em Proc. SIAM Symposium on Algorithmic Principles of Computer
  Systems (APOCS)}, 2021.

\bibitem{spaa21rdcn}
Janardhan Kulkarni, Stefan Schmid, and Pawel Schmidt.
\newblock Scheduling opportunistic links in two-tiered reconfigurable
  datacenters.
\newblock In {\em 33rd ACM Symposium on Parallelism in Algorithms and
  Architectures (SPAA)}, 2021.

\bibitem{optimalOblivious}
Daniel Amir, Tegan Wilson, Vishal Shrivastav, Hakim Weatherspoon, Robert
  Kleinberg, and Rachit Agarwal.
\newblock Optimal oblivious reconfigurable networks.
\newblock {\em arXiv}, abs/2111.08780, 2021.

\bibitem{choudhury1998dynamic}
A.K. Choudhury and E.L. Hahne.
\newblock Dynamic queue length thresholds for shared-memory packet switches.
\newblock {\em IEEE/ACM Transactions on Networking}, 6(2):130--140, 1998.

\bibitem{apostolaki2019fab}
Maria Apostolaki, Laurent Vanbever, and Manya Ghobadi.
\newblock Fab: Toward flow-aware buffer sharing on programmable switches.
\newblock In {\em ACM Workshop on Buffer Sizing}, 2019.

\bibitem{codel}
Kathleen Nichols and Van Jacobson.
\newblock Controlling queue delay: A modern aqm is just one piece of the
  solution to bufferbloat.
\newblock {\em Queue}, 10(5):20–34, may 2012.

\bibitem{6602305}
Rong Pan, Preethi Natarajan, Chiara Piglione, Mythili~Suryanarayana Prabhu,
  Vijay Subramanian, Fred Baker, and Bill VerSteeg.
\newblock Pie: A lightweight control scheme to address the bufferbloat problem.
\newblock In {\em 2013 IEEE 14th International Conference on High Performance
  Switching and Routing (HPSR)}, pages 148--155, 2013.

\bibitem{red}
S.~Floyd and V.~Jacobson.
\newblock Random early detection gateways for congestion avoidance.
\newblock {\em IEEE/ACM Transactions on Networking}, 1(4):397--413, 1993.

\bibitem{westphalpacket}
Cedric Westphal, Kiran Makhijani, and Richard Li.
\newblock Packet trimming to reduce buffer sizes and improve round-trip times.
\newblock In {\em ACM Workshop on Buffer Sizing}, 2019.

\bibitem{hong2012finishing}
Chi-Yao Hong, Matthew Caesar, and P.~Brighten Godfrey.
\newblock Finishing flows quickly with preemptive scheduling.
\newblock {\em ACM SIGCOMM Computer Communication Review}, 42(4):127–138, aug
  2012.

\bibitem{sharma2020programmable}
Naveen~Kr. Sharma, Chenxingyu Zhao, Ming Liu, Pravein~G Kannan, Changhoon Kim,
  Arvind Krishnamurthy, and Anirudh Sivaraman.
\newblock Programmable calendar queues for high-speed packet scheduling.
\newblock In {\em 17th USENIX Symposium on Networked Systems Design and
  Implementation (NSDI 20)}, pages 685--699, Santa Clara, CA, February 2020.
  USENIX Association.

\bibitem{timely}
Radhika Mittal, Vinh~The Lam, Nandita Dukkipati, Emily Blem, Hassan Wassel,
  Monia Ghobadi, Amin Vahdat, Yaogong Wang, David Wetherall, and David Zats.
\newblock Timely: Rtt-based congestion control for the datacenter.
\newblock In {\em Proceedings of the ACM SIGCOMM 2015 Conference}, page
  537–550, 2015.

\bibitem{montazeri2018homa}
Behnam Montazeri, Yilong Li, Mohammad Alizadeh, and John Ousterhout.
\newblock Homa: A receiver-driven low-latency transport protocol using network
  priorities.
\newblock In {\em Proceedings of the ACM SIGCOMM 2018 Conference}, page
  221–235, 2018.

\bibitem{li2019hpcc}
Yuliang Li, Rui Miao, Hongqiang~Harry Liu, Yan Zhuang, Fei Feng, Lingbo Tang,
  Zheng Cao, Ming Zhang, Frank Kelly, Mohammad Alizadeh, and Minlan Yu.
\newblock Hpcc: High precision congestion control.
\newblock In {\em Proceedings of the ACM SIGCOMM 2019 Conference}, page
  44–58, 2019.

\bibitem{powertcp}
Vamsi Addanki, Oliver Michel, and Stefan Schmid.
\newblock {PowerTCP}: Pushing the performance limits of datacenter networks.
\newblock In {\em 19th USENIX Symposium on Networked Systems Design and
  Implementation (NSDI 22)}, Renton, WA, April 2022. USENIX Association.

\bibitem{newman2004analysis}
Mark~EJ Newman.
\newblock Analysis of weighted networks.
\newblock {\em Physical review E}, 70(5):056131, 2004.

\end{thebibliography}

\appendix
\section*{Appendices}

\section{Preliminaries: Throughput of Static Topologies}
\label{sec:motivation-throughput}

Our goal in this section is to formally introduce throughput for static topologies and the relevant definitions, including the limitations of existing bounds. This section builds the intuition and motivation for our throughput analysis in the context of periodic reconfigurable topologies~(\S\ref{sec:throughput-reconf}).
Much of the prior work~\cite{highthroughputSingla,jyothi2016measuring,tubSigcomm2021} focused on static topologies and it remains unclear how the existing methodologies can be used to study the throughput problem in the context of reconfigurable topologies which is the focus of this paper. 
Initial studies on reconfigurable topologies~\cite{sigmetrics22cerberus} informally define throughput of specific existing systems (\eg RotorNet) which emulate a complete graph.
We emphasize that our definitions hold for \emph{any} periodic reconfigurable topology.

The predominant throughput metrics for static topologies are defined via demand matrices (Definition~\ref{def:demand-matrix}, below), and in particular, the worst-case demand matrix.
Following the definition of Jyothi \etal~\cite{jyothi2016measuring}, we formally introduce throughput given a demand matrix for static topologies. In a nutshell, given a demand matrix $\mathcal{M}$, throughput is the highest scaling factor $\theta(\mathcal{M})$ such that the scaled demand matrix $\theta(\mathcal{M})\cdot \mathcal{M}$ is feasible in the topology \ie there exists a feasible flow that satisfies the demand. The throughput $\theta^*$ under a worst case demand matrix is the minimum $\theta(\mathcal{M})$ over the set of all saturated demand matrices \ie $\theta^* = \min_{\mathcal{M}} \theta(\mathcal{M})$.  Before formally defining throughput, we first define certain preliminaries which build the intuition for our definition of throughput in dynamic topologies~(\S\ref{sec:throughput-reconf}).

Consider any static topology (at switch level) represented as a static directed multigraph $G=(V,E)$ where $V$ is the set of vertices (switches) and $E$ is the set of \textbf{labelled} edges (links). Labelled edges allow for distinguishing parallel links. Let $c(e,\ell)$ denote the capacity of an edge $(e,\ell)\in E$ where $\ell$ is the corresponding label and $e$ is directed edge of the form $(u,v)$ connecting two vertex $u$ to vertex $v$. We denote by $c(u)$, the sum of capacities of all outgoing (correspondingly incoming) edges of a node $u$. We assume that each node has equal incoming and outgoing total capacity. A demand matrix specifies the demand in bps (bits/second) between every pair of vertices. 
\begin{definition}[Demand matrix]\label{def:demand-matrix}
    Given a set of vertices $V$, a demand matrix specifies the demand rate between every pair of vertices in bps defined as $\mathcal{M}=\{ m_{u,v} \mid u\in V, v\in V \}$ where $m_{u,v}$ is the demand between the pair $u,v$. A saturated demand matrix is such that the total demand originating at a source~$s$ equals its outgoing capacity and the total demand terminating at a destination~$d$ equals its incoming capacity \ie $\sum_{u\in V} m_{s,u} = c(s)$ and $\sum_{u\in V} m_{u,d} = c(d)$.
\end{definition}
We consider \emph{saturated} demand matrices since we focus on the maximum achievable throughput. Specifically, saturated demand matrices allow for studying the maximum demand that can be routed in a topology within the capacity constraints.

\subsection{Paths and Flow in Static Graphs}
Given a demand matrix $\mathcal{M}$, the graph $G$ has a set of $s\hyphen d$ paths for transmitting the demand $m_{s,d}$ from source~$s$ to destination~$d$ for all $s\hyphen d$ pairs. We consider simple paths \ie without cycles.
\begin{definition}[Simple paths in static graphs (Standard)]\label{def:pathstd}
    Given a static directed graph $G=(V,E)$ with set of vertices $V$ and the set of all labelled edges $E$, a path $p^{*}$ of length \length connecting a source $s$ and a destination $d$  is a sequence of \length labelled edges $\left\langle (e_1,\ell_1), (e_2,\ell_2) \ ...\  (e_\length,\ell_\length) \right\rangle$ where $(e_i,\ell_i)\in E$; the corresponding sequence of vertices is $\left\langle v_1, v_2 \ ... \ v_{\length},v_{\length+1}\right\rangle$ where $v_i\in V$, $v_1$ is the source $s$, $v_{\length+1}$ is the destination $d$ and $e_i$ is an edge between $v_i$ and $v_{i+1}$ \ie $e_i=(v_i,v_{i+1})$ for all $i\in[1,\length]$. We consider \emph{simple} paths where each vertex appears only once in the path, so there are no cycles. 
\end{definition}

Similar to edge labels in a multigraph, we generalize the definition of paths to facilitate distinguishing between two paths with the same sequence of labelled edges (crucial for our analysis later). From here on, we will use an extended definition of paths in static graphs, formally defined below. We associate each path $p$ with a unique identifier. As a result, notice that by associating each path with a unique identifier, the set of all extended paths $P$ may contain more than one path with the same sequence of labelled edges and vertices. For example, $(p^*,i)$ and $(p^*,j)$ are treated as two different paths unless $i=j$ even though $p^*$ from Definition~\ref{def:pathstd} has the same sequence of labelled edges and vertices. We define paths with unique identifiers deliberately for ease of analysis later. Given  a set of (extended) paths $P$ and a path $p\in P$, we denote by $I(p)$, the set of all (extended) paths which have the identical sequence of labelled edges and vertices as that of $p$.

\begin{definition}[Extended paths in static graphs]\label{def:path}
Given a static directed graph $G=(V,E)$, an extended path $p$ is a pair $p=(p^{*}, i)$,
where $p^{*}$ is a standard path (Definition~\ref{def:pathstd}) 
and $i$ is a unique identifier. 
An extended set of paths $P$ is a set of extended paths.
Given such $P$, the set of all $s\hyphen d$ (extended) paths in $P$ is denoted by $P_{s,d}$.
\end{definition}
Intuitively, the transmission of $s\hyphen d$ demand from $s$ to $d$ in a graph is called an $s\hyphen d$ \textit{flow}. The set of all $s\hyphen d$ flows is called a \textit{flow}. We consider that $s\hyphen d$ flow is splittable over multiple edges along all $s\hyphen d$ paths. A legal (or feasible) flow in a static graph must be conserved along the path and must obey capacity constraints on every edge. Given a static graph and the set of \emph{all} paths (Definition~\ref{def:pathstd}), there are infinite possibilities for an extended set of paths from Definition~\ref{def:path}. Note that, flow is a map from a \emph{finite} set of extended paths. We explicitly assume that for each path $p \in P$, and any $p' \in I(p)$, we have $F(p)=F(p')$.
\begin{definition}[Flow in static graphs]\label{def:flow}
    Given static graph $G=(V,E)$ and a finite set of (extended) paths $P$, 
    a Flow $F$ is a map $F: P \rightarrow \mathbb{R}^+ $ where $\mathbb{R}^{+}$ denotes non-negative real numbers. A \emph{legal} flow further has the following constraints: \first flow is conserved along the path and \second it obeys the capacity constraints.
    The amount of flow on a path $p \in P$ is denoted by $F(p)$, namely $F(p)$ is the amount of flow in each edge $e \in p$, denoted by $F(p,e)$. Formally, $F(p,e)=F(p)$ for all $e\in p$. 
    Additionally, w.l.o.g, we assume that for each path $p \in P$, and any $p' \in I(p)$, we have $F(p)=F(p')$. The capacity constraint for a legal flow is given by the following inequality, where~~$\mathbb{I}(\cdot)$ is the indicator function.
    \begin{equation}\label{eq:capacity-constraint-static}
    \sum_{p\in P} F(p)\cdot \mathbb{I}((e,\ell)\in p) \le c(e,\ell) \quad \quad \forall e, \ \forall \ell
    \end{equation}
\end{definition}

\subsection{Throughput of Static Graphs}

Following the definition of Jyothi \etal~\cite{jyothi2016measuring}, throughput given a demand matrix $\mathcal{M}$ is the highest scaling factor $\theta$ such that the scaled demand matrix $\theta\cdot\mathcal{M}$ is feasible in the topology \ie there exists a legal flow which can serve the scaled demand matrix $\theta\cdot\mathcal{M}$. The Throughput $\theta^*$ of a static topology is defined as the throughput under a worst-case demand matrix. Importantly, the following formal definition builds intuition for our context of periodic reconfigurable topologies.

\begin{definition}[Throughput of static graphs]\label{def:throughput}
    Given a demand matrix $\mathcal{M}$ and a graph $G$, a legal flow $F$ has a throughput $\theta(\mathcal{M},F)$ if it satisfies the scaled demand matrix $\theta(\mathcal{M},F)\cdot \mathcal{M}$. Formally, for each $s,d$ pair, the $s\hyphen d$ flow sent from the source $s$ to destination $d$ is greater than the scaled demand $\theta(\mathcal{M},F)\cdot m_{s,d}$.
    \begin{equation}\label{eq:static-throughput}
        \sum_{p \in P_{s,d}} F(p) \ge \theta(\mathcal{M},F)\cdot m_{s,d} \quad\quad\quad \forall s\in V,\ \forall d\in V
    \end{equation}
    The graph $G$ has throughput $\theta(\mathcal{M}) = \max_{F} \theta(\mathcal{M},F)$ and
    the throughput $\theta^*$ under a worst-case demand matrix is the minimum $\theta(\mathcal{M})$ \ie $\theta^* = \min_{\mathcal{M}} \theta(\mathcal{M})$.
\end{definition}

Given a demand matrix $\mathcal{M}$, the throughput maximization problem has the objective to maximize $\theta$, subject to Equation~\ref{eq:capacity-constraint-static} (capacity constraint) and Equation~\ref{eq:static-throughput} (demand constraint). Unfortunately, a linear program approach does not scale well to large topologies. 

\subsection{TUB and its Limitations}\label{sec:tub-limitations}
Recently, Namyar \etal, proposed TUB~\cite{tubSigcomm2021}, a scalable throughput upper bound for static topologies. 
The focus of this paper is not on the throughput of static topologies but rather on  periodic reconfigurable topologies. 
However, even the most recently proposed throughput upper bound for static topologies has key limitations. Specifically, Namyar \etal propose the following throughput upper bound (Theorem 2.2 in~\cite{tubSigcomm2021}),
\[
\theta^* \le \underset{\mathcal{M}}{min} \frac{\displaystyle \sum_e c(e)}{\displaystyle \sum_{s\in V} \sum_{d\in V} m_{s,d}\cdot L_{s,d}}
\]
where $c(e)$ is the capacity of an edge $e\in E$ and $L_{s,d}$ is the \emph{shortest path length} from $s$ to $d$.

\myitem{Key limitations of TUB:}
Consider a topology represented as a complete graph $K_n$ with edge capacity $c$. Every node connects to every other node and hence the shortest path length for all pairs is one. In this case, TUB suggests that the topology has full throughput under a saturated permutation demand matrix, i.e., $\theta^*\le 1$. However, the actual throughput is only $\frac{n}{2\cdot n -1}\approx \frac{1}{2}$. This is since the saturated permutation demand matrix specifies a demand of $(n-1)\times c$ but the shortest path can only accommodate $c$ demand. The remaining demand must be routed through multi hop paths resulting in lower throughput. Unfortunately, TUB does not capture this effect. In general we observe that TUB does not converge even for large scale topologies when the shortest path length significantly differs from the average route length. We show in our analysis that the throughput upper bound for any static graph is indeed related to the \emph{average route length} (Theorem~\ref{th:throughput-upper-bound}) which gives a much tighter bound $\theta^*\le \frac{1}{2}$ in the above example.

\section{Preliminaries: Throughput of Periodic RDCNs}
\label{sec:throughput-reconf}

We now study the throughput of periodic reconfigurable topologies. Similar to static topologies, we first need to formally define paths and flow in the context of periodic reconfigurable topologies. 
In this section, we follow the periodic evolving graph model introduced in \S\ref{sec:evolving-model}. 
Recall that the period of the periodic evolving graph is denoted by $\Gamma$, timeslot is denoted by $\Delta$ and the reconfiguration time is denoted by $\Delta_r$. Since circuit switches do not route traffic during reconfiguration, links remain idle for $\Delta_r$ amount of time in every timeslot~$\Delta$. We denote this fraction of time spent in reconfiguration by~\reconf. We next formally define periodic evolving~graph.

\begin{definition}[Periodic Evolving Graph]\label{def:evolving-graph}
    A periodic evolving graph (henceforth for simplicity referred to as evolving graph) denoted by $\mathcal{G}=(V,\mathcal{E})$ is a periodic sequence of directed graphs with a period of $\Gamma$~timeslots, defined for each timeslot $t\in \mathbb{W}$. The directed graph at time $t\in \mathbb{W}$ is defined as $\mathcal{G}_t = (V, \mathcal{E}_t)$, where $V$ is the set of vertices and $\mathcal{E}_t\subseteq V\times V$ is the set of directed edges at time $t$ and $\mathbb{W}$ is the set of whole numbers. Note that an edge $e\in V\times V$ may appear in multiple edge sets $\mathcal{E}_t$ at different times $t$. The evolving graph $\mathcal{G}$ has the following properties:
    \begin{itemize}
        \item The sequence of graphs starts at $t=0$ and is defined for each timeslot in the interval $t\in[0,\infty)$.
        \item The edge set $\mathcal{E}_t$ at time $t$ is periodic with period $\Gamma$ \ie $\mathcal{E}_{t+\Gamma}=\mathcal{E}_{t}$ and consequently the graph $\mathcal{G}_t$ is periodic with period $\Gamma$.
        \item An edge $e \in V\times V$  has a capacity $c_t(e)$ at time $t$. Since the graph is periodic, the capacity of an edge is also periodic: $c_{t+\Gamma}(e) = c_{t}(e)$. For notational convenience we set $c_t(e)=0$ whenever $e\not\in \mathcal{E}_t$.
    \end{itemize}
\end{definition}

In this section, we first define the notions of path and flow in the periodic evolving graph. Using these definitions, we then formally define throughput.

\subsection{Temporal Paths and Temporal Flow}

Recall from Definition~\ref{def:evolving-graph} that a periodic evolving graph $\mathcal{G}$ is a periodic sequence of directed graphs with period $\Gamma$. The graph at time $t$ is denoted by $\mathcal{G}_t=(V,\mathcal{E}_t)$ where $V$ is the set of vertices (switches) and $\mathcal{E}_t$ is the set of edges (links) at time $t$. Note that the edge set evolves over time.  As a result, paths are formed over time as opposed to static paths. To this end, we define a temporal path in the evolving graph as a sequence of $(e,t)$ edge-time pairs, where an edge $e$ is accessed at time $t$. Consecutive edges along the path are accessed in non-decreasing order of time. We assume that the next edge along a path is accessed within $\Gamma$ timeslots (the period). Further, temporal paths are periodic. We define the set of all temporal paths which start within the first period as the \emph{foundation set} of temporal paths denoted by $\mathcal{P}^0$. The set of \emph{all} temporal paths is denoted by $\mathcal{P}^*$. The foundation set is a crucial part of our analysis in \S\ref{sec:throughput-analysis}, especially since temporal paths are periodic. We define temporal paths formally in the following.

\begin{definition}[Temporal paths in evolving graphs]\label{def:time-path}
    Given a periodic evolving graph $\mathcal{G}$ with set of vertices $V$, a temporal  path $\delta$ of length $\length$ connecting a source $s$ and a destination $d$  is a sequence of $\length$ edges and corresponding time values $\left\langle (e_1,t_1), (e_2,t_2) \ ...\  (e_\length,t_\length) \right\rangle$ where $t_i$ is the time when the edge $e_i$ is accessed along the path and $e_i$ is an edge in $\mathcal{E}_{t_i}$; the corresponding sequence of vertices is $\left\langle v_1, v_2 \ ... \ v_{\length},v_{\length+1}\right\rangle$ where $v_1$ is the source $s$, $v_{\length+1}$ is the destination $d$ and $e_i$ is an edge between $v_i$ and $v_{i+1}$ \ie $e_i=(v_i,v_{i+1})$ for all $i\in[1,\length]$.
    A \emph{legal} temporal path additionally has the following properties:
    \begin{itemize}
        \item An edge $e_{i+1}$ is accessed at time $t_{i+1}$ within a period time $(\Gamma)$ after accessing $e_{i}$ at time $t_i$ \ie $t_i < t_{i+1} \le t_{i}+\Gamma$
        \item $v_i \ne v_j$ if $i\not=j$ \ie the temporal path is simple and has no cycles
    \end{itemize}
    The \emph{foundation set} of temporal paths, denoted as $\mathcal{P}^0$ includes \emph{all} legal temporal paths in $\mathcal{G}$
    that start within the first period (i.e., $t_1 \in [0,\Gamma)$).
    For the \emph{foundation set} $\mathcal{P}^0$, we define the set $\mathcal{P}^*$ of all \emph{legal} temporal paths in $\mathcal{G}$ as follows:
    \begin{itemize}
        \item $\mathcal{P}^0 \subset \mathcal{P}^*$ and temporal paths are periodic with period $\Gamma$.
        \item A function $periodic(\delta)$ is defined as follows for every temporal path $\delta$: the set of all temporal paths which have the same sequence of edges with the time sequence shifted by an integer multiple of~$\Gamma$. Precisely, for any temporal path
        $\timedP \in \mathcal{P}$, 
        $periodic(\timedP)=\{ \left\langle (e_1,t_1+k\cdot\Gamma), (e_2,t_2+k\cdot\Gamma) \ ...\  (e_\ell,t_\ell+k\cdot\Gamma) \right\rangle \mid k \in \mathbb{W}\}$,
        where $\timedP = \left\langle (e_1,t_1), (e_2,t_2) \ ...\  (e_\ell,t_\ell) \right\rangle$ and $\mathbb{W}$ is the set of whole numbers.
        \item $\mathcal{P}^* = \bigcup_{\delta\in \mathcal{P}^0} periodic(\timedP)$ \ie the set of all temporal paths $\mathcal{P}^*$ is the union of periodic temporal paths for each temporal path in the foundation set $\mathcal{P}^0$. For simplicity of notation, we say $\mathcal{P}^* = periodic(\mathcal{P}^0)$. 
    \end{itemize}
The sets $\mathcal{P}^0$ and $\mathcal{P}^*$ are unique given $\mathcal{G}$.
Let the set of all $s\hyphen d$ temporal paths in $\mathcal{P}^0$ and $\mathcal{P}^*$ be denoted by $\mathcal{P}^0_{s,d}$ and $\mathcal{P}^*_{s,d}$, respectively.
\end{definition}

Similar to paths and flow in static graphs (\S\ref{sec:motivation-throughput}), in our context, the evolving graph and the set of $s\hyphen d$ paths (Definition~\ref{def:time-path}) in the evolving graph allow for transmitting the demand of all $s\hyphen d$ pairs. Intuitively, the transmission of $s\hyphen d$ demand along $s\hyphen d$ paths is called flow. In contrast to the flow in static graphs,
the evolving graph $\mathcal{G}$ forces for the flow (Definition~\ref{def:flow}) to be additionally split in time since temporal paths are spread over time. 
Similar to the flow in static graphs, in the following we define a \textit{Temporal-Flow} and its constraints in periodic evolving graphs. Specifically, a legal temporal flow must obey capacity constraints \ie the sum of all flows on an edge at time $t$ is upper bounded by the edge capacity $c_t(e)$. 

\begin{definition}[Temporal flow in evolving graphs]\label{def:time-flow}
    Given a periodic evolving graph $\mathcal{G}$ a legal set of temporal paths $\mathcal{P}$, 
    a temporal flow $\mathcal{F}$ is a map $\mathcal{F}: \mathcal{P} \rightarrow \mathbb{R}^+ $. 
    The amount of flow on a path $\timedP \in \mathcal{P}$ is denoted by $\mathcal{F}(\timedP)$, namely $\mathcal{F}(\timedP)$ is the amount of flow in each edge $(e,t) \in \timedP$, denoted by $\mathcal{F}(\timedP,e,t)$. Formally, $\mathcal{F}(\timedP,e,t)=\mathcal{F}(\timedP)$ for all $(e,t) \in \timedP$ and otherwise $\mathcal{F}(\timedP,e,t)=0$. Temporal-Flow is periodic on periodic paths \ie $\mathcal{F}(\delta) = \mathcal{F}(\delta^\prime)$ if $\delta^\prime \in periodic(\delta)$.
     A temporal flow is said to be \emph{legal} if it obeys the capacity constraints at any time $t$, formally 
    \begin{align}
        \sum_{\timedP \in \mathcal{P}} \mathcal{F}(\timedP,e,t) \le c_t(e) \quad\quad\quad  \forall e , \  \forall t 
    \end{align}
    where $c_t(e)$ is the capacity of an edge $e$ at time $t$.
\end{definition}


\subsection{Throughput of Evolving Graphs}
\label{sec:throughput-defintion-temporal}
Based on the definition of temporal-paths and temporal-flow, we are now ready to define throughput in evolving graphs.  Intuitively, given a demand matrix $\mathcal{M}$, we define throughput as the highest scaling factor $\theta(\mathcal{M})$ such that $\theta(\mathcal{M}) \cdot \mathcal{M}$ amount of demand can be routed in the evolving graph \emph{over one period} on average. 
Specifically, a temporal path $\delta$ sends $\mathcal{F}(\delta)\cdot(\Delta-\Delta_r)$ amount of demands (in bits) over $\Gamma\cdot \Delta$ time before its next periodic flow begins. Hence, $\frac{(\Delta - \Delta_r)}{\Gamma\cdot \Delta} \cdot \mathcal{F}(\delta) = (\frac{1-\reconf}{\Gamma}) \cdot \mathcal{F}(\delta)$ is the average bits per second transmitted on a temporal path $\delta$. Following this intuition, we define throughput in the evolving graph.
\begin{definition}[Throughput in the evolving graph]\label{def:time-throughput}
    Given a demand matrix $\mathcal{M}$ and an evolving graph $\mathcal{G}$, a legal temporal flow $\mathcal{F}$ has throughput $\theta(\mathcal{M}, \mathcal{F})$ if it satisfies the scaled demand matrix over $\Gamma$ timeslots. For each $s,d$ pair, the $s \hyphen d$ temporal flow sent from the source $s$ to destination $d$ over $\Gamma$ timeslots is greater than the scaled demand $\theta(\mathcal{M},\mathcal{F})\cdot m_{s,d}$. Formally, let $\mathcal{P}^0$ be the foundation set of $\mathcal{G}$ and $\mathcal{P}^0_{s,d}$ is the set of all $s\hyphen d$ temporal flows starting in the first period.
    \[
        \left(\frac{1-\reconf}{\Gamma}\right) \cdot \sum_{\timedP \in \mathcal{P}_{s,d}^0} \mathcal{F}(\timedP) \ge \theta(\mathcal{M},\mathcal{F})\cdot m_{s,d} \quad\quad\quad \forall s\in V,\ \forall d\in V
    \]
    The graph $\mathcal{G}$ has throughput $\theta(\mathcal{M}) = \max_{\mathcal{F}} \theta(\mathcal{M},\mathcal{F})$ and
    the throughput $\theta^*$ under a worst-case demand matrix is the minimum $\theta(\mathcal{M})$ \ie $\theta^* = \min_{\mathcal{M}} \theta(\mathcal{M})$.
\end{definition}

Throughput in the evolving graph can be calculated over \emph{any} one period time-interval. For simplicity, we consider the first period and the foundation set of temporal paths $\mathcal{P}^0$ especially since the summation of $s\hyphen d$ flows can be converted to the first period using the periodic property of temporal paths and flow \ie for every $\delta \in \mathcal{P}^0_{s,d}$, $\mathcal{F}(\delta) = \mathcal{F}(\delta^\prime)$ for $\delta^\prime \in periodic(\delta)$.
However, note that it takes at most the duration of the longest temporal path, in order for Definition~\ref{def:time-throughput} to hold at the \emph{destination}. In essence, Definition~\ref{def:time-throughput} for periodic evolving graphs, follows the reasoning behind throughput of static topologies \ie the maximum scaling factor $\theta$ such that the scaled demand matrix is feasible to route continuously in the periodic topology.

We now begin to analyze the throughput maximization problem in periodic evolving graphs. Notice that the problem is much more complicated compared to static graphs. First, the set of all temporal paths is infinite. Second, even the foundation set of temporal paths is exponential in size and grows with larger $\Gamma$ (period).

To this end, our technique involves a static graph which we call the \emph{Emulated graph} corresponding to a periodic evolving graph. We then show that a periodic evolving graph and its corresponding \emph{static} emulated graph have the same throughput.

\subsection{Emulated Graph: An Equivalent Static Topology}\label{sec:union-graph}

Given a periodic evolving graph $\mathcal{G}$, we define the corresponding emulated graph $G(\mathcal{G})$ as a function of $\mathcal{G}$. Specifically, edge set of the emulated graph is obtained from the union of edge sets of the periodic evolving graph taken over any \emph{one period} time interval and the edges are \textbf{labelled} with the corresponding time of the edge in the evolving graph. The edge capacities are such that the emulated graph has the same amount of average capacity between all pairs compared to the evolving graph, including the overhead of reconfiguration ($\Delta_r$). In the following, we formally define the emulated graph.

\begin{definition}[Emulated graph]\label{def:union-graph}
    The emulated graph $G(\mathcal{G})$ of periodic graph $\mathcal{G}$ is defined as a static directed multigraph $G=(V,E)$, where $V$ is the set of vertices of $\mathcal{G}$ and $E= \{ (e,\ell)\mid e\in \mathcal{E}_\ell,\ \ell\in[0,\Gamma) \}$ is the set of directed edges obtained from the union of edges of the evolving graph over the $[0,\Gamma)$ time interval and every edge $(e,\ell)$ is labelled with $\ell$ if $e\in \mathcal{E}_\ell$. An edge $(e,\ell)\in E$ has capacity $\hat{c}(e,\ell)$ where the relation between $\hat{c}(e,\ell)$ and the original capacity $c_\ell(e)$ in the evolving graph is given by,
    \begin{equation}
        \label{eq:bandwidth-relation-appendix}
        \hat{c}(e,\ell) = \left(\frac{1-\reconf}{\Gamma}\right) \cdot c_\ell(e) \quad\quad \forall e, \ \forall \ell \in[0,\Gamma)
    \end{equation}
\end{definition}

We now precisely specify the extended set of paths (see Definition~\ref{def:path}) in the emulated graph. Specifically, the extended set of paths with unique identifiers allow for a relation between a path in the emulated graph to a temporal path in the evolving graph. We define a function \emph{static} which converts a temporal path to static extended path and a function \temporal which converts a static extended path to a temporal path. In essence, the set of all extended paths in the emulated graph is $P^* = \static(\mathcal{P}^0)$. The \temporal function allows for backward conversion \ie $\mathcal{P}^0 = \temporal(P^*)$.

\begin{definition}[Extended paths in Emulated graph]\label{def:union-paths}
Given a periodic evolving graph $\mathcal{G}$ with the foundation set of paths $\mathcal{P}^0$, the corresponding emulated graph $G(\mathcal{G})$ has a set of extended paths $P^*$ given by $\static(\mathcal{P}^0)$. The $\static$ function is defined as follows:
for every temporal path $\timedP$, where $\timedP = \left\langle (e_1,t_1), (e_2,t_2) \ ...\  (e_n,t_n) \right\rangle$, the extended static path $p=\static(\timedP)$ is given by $p=(\left\langle (e_1,\ell_1), (e_2,\ell_2) \ ...\  (e_\length,\ell_\length) \right\rangle, \timedP)$, where $\ell_i= t_i \ \mathbf{mod} \ \Gamma$ and $\timedP$ is the unique identifier of the extended static path; recall that $\timedP$ is unique. 
The inverse of \static function is defined as:
$\temporal(p) = \static^{-1}(p) = \delta$, for all $p\in P^*$, where $\delta$ is the unique identifier associated with the extended path $p$.
\begin{align}
    P^* &= \static(\mathcal{P}^0) = \{ \static(\timedP) \mid \timedP \in \mathcal{P}^0 \}\nonumber \\
    \mathcal{P}^0 &= \temporal(P^*) = \{ \temporal(p) \mid p \in P^* \}
\end{align}

\end{definition}

Note that, given a periodic evolving graph $\mathcal{G}$, its foundation set of temporal paths $\mathcal{P}^0$ is unique and includes \emph{all} temporal paths which start in the first period. Since the emulated graph is a function of the periodic evolving graph, the set of extended static paths (Definition~\ref{def:union-paths}) is also unique.
The extended set of paths and the \static, \temporal functions are a crucial part of our throughput analysis in the next section.

\section{Analysis: Throughput of Periodic RDCNs}
\label{sec:throughput-analysis}

Before presenting our analysis, we first state our main result in this section. 
Consider a periodic periodic evolving graph (see Definition~\ref{def:evolving-graph}) $\mathcal{G}=(V,\mathcal{E})$ with edge capacities $c_t(e)$ for every edge~$e$ at time~$t$. Consider a static graph which is a function of the evolving graph $\mathcal{G}$ obtained from Definition~\ref{def:union-graph} represented as $G(\mathcal{G})=(V,E)$. We prove that the throughput of a periodic evolving graph is equivalent to the corresponding emulated graph.

\relationtoStatic*

A sketch of our approach is as follows:  \first we prove in Lemma~\ref{lem:static-to-evolving} that if a legal flow in the emulated graph has throughput $\theta$ over set of extended paths $P^* = \static(\mathcal{P}^0)$, then there exists a legal temporal flow in the periodic evolving graph with the same throughput; \second we prove in Lemma~\ref{lem:evolving-to-static} that if a legal temporal flow achieves throughput $\theta$ in the evolving graph, then there exists a legal flow in the emulated graph with the same throughput over the set of extended paths $P^* = \static(\mathcal{P}^0)$; \third finally, we prove in Lemma~\ref{lem:extendedJustif} that a flow in emulated graph over the set of \emph{all paths} $P$ has the same throughput upper bound as that of a flow over the set of extended paths $P^* = \static(\mathcal{P}^0)$. Using the above three results, we prove our main claim.

\begin{Lemma}\label{lem:static-to-evolving}
    Let $\mathcal{M}$ be a demand matrix, $\mathcal{G}$ be a periodic evolving graph, $G(\mathcal{G})$ be its emulated graph and $P^* = \static(\mathcal{P}^0)$. 
    If $F: P^* \rightarrow \mathbb{R}^+$ is a legal flow in $G(\mathcal{G})$ with throughput $\theta(\mathcal{M},F)$ then there exists a legal temporal flow $\mathcal{F}$ in the evolving graph $\mathcal{G}$ with the same throughput $\theta(\mathcal{M},\mathcal{F})=\theta(\mathcal{M},F)$. 
    The temporal flow $\mathcal{F}$ in the evolving graph can be constructed as follows:
    \begin{align}
    \label{eq:flow-relation}
    \quad \quad \mathcal{F}(\timedP)  &= \left( \frac{\Gamma}{1-\reconf}\right) \cdot F(p) \nonumber \\
     & \quad \quad  \forall p \in P^*;\  \timedP\in periodic(\temporal(p))
    \end{align}
    where $\mathcal{F}$ is periodic; 
    for any temporal path $\delta \in \temporal(P^*)$, $\mathcal{F}(\delta,e,t)=\mathcal{F}(\delta)$ for all $(e,t)\in \delta$ and $\mathcal{F}(\delta,e,t)=0$ if $(e,t)\not\in \delta$. $\mathcal{F}$ is then a map $\mathcal{F}:\mathcal{P}^* \rightarrow \mathbb{R}^{+}$, where $\mathcal{P}^* = periodic(\temporal(P^*))$.
\end{Lemma}

\begin{proof}
    From Definition~\ref{def:flow}, since $F$ is a legal flow in the emulated graph, $F$ obeys capacity constraints.
    \[
        \displaystyle \sum_{p\in P^*} F(p)\cdot\mathbb{I}((e,\ell)\in p) \le \hat{c}(e,\ell) \quad\quad\quad \forall e, \  \forall \ell\in [0,\Gamma)
    \]
    We first substitute $\hat{c}(e,\ell) = \frac{(1-\reconf)}{\Gamma}\cdot c_\ell (e)$ and change $\ell$ to $\ell \ \mathbf{mod}\  \Gamma$ without changing the value of both sides of the above inequality.
    \begin{align}
        \displaystyle \sum_{p\in P^*} F(p)\cdot\mathbb{I}((e,\ell \ \mathbf{mod}\ \Gamma)\in p) \le \left(\frac{1-\reconf}{\Gamma}\right)\cdot c_{(\ell \ \mathbf{mod}\  \Gamma)}(e) \nonumber \\
        \quad\quad\quad \forall e, \  \forall \ell\in [0,\Gamma)\nonumber
    \end{align}
    We now expand the above inequality as follows for all $\ell \in [0,\infty)$, without changing the value on both sides of the inequality. Recall that $c_\ell(e)=c_{\ell + k\cdot \Gamma}$ for any integer $k\ge 0$ ($c_\ell(e)$ is periodic).
    \begin{align}
        \displaystyle \sum_{p\in P^*} F(p)\cdot\mathbb{I}((e,\ell \ \mathbf{mod}\ \Gamma) \in p) \le \left(\frac{1-\reconf}{\Gamma}\right)\cdot c_\ell(e) \nonumber \\
        \quad\quad\quad \forall e, \  \forall \ell\in[0,\infty) \nonumber
    \end{align}
    The above inequality holds since $c_\ell(e)$ is periodic and $\ell \ \mathbf{mod} \ \Gamma$ always ranges between $[0,\Gamma)$. Substituting the temporal flow $\mathcal{F}$ for the static flow $F$, using Equation~\ref{eq:flow-relation},
    \begin{align}
        \displaystyle \sum_{p\in P^*} \mathcal{F}(\temporal(p))\cdot\mathbb{I}((e,\ell \ \mathbf{mod}\ \Gamma) \in p) \le c_\ell(e) \nonumber \\
        \quad\quad\quad \forall e, \  \forall \ell\in[0,\infty) \nonumber
    \end{align}
    From Definition~\ref{def:union-paths} and from the periodic property of temporal paths (Definition~\ref{def:time-path}), for every path $p\in P^*$ and for every $(e,\ell\ \mathbf{mod}\ \Gamma)\in p$, there exists exactly one path $\delta \in periodic(\temporal(p))$ such that $(e,\ell)\in \delta$. Using this relation, we convert the above inequality as follows,
    \begin{align}
        \displaystyle \sum_{\delta\in periodic(\temporal(P^*))} \mathcal{F}(\delta)\cdot\mathbb{I}((e,t)\in \delta) \le c_t(e) \nonumber \\
        \quad\quad\quad \forall e, \  \forall t\in[0,\infty) \nonumber 
    \end{align}
    \noindent Since $periodic(\temporal(P^*))=\mathcal{P}^*$, and since $c_t(e)=0$ if $e\not\in \mathcal{E}_t$, we obtain the following relation.
    \[
        \sum_{\timedP \in \mathcal{P}^*} \mathcal{F}(\timedP,e,t) \le
        \begin{cases}
            c_t (e) & e\in \mathcal{E}_t      \\
            0       & e\not \in \mathcal{E}_t
        \end{cases}
    \]
    From Definition~\ref{def:time-flow}, we conclude that $\mathcal{F}$ (from Equation~\ref{eq:flow-relation}) obeys capacity constraints. Further since temporal flow is constant and equal for all $(e,t) \in \timedP$ for all temporal paths, it obeys flow conservation rules. It remains to prove that $\mathcal{F}$ also achieves throughput $\theta$.
    Since $F$ has a throughput $\theta(\mathcal{M})$, from Definition~\ref{def:throughput} we have that,
    \[
        \sum_{p \in P^*_{s,d}} F(p) \ge \theta(\mathcal{M},F)\cdot m_{s,d} \quad\quad\quad \forall s\in V,\ \forall d\in V
    \]
    \noindent using the relation between $F$ and $\mathcal{F}$ from  Equation~\ref{eq:flow-relation}, we obtain the following inequality. From Definition~\ref{def:time-throughput}, we conclude that $\mathcal{F}$ also achieves throughput $\theta(\mathcal{M},\mathcal{F}) = \theta(\mathcal{M},F)$.
    \[
        \left(\frac{1-\reconf}{\Gamma}\right) \cdot \sum_{\timedP \in \mathcal{P}^0_{s,d}} \mathcal{F} (p) \ge \theta(\mathcal{M},F)\cdot m_{s,d} \quad\quad\quad \forall s\in V,\ \forall d\in V
    \]
\end{proof}
From Lemma~\ref{lem:static-to-evolving}, we have that, given a demand matrix $\mathcal{M}$, if the static emulated graph has throughput $\theta(\mathcal{M},F)$ for a specific flow $F$, then the scaled demand matrix $\theta(\mathcal{M},F)\cdot \mathcal{M}$ is feasible in the periodic evolving graph. In the following, we state the reverse \ie given a demand matrix $\mathcal{M}$, if the periodic evolving graph has throughput $\theta(\mathcal{M},\mathcal{F})$ for a specific temporal flow $\mathcal{F}$, then the scaled demand $\theta(\mathcal{M},\mathcal{F})\cdot \mathcal{M}$ is feasible in the static emulated graph.

\begin{Lemma}\label{lem:evolving-to-static}
    Given a demand matrix $\mathcal{M}$, if $\mathcal{F}: \mathcal{P}^* \rightarrow \mathbb{R}^+$ is a legal temporal flow in the evolving graph $\mathcal{G}$ with throughput $\theta(\mathcal{M},\mathcal{F})$ then there exists a legal flow $F$ in the emulated graph $G(\mathcal{G})$ with same throughput $\theta(\mathcal{M},F)=\theta(\mathcal{M},\mathcal{F})$. $F$~is obtained as follows: 
    \begin{equation}\label{eq:static-time-flow}
        F(\static(\timedP)) = \left(\frac{1-\reconf}{\Gamma}\right) \cdot \mathcal{F}(\timedP) \quad \forall \delta \in \mathcal{P}^0
    \end{equation}
$F$ is then a map $F: P^* \rightarrow \mathbb{R}^{+}$, where $P^*=\static(\mathcal{P}^0)$.
\end{Lemma}

\begin{proof}
    From Definition~\ref{def:flow}, for $F$ to be a legal flow in the emulated graph $G$, $F$ must obey capacity and flow conservation constraints. Further in order to achieve throughput $\theta(\mathcal{M},F)$, we require that $F\ge \theta(\mathcal{M},F)\cdot \mathcal{M}$. Since $\mathcal{F}$ is a legal flow in the evolving graph, $\mathcal{F}$ obeys capacity constraints. From Definition~\ref{def:time-flow}, for all $e\in \mathcal{E}_{t}$ at any time $t$ we have that,
    \[
        \sum_{\timedP \in \mathcal{P}^*} \mathcal{F}(\timedP)\cdot \mathbb{I}((e,t)\in \delta) \le c_t(e) \quad\quad\quad  \forall e ,\ \forall t\in[0,\infty)
    \]
    \noindent The remainder of the proof follows similar logic as that of the proof of Lemma~\ref{lem:static-to-evolving}. Since for any path $\delta \in \mathcal{P}^*$, $\mathcal{P}^* = periodic(\mathcal{P}^0) = periodic(\temporal(P^*))$, if $(e,t)\in \delta$, then there exists no other path $\delta^\prime \in periodic(\delta)$ where $(e,t)\in \delta^\prime$, except for the path $\delta$ itself. For every such path $\delta$, there exists exactly one path $p\in P^*$ where $(e,\ell) \in p$ and $\ell = t\ \mathbf{mod} \ \Gamma$. Using the relation between $\mathcal{F}$ and $F$ from Equation~\ref{eq:static-time-flow} and substituting the capacity relation $\hat{c}(e,\ell) = \frac{(1-\reconf)}{\Gamma}\cdot c_\ell (e)$, we obtain the following:
    \[
        \sum_{p \in P^*} F(p)\cdot \mathbb{I}((e,\ell)\in p) \le \hat{c}(e,\ell) \quad\quad\quad  \forall e ,\ \forall \ell\in[0,\Gamma)
    \]
    From Definition~\ref{def:flow}, the above inequality implies that the flow $F$ obeys capacity constraints. Further, flow is conserved since $F(p)$ is constant and equal for all $e\in p$ for all paths. It remains to prove that $F$ achieves throughput $\theta(\mathcal{M},F)=\theta(\mathcal{M},\mathcal{F})$.

    \noindent Since $\mathcal{F}$ has a throughput $\theta(\mathcal{M},\mathcal{F})$ for a demand matrix $\mathcal{M}$, from Definition~\ref{def:time-throughput} we have that,
    \[
        \left(\frac{1-\reconf}{\Gamma}\right) \cdot \sum_{\timedP \in \mathcal{P}_{s,d}^0} \mathcal{F}(\timedP) \ge \theta(\mathcal{M},\mathcal{F})\cdot m_{s,d} \quad\quad\quad \forall s\in V,\ \forall d\in V
    \]
    We now substitute static variables using Equation~\ref{eq:static-time-flow} and obtain the throughput relation,
    \[
        \sum_{p \in \static(\mathcal{P}_{s,d}^0)} F(p) \ge \theta(\mathcal{M},\mathcal{F})\cdot m_{s,d} \quad\quad\quad \forall s\in V,\ \forall d\in V
    \]

    \noindent Using the above inequality and since $static(\mathcal{P}^0_{s,d})=P^*_{s,d}$, from Definition~\ref{def:throughput},  we conclude that $F$ also achieves throughput $\theta(\mathcal{M},F)=\theta(\mathcal{M},\mathcal{F})$.
\end{proof}

From Lemma~\ref{lem:static-to-evolving} and Lemma~\ref{lem:evolving-to-static}, given a demand matrix $\mathcal{M}$, the periodic evolving graph and the emulated graph with extended set of paths $P^*=\static(\mathcal{P}^0)$, have the same throughput. It remains to show that the emulated graph with set of extended paths $P^*=\static(\mathcal{P}^0)$ has the same throughput compared to the emulated graph with set of all possible paths (not extended).

\begin{Lemma}\label{lem:extendedJustif}
    Let $\mathcal{M}$ be a demand matrix, $\mathcal{G}$ be a periodic evolving graph, $G(\mathcal{G})$ be its emulated graph and $P^*=\static(\mathcal{P}^0)$.
    If $F: P \rightarrow \mathbb{R}^+$ is a legal flow in $G(\mathcal{G})$ with throughput $\theta(\mathcal{M},F)$ where $P$ is the set of all simple paths then there exists a legal flow $F^\prime: P^* \rightarrow \mathbb{R}^+$  in the emulated graph $\mathcal{G}$ with the same throughput $\theta(\mathcal{M},F^\prime) = \theta(\mathcal{M},F)$ 
    where $F^\prime$ is as follows:
    \begin{equation}\label{eq:extended-std-flow}
        F^\prime (p^\prime) = 
        \frac{F(p)}{|I(p)|} \quad \forall p^\prime \in I(p) \\ 
    \end{equation}
    where $I(p)$ is the set of all extended paths which have the same sequence of labelled edges as that of a path $p\in P$.
\end{Lemma}

\begin{proof}
We have that $F: P \rightarrow \mathbb{R}^+$ is a legal flow in the emulated graph and has a throughput $\theta(\mathcal{M},F)$. The proof follows by showing that $F^\prime: P^* \rightarrow \mathbb{R}^+$ also obeys capacity constraints and has throughput $\theta(\mathcal{M},F^\prime) = \theta(\mathcal{M},F)$.
We will first prove that the following capacity constraint (from Definition~\ref{def:flow}) for $F^\prime$ holds. We prove this using the relation between $F$ and $F^\prime$ from Equation~\ref{eq:extended-std-flow} and the fact that $F$ obeys capacity constraints. Since $F$ obeys capacity constraints, the following inequality holds.
\[
\sum_{p\in P} F(p) \cdot \mathbb{I}((e,\ell)\in p) \le c(e) \quad \quad \forall e, \ \forall \ell
\]
For every path $p\in P$, the set of extended paths $P^*$ consists of a set $I(p)$ of extended paths. Note that $I(p)$ is strictly greater than zero \ie $I(p) > 0$. The argument is that \first $P$ is the set of all paths; \second $P^*$ is obtained from the foundation set of all temporal paths $\mathcal{P}^0$ as $P^*=static(\mathcal{P}^0)$; \third for any path $p\in P$ with the sequence of edges $\langle e_1, e_2,...e_n\rangle$, there exists at least one legal temporal path $\delta \in \mathcal{P}^0$ where $\timedP = \left\langle (e_1,t_1), (e_2,t_2) \ ...\  (e_n,t_n) \right\rangle$ such that $t_1$ is the time when the edge $e_1$ appears for the first time in the first period, $t_2$ is the time when the edge $e_2$ appears for the first time after $t_1$ and so on; each edge appears within $\Gamma$ timeslots due to the periodicity of the edge set and such a temporal path $\delta$ belongs to $\mathcal{P}^0$ by definition. 
Further, for any path $p\in P$ and a labelled edge $(e,\ell)\in p$, by definition, the labelled edge $(e,\ell)$ also belongs to any path $p^\prime \in I(p)$. Using this relation, we expand the summation in the above inequality as follows: 
\[
\sum_{p\in P} \sum_{p^\prime \in I(p)} \frac{F(p)}{|I(p)|} \cdot \mathbb{I}((e,\ell)\in p^\prime) \le c(e) \quad \quad \forall e, \ \forall \ell
\] 
Substituting $F^\prime$ using Equation~\ref{eq:extended-std-flow}, we obtain the following relation.
\[
\sum_{p\in P^*} F^\prime(p^\prime) \cdot \mathbb{I}((e,\ell)\in p^\prime) \le c(e) \quad \quad \forall e, \ \forall \ell
\]
From Definition~\ref{def:flow}, the above inequality suggests that $F^\prime$ obeys capacity constraints. Similarly, it is easy to show that $\sum_{p^\prime \in P^*_{s,d}} F^\prime(p^\prime) \ge \theta(\mathcal{M},F)\cdot m_{s,d}$ using the fact that $F$ has a throughput $\theta(\mathcal{M})$ \ie $\sum_{p \in P_{s,d}} F(p) \ge \theta(\mathcal{M},F)\cdot m_{s,d}$, for all $s,d$ source-destination pairs. This concludes that the flow $F^\prime$ has throughput $\theta(\mathcal{M},F^\prime) = \theta(\mathcal{M},F)$.
\end{proof}

Using Lemma~\ref{lem:static-to-evolving}, Lemma~\ref{lem:evolving-to-static} and Lemma~\ref{lem:extendedJustif}, it is now straight-forward that our claim in Theorem~\ref{th:evolving-union} holds.

\begin{proof}[\textbf{Proof of Theorem}~\ref{th:evolving-union}]
    From Lemma~\ref{lem:static-to-evolving} and Lemma~\ref{lem:evolving-to-static}, for any feasible temporal flow $\mathcal{F}$ in the evolving graph with throughput $\theta(\mathcal{M},\mathcal{F})$, there exists a feasible flow $F$ in the emulated graph with the same throughput $\theta(\mathcal{M},F)=\theta(\mathcal{M},\mathcal{F})$ and vice versa. From Lemma~\ref{lem:extendedJustif}, we have that, our definition of extended paths and its use in our analysis does not impact the throughput of emulated graph \ie throughput with our definition of extended set of paths is same as the throughput with the standard definition of paths. This concludes that the evolving graph and the corresponding emulated graph have the same throughput $\theta(\mathcal{M})$: the maximum scaling factor given a demand matrix and the same throughput $\theta^*$ under a worst-case demand matrix.
\end{proof}

\begin{corollary}[Reduction to simple graph]\label{cor:reduction}
Throughput of a periodic evolving graph $\mathcal{G}=(V,\mathcal{E})$ is equivalent to the simple static graph $G(\mathcal{G})=(V,E)$, where $V$ is the set of vertices (same as the evolving graph) and $E=\bigcup_{t\in[0,\Gamma)} \mathcal{E}_t$ is the union of edges over one period of time (without any labels). The capacity of an edge $e\in E$ is given by,
\[
\hat{c}(e) = \left(\frac{1-\reconf}{\Gamma}\right)\cdot \sum_{t\in[0,\Gamma)} c_t(e)
\]
where $c_t(e)$ is the capacity of an edge $e\in \mathcal{E}_t$ at time $t$ in the evolving graph.
\end{corollary}

Notice that the simple static graph in Corollary~\ref{cor:reduction} is a weighted simple graph corresponding to the emulated graph (Definition~\ref{def:union-graph}) with the same amount of capacity between any $u,v\in V$ for every $(u,v)\in E$. It is standard in the literature that a multigraph and the corresponding weighted graph have the same max-flow~\cite{newman2004analysis}. This allows us to use the simple static graph in order to obtain the throughput of a periodic evolving graph.  

\section{Revisiting Throughput of Static Topologies}
\label{sec:improved-bound}

\begin{definition}[Average route length]\label{def:arl}
Given a demand matrix $\mathcal{M}$, a static graph $G=(V,E)$ and a flow $F$ that satisfies the demand, the average route length denoted by $\arl(\mathcal{M},F)$ is defined by, 
\[
\arl(\mathcal{M},F) = \sum_{s,d\in V}\sum_{p \in P_{s,d}}  \frac{m_{s,d}}{M} \cdot r_p \cdot \len(p)
\] 
where $M=\sum_{s\in V}\sum_{d\in V} m_{s,d}$; $r_p$ is the fraction of $s\hyphen d$ demand transmitted on the path $p$ and $\len(p)$ is the length of the path~$p$.

\end{definition}

\throughputUpperBound*

\begin{proof}
We restrict our analysis to saturated demand matrices. Let the static topology under consideration be a graph $G=(V^\prime, E)$, where $V^\prime$ is the set of all vertices and $E$ is the set of all edges. A set $V\subseteq V^\prime$ denotes the set of all vertices generating traffic (demand). Specifically, given any set of vertices $V$, a demand matrix $\mathcal{M}$ from Definition~\ref{def:demand-matrix} has elements $m_{u,v}$ such that $\sum_{v\in V} m_{u,v} = c(u)$ for all $u\in V$ where $c(u)$ is the total \emph{physical} bandwidth of the vertex $u$. We denote the capacity of an edge $e$ by $\hat{c}(e)$.

Given a graph $G=(V^\prime,E)$, the set of vertices generating traffic (demand) $V\subseteq V^\prime$, a~demand matrix $\mathcal{M}$ for $V$, and a flow $F$, the relation between the throughput as a function of demand matrix $\mathcal{M}$ is given by Definition~\ref{def:throughput}.
From Definition~\ref{def:flow}, for $F$ to be legal we have that,
\[
\sum_{p\in P} F(p)\cdot \mathbb{I}(e\in p) \le \hat{c}(e) \quad \forall e
\]
\noindent summing over all the edges $e\in E$, we obtain the following:
\begin{equation}\label{eq:flow-capacity-inequality}
\sum_{e\in E} \sum_{p\in P} F(p)\cdot \mathbb{I}(e\in p) \le \sum_{e\in E} \hat{c}(e)
\end{equation}
From Definition~\ref{def:throughput}, a flow $F$ achieves throughput $\theta(\mathcal{M},F)$ if it obeys the following constraint:
\[
\sum_{p\in P_{s,d}} F(p) \ge \theta(\mathcal{M},F)\cdot m_{s,d} \quad \forall s\in V,\ d\in V
\]

Let $r_p = \frac{F(p)}{\sum_{p\in P_{s,d}} F(p) }$ for all $p\in P_{s,d}$ for each $s,d$ pair. We now expand the inequality in Equation~\ref{eq:flow-capacity-inequality} as follows:

\[
\sum_{s\in V}\sum_{d\in V} \sum_{e\in E} \sum_{p\in P_{s,d}} F(p)\cdot \mathbb{I}(e\in p) \le \sum_{e\in E} \hat{c}(e)
\]
\noindent we convert the summation over all edges $e\in E$ to a summation over only edges of each path $e\in p$ and drop the identifier $\mathbb{I}$ without changing the value of the LHS in the above inequality:
\[
\sum_{s\in V}\sum_{d\in V} \sum_{p\in P_{s,d}} \sum_{e\in p} F(p) \le \sum_{e\in E} \hat{c}(e)
\]
\noindent substituting $F(p)=r_p\cdot \sum_{p\in P_{s,d}} F(p)$ and since $\sum_{p\in P_{s,d}} F(p) \ge \theta(\mathcal{M},F)\cdot m_{s,d}$: 
\[
\sum_{s\in V}\sum_{d\in V} \sum_{p\in P_{s,d}} \sum_{e\in p} \theta(\mathcal{M},F)\cdot m_{s,d}\cdot r_p \le \sum_{e\in E} \hat{c}(e)
\]
Since $\theta(\mathcal{M},F)\cdot m_{s,d}\cdot r_p $ is constant for all edges of a given path, the summation over all edges $e\in p$ gives $\len(p)\cdot\theta(\mathcal{M},F)\cdot m_{s,d}\cdot r_p $ where $\len(p)$ denotes the number of edges in the path $p$ or simply the length of the path.
\[
\sum_{s\in V}\sum_{d\in V} \sum_{p\in P_{s,d}} \len(p)\cdot \theta(\mathcal{M},F)\cdot m_{s,d}\cdot r_p \le \sum_{e\in E} \hat{c}(e)
\]
Finally from the inequality, we obtain the throughput upper bound given a demand matrix as follows,
\begin{equation*}
\theta(\mathcal{M},F) \le \frac{\sum_{e\in E} \hat{c}(e)}{\sum_{s\in V}\sum_{d\in V} m_{s,d}\cdot \left(\sum_{p\in P_{s,d}} \len(p) \cdot r_p\right)}
\end{equation*}
where $\sum_{p\in P_{s,d}} \len(p) \cdot r_p$ is conceptually the average route length from $s$ to $d$. Let $M=\sum_{s\in V}\sum_{d\in V} m_{s,d}$ and $\hat{C}=\sum_{e\in E} \hat{c}(e)$,
\begin{equation}
\theta(\mathcal{M},F) \le \frac{\hat{C}}{M\cdot \arl(\mathcal{M},F)}
\end{equation}
where $\arl(\mathcal{M},F) = \sum_{s,d\in V}\sum_{p \in P_{s,d}}  \frac{m_{s,d}}{M} \cdot r_p \cdot \len(p)$ is the \emph{average route length} for $\mathcal{M}$ and $F$.

\noindent Given a demand matrix, the above inequality varies based on the flow $F$. In order to maximize throughput, we rewrite the above inequality as a maximum over possible flow $F$.
\begin{equation}
\theta(\mathcal{M}) = \max_{F} \frac{\hat{C}}{M\cdot \arl(\mathcal{M},F)}
\end{equation}
The throughput under a worst-case demand matrix is then the minimum $\theta(\mathcal{M})$ over the set of all saturated demand matrices.
\begin{equation}
\theta^* = \min_{\mathcal{M}} \max_{F} \frac{\hat{C}}{M\cdot \arl(\mathcal{M},F)}
\end{equation}
\end{proof}

We note that providing a \emph{lower bound} for the $\arl(\mathcal{M},F)$, even when $M$ and/or $F$ are not known would provide an upper bound for $\theta^*$. One such example is to take the shortest paths in the topology, as was done in~\cite{tubSigcomm2021}. But this will not always lead to a tight bound (see Appendix~\ref{sec:tub-limitations}), for example in the complete graph $K_n$. While efficiently computing the throughput upper bound is an on-going research~\cite{tubSigcomm2021,jyothi2016measuring,highthroughputSingla}, Theorem~\ref{th:throughput-upper-bound} shows that computing throughput essentially boils down to computing average route lengths.

\section{Analysis: Delay \& Buffer Requirements}\label{sec:buffer-analysis}
In this section, we determine the amount of buffer required at each node in order to achieve a throughput $\theta(\mathcal{M})$ which reveals an interesting relation between throughput and buffer in periodic reconfigurable topologies. We first formally define the delay of a temporal path for a better understanding later on the flow that needs to  be ``stored'' at intermediate nodes.

\begin{definition}[Temporal path delay]\label{def:time-pathDelay}
    Given a periodic evolving graph $\mathcal{G}$ and the set of all temporal paths $\mathcal{P}$, the delay of a temporal path $\delta \in \mathcal{P}$ denoted by $\delay(\delta)$ is time it takes when the first edge $e_1 \in \delta$ is accessed until the time the flow reaches the destination on the last edge $e_n \in \delta$. Formally, for a path $\delta \in \mathcal{P}$, if $\left\langle (e_1,t_1), (e_2,t_2) \ ...\  (e_\length,t_\length) \right\rangle$ is the sequence of edge-time values, then $\delay(\delta) = (t_n-t_1+1)\cdot \Delta + (\Gamma-1)\cdot \Delta$ where $\Delta$ is the absolute time of each timeslot and $\Gamma$ is the period of the evolving graph.
\end{definition}

Clearly, the delay is lower bounded by $\len(\delta)\cdot \Delta + (\Gamma-1)\cdot\Delta $ since it takes $\Delta $ (timeslot) amount of time to send over each edge $e\in \delta$ and there is always an inherent delay of $(\Gamma-1)$ timeslots due to the periodic nature. Similarly, the delay is upper bounded by $\len(\delta)\cdot \Gamma \cdot \Delta$ since the delay between accessing each edge along a path can be at most $\Gamma$ timeslots. 

We first, define \emph{average route delay} similar to average route length (Definition~\ref{def:arl}, which plays a key role in the analysis of the maximum latency and the required buffer at each node.

\begin{definition}[Average route delay]\label{def:ard}
Given a demand matrix $\mathcal{M}$, a periodic evolving graph $\mathcal{G}$ and a flow $\mathcal{F}$ that satisfies the demand, the average route delay denoted by $\ard(\mathcal{M},\mathcal{F})$ is defined by, 
\[
\ard(\mathcal{M},\mathcal{F}) = \sum_{s,d\in V} \frac{m_{s,d}}{M} \cdot L_{s,d}
\]
where $M=\sum_{s\in V}\sum_{d\in V} m_{s,d}$; $L_{s,d} = \sum_{\delta \in \mathcal{P}^0_{s,d}} r_\delta \cdot L(\delta)$ is the $s\hyphen d$ delay; $r_\delta$ is the fraction of $s\hyphen d$ demand transmitted on the temporal path $\delta$ and $\delay(\delta)$ is the delay of the temporal path~$\delta$.
\end{definition}

We now state the lower bound for the maximum delay in a periodic evolving graph.

\delayTheorem*

\begin{proof}
Based on Definition~\ref{def:time-throughput}, for every $s\hyphen d$ source-destination pair, the source generates $\theta(\mathcal{M},\mathcal{F})\cdot m_{s,d}$ flow on average in bits per second in every period. From Definition~\ref{def:ard}, $L_{s,d}$ denote the average delay between a source $s$ and destination $d$. Then the destination only receives its first data at $L_{s,d}$ time later, implying that the source has already generated $\theta(\mathcal{M},\mathcal{F})\cdot m_{s,d}\cdot (L_{s,d})$ amount of data. From then on, the source transmits $\theta(\mathcal{M},\mathcal{F})\cdot m_{s,d}$ average flow in bits per second and the destination receives $\theta(\mathcal{M},\mathcal{F})\cdot m_{s,d}$ average flow in bits per second in every period. Due to conservation of flow, for every source-destination pair, the data generated until $L_{s,d}$ delay \ie $\theta(\mathcal{M},\mathcal{F})\cdot m_{s,d}\cdot (L_{s,d})$ amount of data, circulates (moves between source-destination) in the network in every period. As a result, the capacity consumed by each $s\hyphen d$ flow is $\theta(\mathcal{M},\mathcal{F})\cdot m_{s,d}\cdot \frac{L_{s,d}}{\Gamma \cdot \Delta}$. The total capacity utilized is then $\sum_{s\in V} \sum_{d\in V} \theta(\mathcal{M},\mathcal{F})\cdot m_{s,d}\cdot \frac{L_{s,d}}{\Gamma \cdot \Delta}$. However, the total utilized capacity can also be written as $\theta(\mathcal{M},\mathcal{F})\cdot M \cdot \arl(\mathcal{M},\mathcal{F})$, where $\theta(\mathcal{M},\mathcal{F})\cdot M$ is the total bits per second generated by the sources and $\arl(\mathcal{M},\mathcal{F})$ is the average route length. Equating the above two, we have that,
\[
\sum_{s\in V} \sum_{d\in V} \theta(\mathcal{M},\mathcal{F})\cdot m_{s,d}\cdot \frac{L_{s,d}}{\Gamma \cdot \Delta} = \theta(\mathcal{M},\mathcal{F})\cdot M \cdot \arl(\mathcal{M},\mathcal{F})
\]
\[
\frac{1}{M}\sum_{s\in V} \sum_{d\in V} m_{s,d}\cdot L_{s,d} = \ard(\mathcal{M},\mathcal{F}) = \arl(\mathcal{M},\mathcal{F}) \cdot \Gamma \cdot \Delta
\]
Since the maximum delay $L_{max}$ is atleast the average delay,
\begin{align*}
L_{max} \ge \ard(\mathcal{M},\mathcal{F}) = \arl(\mathcal{M},\mathcal{F})\cdot \Gamma\cdot \Delta
\end{align*}
The period $\Gamma$ is at least $\frac{d}{\NU}$ timeslots since the node degree is limited to the number of uplinks $\NU$ in each timeslot. Further, since the total capacity $\hat{C}$ equals the total demand $M$ for saturated demand matrices; and using the relation between $\arl$ and $\theta^*$ from Theorem~\ref{th:throughput-upper-bound}, for a worst-case demand matrix we obtain the following:
\[
L_{max} \ge \Omega\left(\frac{d\cdot \Delta}{\NU\cdot \theta^*}\right) 
\]
\end{proof}

Notice from Definition~\ref{def:time-path} and Definition~\ref{def:time-flow} that it requires to ``store'' flow for a certain time and ``forward'' the flow when the next edge along a path is available. As a result, every node in the evolving graph requires certain amount of buffer space to store and forward flow on temporal paths in the evolving graph. 

In the following, we extend our model to capture the required buffer space at each node. 

\bufferBound*
\begin{proof}
Let $B(u)$ be the amount of available (and used) buffer at a node $u\in V$. Then at any time $T\ge 0$, the difference between the total flow arrived and departed from node $u$ (except the flow originating and terminating at $u$) is the amount of flow stored at $u$, formally expressed below. For simplicity, we set $\Delta_t = \Delta -\Delta_r$ and $\frac{\Delta_t}{\Delta} = 1 - \frac{\Delta_r}{\Delta} = 1 - \Delta_u$. We denote the set of incoming (outgoing) edges of a node $u$ at any time $t$ by $\mathcal{E}_t^-(u)$ ($\mathcal{E}_t^+(u)$).
\begin{align}
& B(u)\ge \nonumber \\
&\displaystyle \sum_{t=0}^T \sum_{s,d\in V\backslash \{u\}} \sum_{\delta\in \mathcal{P}_{s,d}} \left( \sum_{e\in \mathcal{E}_t^-(u)} \mathcal{F}(\delta,e,t) - \sum_{e\in \mathcal{E}_t^+(u)} \mathcal{F}(\delta,e,t) \right) \cdot \Delta \label{eq:bufferAtU}
\end{align}
For simplicity, we define the sum of all flows arriving and departing from a node $u$ at any time $t$ as follows:
\[
R^-_t(u) = \sum_{s,d \in V} \sum_{\delta \in \mathcal{P}_{s,d}} \sum_{e\in\mathcal{E}^-_t(u)} \mathcal{F}(\delta,e,t)
\]
\[
R^+_t(u) = \sum_{s,d \in V} \sum_{\delta \in \mathcal{P}_{s,d}} \sum_{e\in\mathcal{E}^+_t(u)} \mathcal{F}(\delta,e,t)
\]
Notice that at any time $t$, $\sum_{u\in V} R^-_t(u) = \sum_{u\in V} R^+_t(u)$. Since we take the summation over all nodes, for every outgoing flow on an edge, there is a corresponding equal flow incoming at the other end of the edge.

\noindent We now simplify Equation~\ref{eq:bufferAtU} using the above notation. Note that $\sum_{s\in V\backslash \{ u\}} \sum_{\delta \in \mathcal{P}_{s,u}} \sum_{e\in E^-(u)} \mathcal{F}(\delta,e,t)$ is same as $\sum_{s\in V} \sum_{\delta \in \mathcal{P}_{s,u}} \sum_{e\in E^-(u)} \mathcal{F}(\delta,e,t)$ since $\mathcal{P}_{u,u}$ is a null set.
\begin{align*}
B(u) \ge & \sum_{t\in[0,T]} \left( R_t^-(u) -  \sum_{s\in V} \sum_{\delta\in \mathcal{P}_{s,u}} \sum_{e\in E^-(u)} \mathcal{F}(\delta,e,t) \right)\cdot \Delta \\
-& \sum_{t\in[0,T]} \left(R^+_t(u)-\sum_{d\in V} \sum_{\delta\in \mathcal{P}_{u,d}}  \sum_{e\in E^+(u)} \mathcal{F}(\delta,e,t) \right) \cdot \Delta
\end{align*}

\noindent Summing over all the nodes $u\in V$, we obtain the following,
\begin{align*}
\sum_{u\in V} B(u) \ge & \overbrace{\sum_{u\in V} \left(\sum_{t\in[0,T]} \sum_{d\in V} \sum_{\delta\in \mathcal{P}_{u,d}}  \sum_{e\in E^+(u)} \mathcal{F}(\delta,e,t) \right) \cdot \Delta}^{\text{Data sent out from each source $u\in V$}} \nonumber \\
-& \underbrace{\sum_{u\in V}  \left( \sum_{t\in[0,T]} \sum_{s\in V} \sum_{\delta\in \mathcal{P}_{s,u}} \sum_{e\in E^-(u)} \mathcal{F}(\delta,e,t) \right) \cdot \Delta}_{\text{Data received at each destination $u\in V$}}
\end{align*}
Let $\delay(\delta)$ denote the temporal path delay. $\mathcal{F}(\delta)\cdot \frac{\delay(\delta)\cdot(1-\reconf)}{\Gamma}$ is the amount of flow sent on a periodic path before the first data arrives at the destination \ie $\mathcal{F}(\delta)\cdot \Delta\cdot (1-\reconf)$ is the data sent over $\Gamma$ timeslots ($\Gamma\cdot\Delta$ time) and $\mathcal{F}(\delta)\cdot \frac{\delay(\delta)\cdot(1-\reconf)}{\Gamma}$ is the data sent over $\delay(\delta)$ time. For each $s\hyphen d$ source-destination, the total data received at the destination $d$ until time $T$ is the total data sent from source $s$ until time $T$ minus the total data sent from source $s$ before the destination $d$ receives its first data.
\begin{align*}
\sum_{u\in V} B(u) \ge & \sum_{u \in V} \left( T\cdot \Delta \cdot \sum_{d\in V} \theta(\mathcal{M},\mathcal{F})\cdot m_{u,d}\right) \\
& - \sum_{u \in V} \left( T\cdot \Delta \cdot \sum_{s\in V} \theta(\mathcal{M},\mathcal{F})\cdot m_{s,u}\right) \nonumber \\
+& \sum_{u \in V} \sum_{s\in V}\sum_{\delta \in \mathcal{P}^0_{s,u}}  \frac{\mathcal{F}(\delta)\cdot\delay(\delta)\cdot(1-\reconf)}{\Gamma}
\end{align*}
Finally, the above inequality reduces to,
\begin{align*}
\sum_{u\in V} B(u) \ge & \sum_{s \in V} \sum_{d\in V} \sum_{\delta \in \mathcal{P}^0_{s,d}} \mathcal{F}(\delta)\cdot \frac{\delay(\delta)\cdot(1-\reconf)}{\Gamma}
\end{align*}
We expand the above summation $\sum_{\delta \in \mathcal{P}^0_{s,d}} \mathcal{F(\delta)}$ by multiplying and dividing by $\theta(\mathcal{M},\mathcal{F})\cdot m_{s,d}$ where $\theta(\mathcal{M},\mathcal{F})\cdot m_{s,d} = \sum_{\delta \in \mathcal{P}^0_{s,d}} \frac{\mathcal{F}(\delta)\cdot(1-\reconf)}{\Gamma}$ from Definition~\ref{def:time-throughput}.
\begin{align*}
& \sum_{u\in V} B(u) \ge \nonumber \\
& \sum_{s \in V} \sum_{d\in V}\left( \frac{\displaystyle\theta(\mathcal{M},\mathcal{F}) \cdot m_{s,d} \cdot \Gamma \cdot \sum_{\delta \in \mathcal{P}^0_{s,d}} \mathcal{F}(\delta)\cdot \frac{\delay(\delta)\cdot(1-\reconf)}{\Gamma}}{\displaystyle \sum_{\delta \in \mathcal{P}^0_{s,d}} \mathcal{F}(\delta)\cdot(1-\reconf)}\right)
\end{align*}

\noindent Let $r_{\delta}$ denote the fraction of temporal flow on a path $\delta\in P^0_{s,d}$ \ie $r_{\delta} = \frac{\mathcal{F}(\delta)}{\sum_{\delta \in \mathcal{P}^0_{s,d}} \mathcal{F}(\delta)}$.

\begin{align*}
\sum_{u\in V} B(u) \ge & \theta(\mathcal{M},\mathcal{F}) \cdot \sum_{s \in V} \sum_{d\in V}\left( m_{s,d} \cdot \sum_{\delta \in \mathcal{P}^0_{s,d}} r_\delta\cdot \delay(\delta)\right)
\end{align*}

\noindent Let $ M=\sum_{s\in V}\sum_{d\in V} m_{s,d}$ and the \emph{average route delay} for $\mathcal{M}$ and $\mathcal{F}$ is denoted by $ \ard(\mathcal{M},\mathcal{F}) = \sum_{s,d\in V}\sum_{\delta\in \mathcal{P}_{s,d}}  \frac{m_{s,d}}{M} \cdot r_\delta \cdot \delay(\delta)$. The total buffer in the network is $\hat{B}=\sum_{u \in V} B(u)$. We obtain the following,
\[
\hat{B} \ge (\theta(\mathcal{M},\mathcal{F})\cdot M)\cdot \ard(\mathcal{M},\mathcal{F})
\]

\end{proof}

\section{Analysis: \name}\label{sec:mars-analyis}

\medskip
We now restate and prove the throughput optimal degree $d$ of \name, given the delay requirement $L$.
\optdegreeTheorem*
\begin{proof}
We begin by equating the delay of \name and the desired delay $L$.
\[
L = \frac{2\cdot \log_d(\NT) \cdot d \cdot \Delta}{\NU}
\]
Rearranging the terms in the above equation, we obtain the following,
\[
\frac{\ln(d)}{d} = \frac{2\cdot \ln(\NT) \cdot \Delta}{\NU\cdot L}
\]
\[
\frac{1}{d}\cdot \ln\left(\frac{1}{d}\right) = - \frac{2\cdot \ln(\NT) \cdot \Delta}{\NU\cdot L}
\]

For simplicity, let $- \frac{2\cdot \ln(\NT) \cdot \Delta}{\NU\cdot L}$ be some constant $k$. 
\[
\frac{1}{d}\cdot \ln\left(\frac{1}{d}\right) = k
\]
The above equation is of the form $y\cdot\ln(y) = k$ whose solution is $y=e^{\mathcal{W}(k)}$ where $\mathcal{W}$ is the lambert $W$ function~\cite{corless1996lambertw}.
\[
\frac{1}{d} = e^{\mathcal{W}(k)}
\]
Substituting $k$ in the above, we obtain the following.
\[
d = e^{-\mathcal{W}(k)} = e^{-\mathcal{W}\left(- \frac{2\cdot \ln(\NT)\cdot \Delta}{\NU\cdot L} \right)}
\]
Here, if $k<-\frac{1}{e}$, there exists no real solution. However, we show that $k$ is always greater than $-\frac{1}{e}$ \ie $k>-\frac{1}{e}$
\[
k = - \frac{2\cdot \ln(\NT) \cdot \Delta}{\NU\cdot L} \ge - \frac{2\cdot \ln(\NT)}{\NU} \ge - 2 \cdot \ln(\NT) > -\frac{1}{e}
\]
The above inequalities hold since the latency $L$ is at least one timeslot \ie $L\ge \Delta$; the topology consists of at least two ToRs \ie $\NT\ge 2$ and $\ln(2) > \frac{1}{e}$.

If $-\frac{1}{e}\le k < 0$, there exists two real solutions for the degree~$d$. In this case, we take the the highest value and round it to the nearest integer. Here, choosing the highest value $d$ lowers the average route length  $2\cdot \log_d(\NT)$ and consequently maximizes throughput within the latency requirement.
\end{proof}

\optDegreeBuffer*
\begin{proof}
From Theorem~\ref{th:buffer}, the required buffer is ($\theta(\mathcal{M},\mathcal{F})\cdot M \cdot ARD(\mathcal{M},\mathcal{F})$). Further, from Theorem~\ref{th:delay} we have that the delay is bounded by $\frac{d\cdot \Delta}{\NU\cdot \theta(\mathcal{M},\mathcal{F})}$. Since $M=\NT\cdot \NU \cdot c$ (total demand and capacity), this gives us the buffer requirement of $\NT \cdot c \cdot d \cdot \Delta$ in the whole network. Given the regularity of the \name, each node would then require $c\cdot d\cdot \Delta$ amount of buffer to hold the demand. Now, given the constraint that only $B$ amount of buffer is available at each node, the optimal degree $d$ is then $\frac{B}{c\cdot \Delta}$ rounded to the nearest integer.
\end{proof}

\begin{figure*}
\centering
\begin{subfigure}{0.48\linewidth}
\centering
\includegraphics[width=1\linewidth]{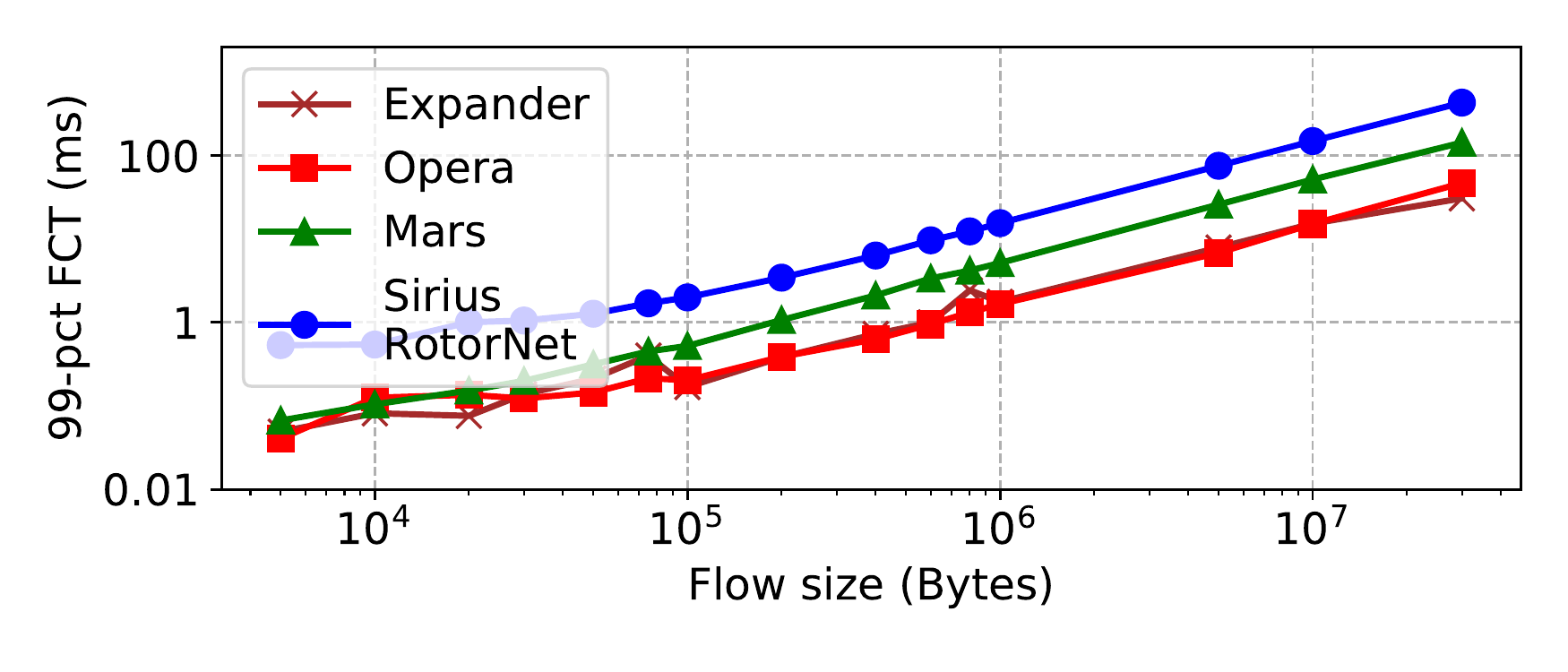}
\subcaption{$1\%$ load}
\label{fig:fcts-0.01-permDatamining}
\end{subfigure}
\begin{subfigure}{0.48\linewidth}
\centering
\includegraphics[width=1\linewidth]{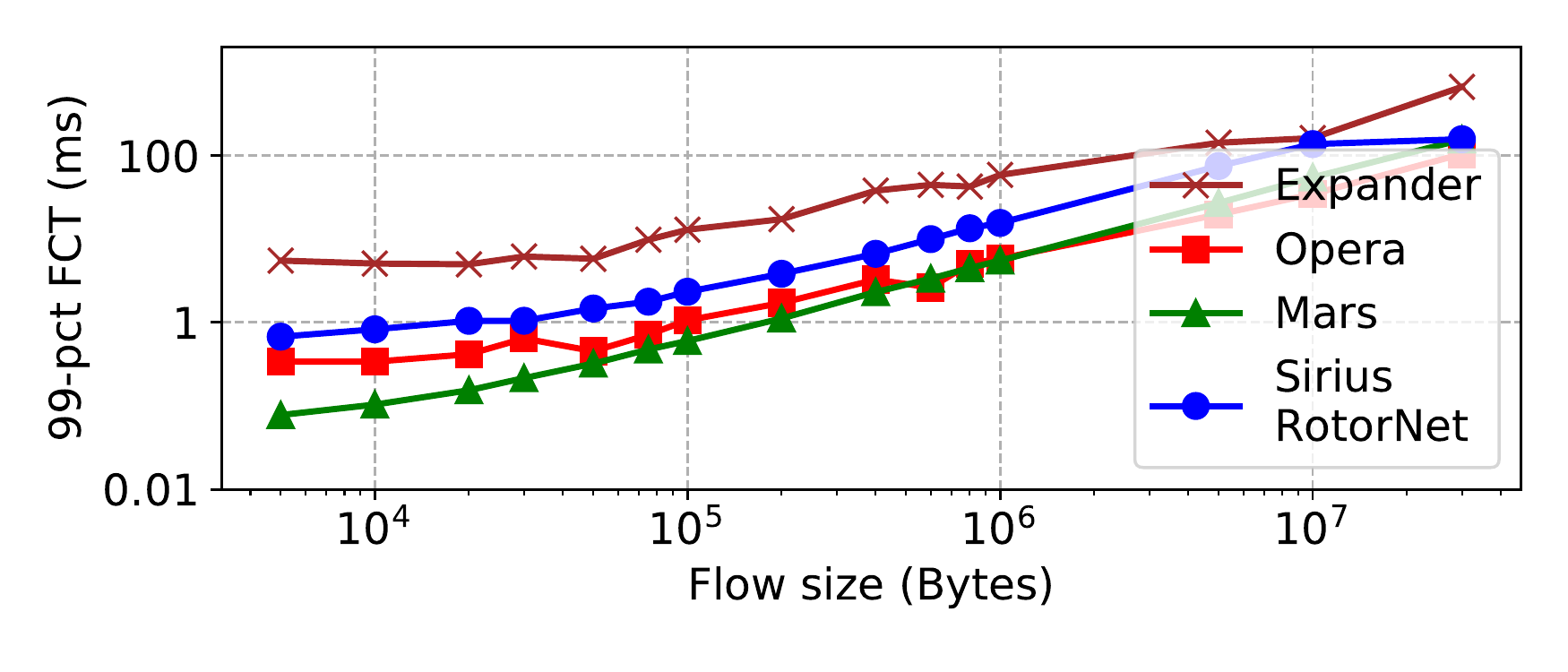}
\subcaption{$5\%$ load}
\label{fig:fcts-0.05-permDatamining}
\end{subfigure}
\begin{subfigure}{0.48\linewidth}
\centering
\includegraphics[width=1\linewidth]{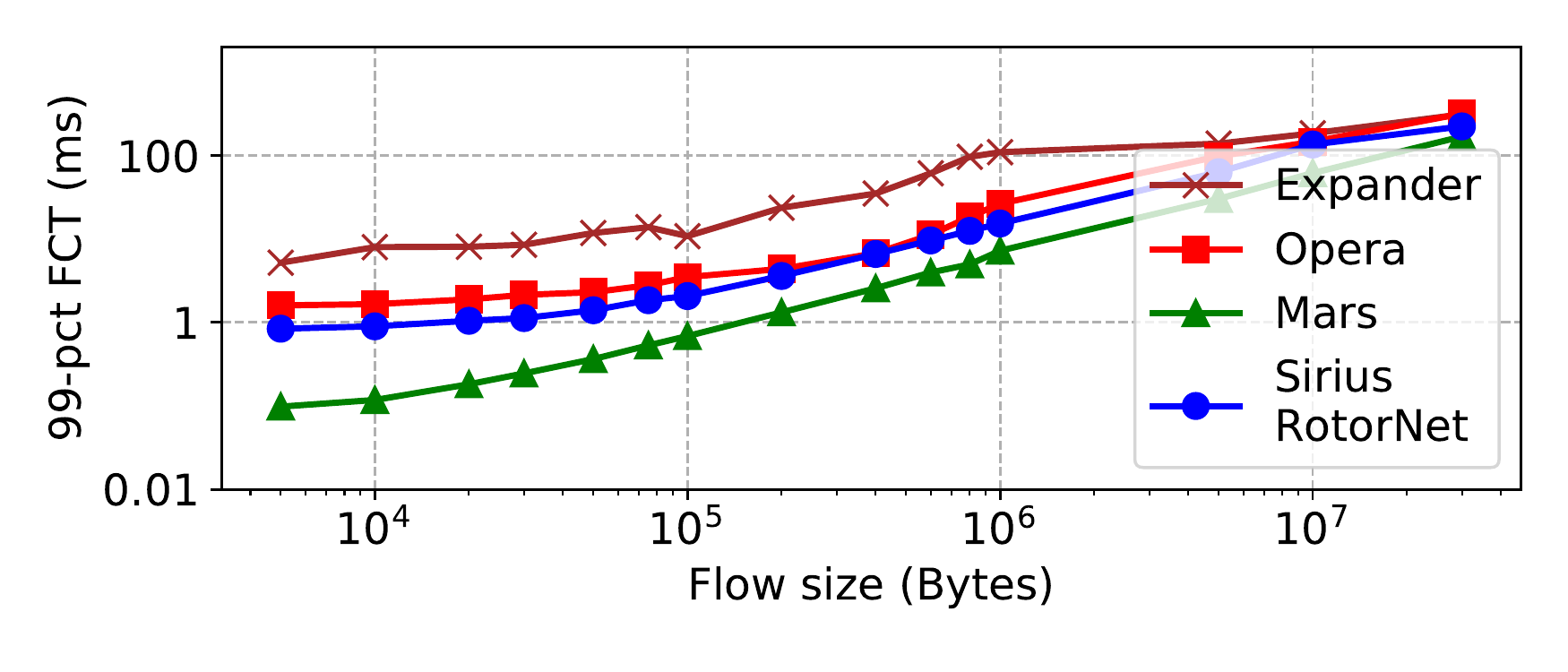}
\subcaption{$10\%$ load}
\label{fig:fcts-0.1-permDatamining}
\end{subfigure}
\begin{subfigure}{0.48\linewidth}
\centering
\includegraphics[width=1\linewidth]{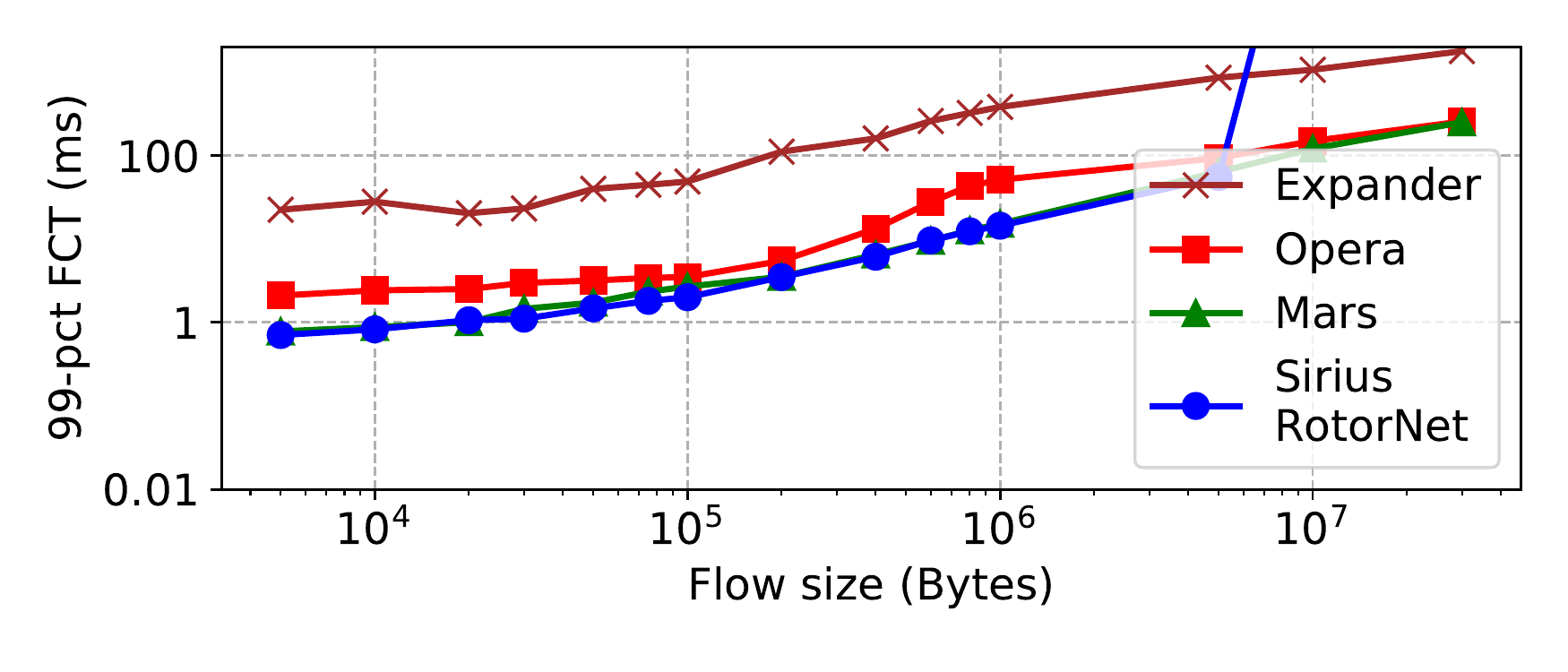}
\subcaption{$20\%$ load}
\label{fig:fcts-0.2-permDatamining}
\end{subfigure}
\caption{$99$-percentile flow completion times for Datamining workload~\cite{alizadeh2013pfabric} under \textbf{random permutation} demand matrix.}
\label{fig:fctspermDatamining}
\end{figure*}

\begin{figure*}
\centering
\begin{subfigure}{0.48\linewidth}
\centering
\includegraphics[width=1\linewidth]{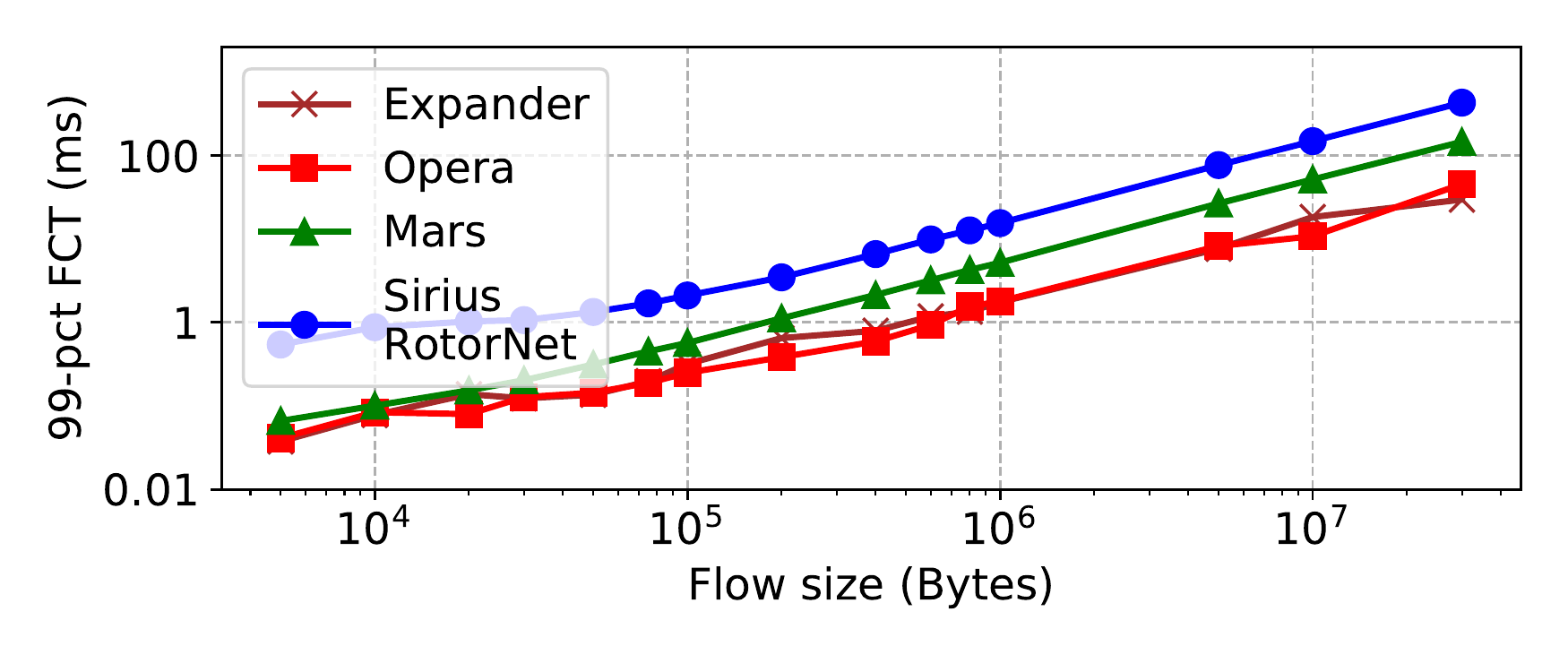}
\subcaption{$1\%$ load}
\label{fig:fcts-0.01-a2aDatamining}
\end{subfigure}
\begin{subfigure}{0.48\linewidth}
\centering
\includegraphics[width=1\linewidth]{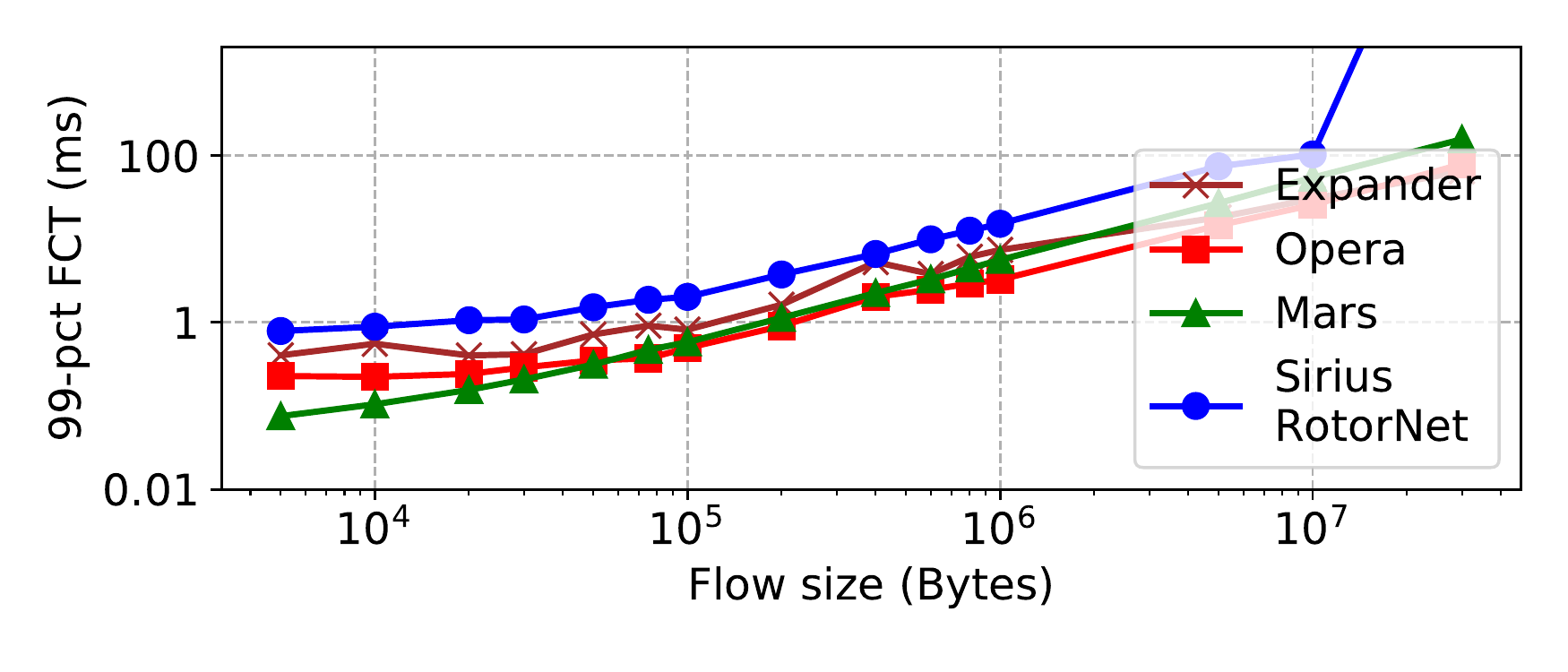}
\subcaption{$5\%$ load}
\label{fig:fcts-0.05-a2aDatamining}
\end{subfigure}
\begin{subfigure}{0.48\linewidth}
\centering
\includegraphics[width=1\linewidth]{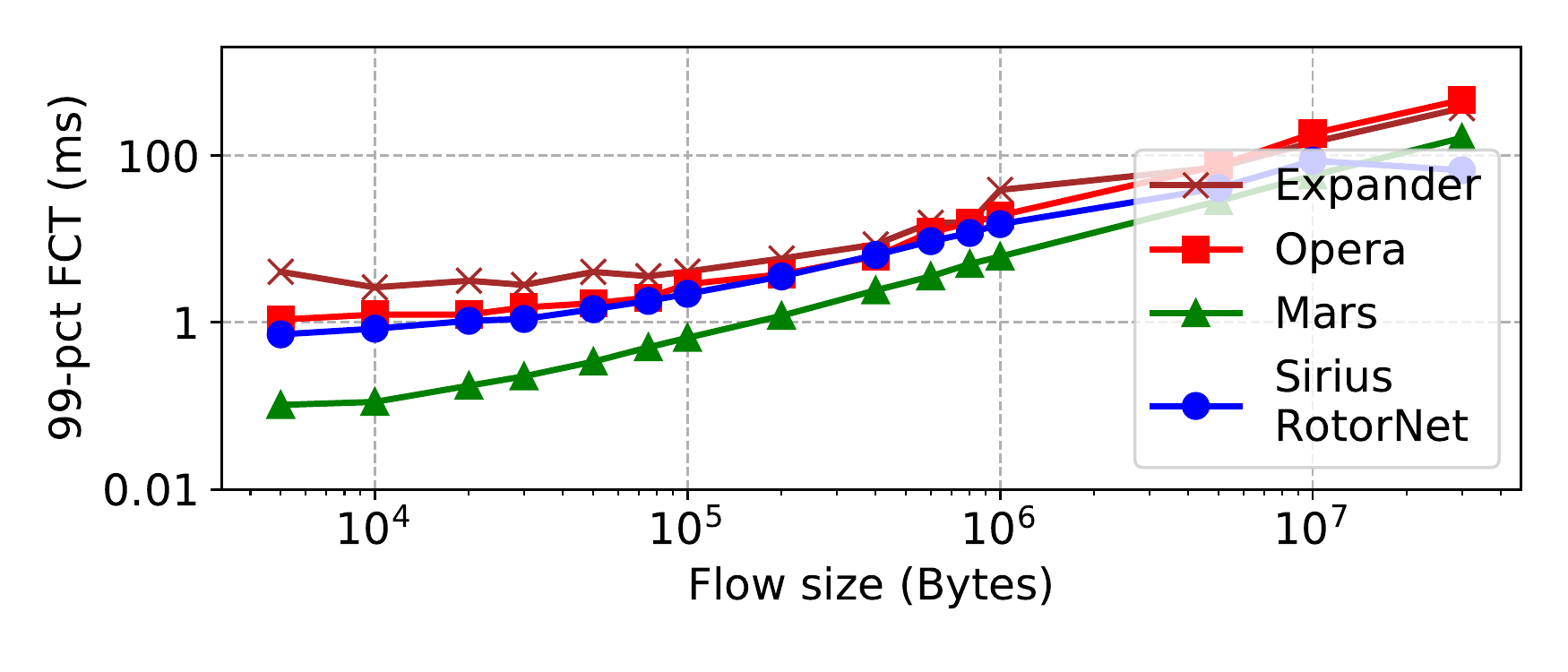}
\subcaption{$10\%$ load}
\label{fig:fcts-0.1-a2aDatamining}
\end{subfigure}
\begin{subfigure}{0.48\linewidth}
\centering
\includegraphics[width=1\linewidth]{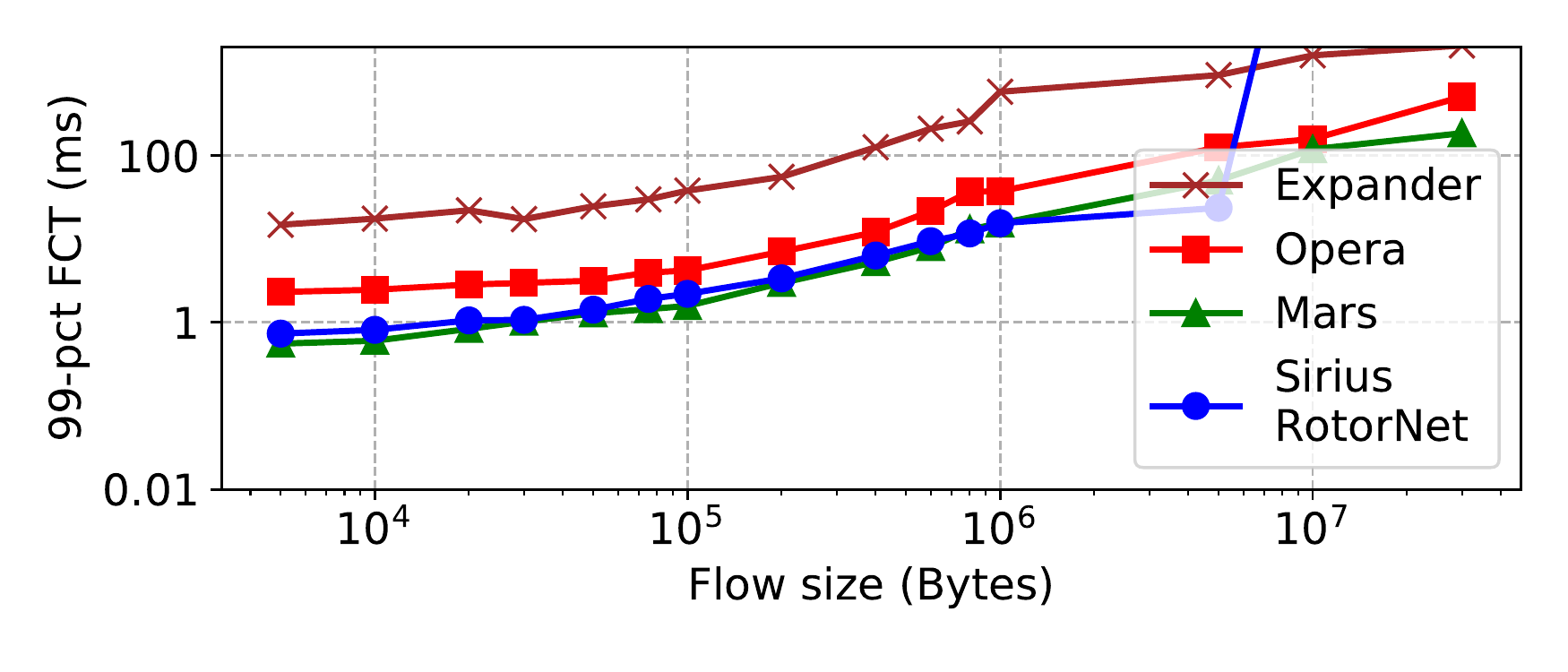}
\subcaption{$20\%$ load}
\label{fig:fcts-0.2-a2aDatamining}
\end{subfigure}
\caption{$99$-percentile flow completion times for Datamining workload~\cite{alizadeh2013pfabric} under \textbf{all-to-all} demand matrix.}
\label{fig:fctsa2aDatamining}
\end{figure*}

\label{LastPage}

\end{document}